\documentclass[twocolumn,twocolappendix]{aastex631}
\usepackage{color}
\usepackage{graphicx}
\usepackage{amsmath}    
\usepackage{amssymb}
\usepackage{mathrsfs}
\usepackage{times}
\usepackage{xcolor}
\usepackage{xspace}
\usepackage{datetime}
\usepackage{version}    

\usepackage{natbib}

\newcommand{\simgt}{\lower.5ex\hbox{$\; \buildrel > \over \sim \;$}}
\newcommand{\simlt}{\lower.5ex\hbox{$\; \buildrel < \over \sim \;$}}

\newcommand{\Om}{\Omega_\mathrm{m}}
\newcommand{\OL}{\Omega_\Lambda}
\newcommand{\Ob}{\Omega_\mathrm{b}}
\newcommand{\zs}{z_s}
\newcommand{\zl}{z_l}
\newcommand{\zeff}{z_\mathrm{eff}}
\newcommand{\percent}{\ensuremath{\%}}
\newcommand{\Mproj}{M_\mathrm{2D}}
\newcommand{\SigmaCrit}{\Sigma_\mathrm{cr}}
\newcommand{\Rein}{\theta_\mathrm{Ein}}
\newcommand{\Npix}{N_\mathrm{pix}}
\newcommand{\Nbin}{N_\mathrm{bin}}
\newcommand{\slim}{s}
\newcommand{\mcut}{m_\mathrm{cut}}
\newcommand{\Mhse}{M_\mathrm{HE}}
\newcommand{\fgas}{f_\mathrm{gas}}
\newcommand{\kpch}{\,h^{-1}\,\mathrm{kpc}}
\newcommand{\Mpch}{\,h^{-1}\,\mathrm{Mpc}}
\newcommand{\Mpc}{\,\mathrm{Mpc}}
\newcommand{\Msunh}{\,h^{-1}\,M_\odot}
\newcommand{\Msun}{\,M_\odot}
\newcommand{\CBI}{$C_\mathrm{BI}$\xspace}
\newcommand{\SBI}{$S_\mathrm{BI}$\xspace}
\newcommand{\LCDM}{$\Lambda$CDM\xspace}
\newcommand{\Chandra}{Chandra\xspace}
\newcommand{\HST}{HST\xspace}
\newcommand{\Planck}{Planck\xspace}
\newcommand{\BRz}{$BR_\mathrm{C}z^\prime$\xspace}
\newcommand{\LePhare}{\textsc{lephare}\xspace}

\newcommand{\FIG}{Figs/}

\def\bc{\mbox{\boldmath $c$}}
\def\btheta{\mbox{\boldmath $\theta$}} 
\def\bnabla{\mbox{\boldmath $\nabla$}}
\def\bSigma{\mbox{\boldmath $\Sigma$}} 
\def\blambda{\mbox{\boldmath $\lambda$}}
\def\bm{\mbox{\boldmath $m$}}
\def\bd{\mbox{\boldmath $d$}}
\def\bp{\mbox{\boldmath $p$}}
\def\bmprof{\mbox{\boldmath $m_\mathrm{1D}$}}
\def\ge{\geqslant}
\def\le{\leqslant}

\shorttitle{Line-of-sight Elongation and Hydrostatic Mass Bias of A370}
\shortauthors{Umetsu et~al.}

\begin{document}
\title{Line-of-sight Elongation and Hydrostatic Mass Bias of the Frontier Fields Galaxy Cluster Abell~370} 

\correspondingauthor{Keiichi Umetsu}
\email{keiichi@asiaa.sinica.edu.tw}

\author[0000-0002-7196-4822]{Keiichi Umetsu}
\affiliation{Academia Sinica Institute of Astronomy and Astrophysics (ASIAA), No.~1, Section~4, Roosevelt Road, Taipei 10617, Taiwan}
\author[0000-0001-6252-7922]{Shutaro Ueda}
\affiliation{Academia Sinica Institute of Astronomy and Astrophysics (ASIAA), No.~1, Section~4, Roosevelt Road, Taipei 10617, Taiwan}
\author[0000-0001-5615-4904]{Bau-Ching Hsieh}
\affiliation{Academia Sinica Institute of Astronomy and Astrophysics (ASIAA), No.~1, Section~4, Roosevelt Road, Taipei 10617, Taiwan}
\author[0000-0001-6342-9662]{Mario Nonino}
\affiliation{INAF-Osservatorio Astronomico di Trieste, Via G.~B. Tiepolo 11, 34143, Trieste, Italy}
\author[0000-0002-5819-6566]{I-Non Chiu}
\affiliation{Tsung-Dao Lee Institute, and Key Laboratory for Particle Physics,
Astrophysics and Cosmology, Ministry of Education, Shanghai Jiao Tong University, Shanghai 200240, China}
\affiliation{Department of Astronomy, School of Physics and Astronomy, and Shanghai Key Laboratory for Particle Physics and Cosmology, Shanghai Jiao Tong University, Shanghai 200240, China}
\affiliation{Academia Sinica Institute of Astronomy and Astrophysics (ASIAA), No.~1, Section~4, Roosevelt Road, Taipei 10617, Taiwan}
\author[0000-0003-3484-399X]{Masamune Oguri}
\affiliation{Center for Frontier Science, Chiba University, 1-33 Yayoi-cho, Inage-ku, Chiba 263-8522, Japan}
\affiliation{Department of Physics, Graduate School of Science, Chiba University, 1-33 Yayoi-Cho, Inage-Ku, Chiba 263-8522, Japan}
\author[0000-0002-4227-956X]{Sandor M. Molnar}
\affiliation{Academia Sinica Institute of Astronomy and Astrophysics (ASIAA), No.~1, Section~4, Roosevelt Road, Taipei 10617, Taiwan}
\author[0000-0002-6610-2048]{Anton M. Koekemoer}
\affiliation{Space Telescope Science Institute, 3700 San Martin Drive, Baltimore, MD~21218, USA}
\author[0000-0002-6724-833X]{Sut-Ieng Tam}
\affiliation{Academia Sinica Institute of Astronomy and Astrophysics (ASIAA), No.~1, Section~4, Roosevelt Road, Taipei 10617, Taiwan}

\begin{abstract}
We present a detailed weak-lensing and X-ray study of the Frontier Fields galaxy cluster Abell~370, one of the most massive known lenses on the sky, using wide-field \BRz Subaru/Sprime-Cam and \Chandra X-ray observations. By combining two-dimensional (2D) shear and azimuthally averaged magnification constraints derived from Subaru data, we perform a lensing mass reconstruction in a free-form manner, which allows us to determine both radial structure and 2D morphology of the cluster mass distribution. In a triaxial framework assuming a Navarro--Frenk--White density profile, we constrain the intrinsic structure and geometry of the cluster halo by forward modeling the reconstructed mass map. We obtain a halo mass $M_{200}=(1.54\pm 0.29)\times 10^{15}\Msunh$, a halo concentration $c_{200}=5.27\pm 1.28$, and a minor--major axis ratio $q_a=0.62\pm 0.23$ with uninformative priors. Using a prior on the line-of-sight alignment of the halo major axis derived from binary merger simulations constrained by multi-probe observations, we find that the data favor a more prolate geometry with lower mass and lower concentration. From triaxial lens modeling with the line-of-sight prior, we find a spherically enclosed gas mass fraction of $\fgas=(8.4\pm 1.0)\percent$ at $0.7\Mpch\sim 0.7r_{500}$.  When compared to the hydrostatic mass estimate ($\Mhse$) from \Chandra observations, our triaxial weak-lensing analysis yields spherically enclosed mass ratios of $1-b\equiv \Mhse/M_\mathrm{WL}=0.56\pm 0.09$ and $0.51\pm 0.09$ at $0.7\Mpch$ with and without using the line-of-sight prior, respectively. Since the cluster is in a highly disturbed dynamical state, this represents the likely maximum level of hydrostatic bias in galaxy clusters. 
\end{abstract}
 



\section{Introduction}
\label{sec:intro}

Galaxy clusters can provide a range of valuable information from the physics driving structure formation to the nature of dark matter and dark energy. Their matter content reflects that of the universe: $\sim 85\percent$ dark matter and $\sim 15\percent$ baryons, with $\sim 90\percent$ of the baryons residing in the hot intracluster medium (ICM). Determining the evolution of the abundance of rare massive clusters provides powerful cosmological constraints, especially on the matter density parameter, $\Om$, and the amplitude of linear density fluctuations, $\sigma_8$ \citep[e.g., see][and references therein]{Mantz2015}. Conversely, an accurate determination of the total mass of galaxy clusters using direct mass probes, such as weak gravitational lensing, is essential to harness the full potential of cluster cosmology \citep[e.g.,][]{Pratt2019,Chiu2021efeds,Tam2022}.

In the context of the standard $\Lambda$ cold dark matter (\LCDM) model, galaxy clusters are non-spherical in shape and better approximated as triaxial halos \citep{2002ApJ...574..538J}, with a preference for prolateness over oblateness and preferentially aligned with surrounding filaments \citep{Bett2007}. Cluster-scale halos can be characterized as triaxial ellipsoids with a typical minor-to-major axis ratio of $0.4$--$0.5$ \citep{Bonamigo2015}, where more massive objects tend to be more prolate. Thus, while the intrinsic shape and orientation of galaxy clusters contain unique cosmological information \citep{Okumura+Taruya2020}, they can also introduce significant scatter and bias in cluster mass estimates due to the unknown orientation of cluster halos. In particular, gravitational lensing is sensitive to such projection effects \citep[e.g.,][]{Becker+Kravtsov2011}.

According to cosmological $N$-body simulations, ``superlens'' clusters characterized by large Einstein radii ($\Rein\ge 30\arcsec$ for a source redshift of $\zs=2$) represent the most lensing-biased population of clusters, with their major axis preferentially aligned with the observer's line of sight \citep{Hennawi2007,Oguri+Blandford2009,Meneghetti+2010MARENOSTRUM,Meneghetti+2011}. A statistical bias in favor of prolate structure pointed close to the observer arises, because such a halo geometry can boost the projected mass density and hence the lensing signal. In particular, major mergers of two clusters colliding nearly along the line of sight provide a possible mechanism for producing a powerful superlens \citep[e.g., see the case of Cl0024+1654;][]{Umetsu+2010CL0024}.

Abell~370 (hereafter A370; a.k.a. PSZ2~G172.98$-$53.55) at $z=0.375$ is known as a prominent strong lens with an Einstein radius of $\Rein\approx 34\arcsec$ (for $\zs=2$; see Table~\ref{tab:cluster}) and is the first galaxy cluster where gravitational lensing has been observed in the form of a giant luminous arc \citep{1987A&A...172L..14S,1988A&A...191L..19S}. A370 is also among the most massive clusters based on weak gravitational lensing, with an estimated virial mass of $M_\mathrm{vir}\sim 2\times 10^{15}\Msunh$ \citep[][all relevant symbols are defined at the end of this section]{Umetsu+2011,Hoekstra2015CCCP}. Because of its large projected mass and exceptional lensing strength, A370 was selected as one of the six Hubble Frontier Fields \citep{Lotz2017hff} and has recently been targeted by the Beyond Ultra-deep Frontier Fields and Legacy Observations \citep[BUFFALO;][]{Buffalo} with the Hubble Space Telescope (\HST), which expands the area coverage of the Frontier Fields in optical and near-infrared pass bands.

Lensing studies of A370 reveal a bimodal mass distribution in the core elongated in the north--south direction \citep[e.g.,][]{Kneib1993,Umetsu1999,Richard2010a370,Diego2018a370,Lagattuta2017,Lagattuta2019a370,Ghosh2021}. \citet{Strait2018} combined strong and weak lensing constraints from Frontier Fields imaging and spectroscopic observations to reconstruct the central mass distribution of A370. Their mass map shows two dominant peaks associated with the two brightest cluster galaxies (BCGs), with the northern peak much less concentrated than the southern one and slightly offset from the stellar mass distribution. These bimodal and offset features are often an indication of recent major merger activity \citep[e.g.,][]{Bradac2008macs0025}.

In contrast to its extreme mass and exceptional lensing properties, A370 is intriguingly faint in both X-ray and Sunyaev--Zel'dovich effect (SZE) signals and does not follow the X-ray/SZE observable--mass scaling relations \citep[see][]{Czakon2015}. The X-ray brightness distribution revealed from \Chandra observations is highly elongated in the north--south direction, showing a disturbed morphology with the brightest X-ray peak located about halfway between the two BCGs \citep{Molnar2020}. The irregular morphology in X-ray emission with a large elongation similar to that of the mass distribution is a strong indication that the cluster is far from hydrostatic equilibrium \citep[][]{Lee+Suto2003}.

Recently, $N$-body hydrodynamical simulations of binary cluster mergers constrained by lensing, X-ray, SZE, and optical spectroscopic observations suggest that A370 is a massive post-major merger viewed after the second core passage in the infalling phase, just before the third core passage \citep{Molnar2020}. In this post-collision phase, the gas has not settled into the gravitational potential of the cluster, which explains the faintness of the X-ray and SZE signals. These results also suggest that the mass distribution of A370 is highly elongated along the current direction of the collision axis, which is oriented close to the line of sight in their best-matching simulation.

In this paper, we present a detailed weak-lensing and X-ray study of A370 using wide-field \BRz imaging taken with Suprime-Cam on the Subaru telescope and high-quality data from the \Chandra X-ray Observatory. The primary aims of this paper are to obtain an accurate inference of the three-dimensional (3D) mass model of A370 from a full triaxial analysis of two-dimensional (2D) weak-lensing data and to determine the level of hydrostatic mass bias and the gas mass fraction as a function of cluster radius. The key for this study is to perform a lensing mass reconstruction in an unbiased manner, from which to constrain both radial structure and 2D morphology of the cluster mass distribution. To this end, we perform an improved joint shear and magnification analysis of 2D Subaru weak-lensing data, revisiting our earlier one-dimensional (1D) work presented in \citet{Umetsu+2011}. Since A370 is extremely massive and in a highly disturbed dynamical state, this analysis will provide a constraint on the likely maximum level of the hydrostatic bias expected in galaxy clusters.

This paper is organized as follows. Section~\ref{sec:method} describes the basic theory of cluster weak lensing and outlines the methodology used to reconstruct the cluster mass distribution. Section~\ref{sec:data} describes details of the Subaru observations, reduction procedures, and weak-lensing analysis. Section~\ref{sec:mrec} presents the results of our mass reconstruction, followed by our triaxial modeling in Section~\ref{sec:model}. Section~\ref{sec:xray} describes the X-ray data analysis. Section~\ref{sec:discussion} compares the weak-lensing and \Chandra mass profiles. Finally, a summary is given in Section~\ref{sec:summary}.

Throughout this paper, we assume a spatially flat \LCDM cosmology with $\Om=0.3$, $\OL=0.7$, and a Hubble constant of $H_0 = 100~h$~km~s$^{-1}$~Mpc$^{-1}$ with $h=0.7$. In this cosmology, $1\arcmin$ corresponds to $216.8\kpch$ at the cluster redshift of $z=0.375$. The reference center of the cluster is taken to be the optical cluster center defined by \citet[][see also \citealt{Buffalo}]{Lotz2017hff}: $\mathrm{R.A.}= \mathrm{02:39:52.9}$, $\mathrm{decl.}=-\mathrm{01:34:36.5}$ (see Table~\ref{tab:cluster}).

We denote the critical density of the universe at a particular redshift $z$ as $\rho_\mathrm{c}(z)=3H^2(z)/(8\pi G)$, with $H(z)$ the Hubble function. We generally denote spherical and projected radii from the cluster center as $r$ and $r_\perp$, respectively, and reserve the symbol $R$ for ellipsoidal cluster radii. We adopt the standard notation $M_\Delta$ (or $M_{\Delta\mathrm{m}}$) to denote the mass enclosed within a sphere of radius $r_\Delta$ (or $r_{\Delta\mathrm{m}}$) within which the mean overdensity equals $\Delta$ (or $\Delta_\mathrm{m}$) times $\rho_\mathrm{c}(z)$ (or the mean background density $\rho_\mathrm{m}(z)$). We compute the virial mass and radius, $M_\mathrm{vir}$ and $r_\mathrm{vir}$, using an expression for $\Delta_\mathrm{vir}$ based on the spherical collapse model \citep{Bryan+Norman1998}. For its ellipsoidal counterpart $R_\Delta$, see Section~\ref{subsec:triNFW}. We use ``$\log$'' to denote the base-10 logarithm and ``$\ln$'' to denote the natural logarithm. All quoted errors are at the $1\sigma$ confidence level unless otherwise stated. The AB magnitude system is used throughout.

\begin{deluxetable}{lc}[tbp]
\centering
\tabletypesize{\footnotesize}
\tablecaption{Properties of the galaxy cluster A370} 
\label{tab:cluster}
\tablehead{ 
 \multicolumn{1}{c}{Parameter} &
 \multicolumn{1}{c}{Value} 
} 
\startdata
ID                                  & A370 \\
Reference center position (J2000.0) & \\
R.A.                        & 02:39:52.9\\
Decl.                       & $-$01:34:36.5\\
X-ray emission centroid (J2000.0)   & \\
R.A.                        & 02:39:53.2\\
Decl.                       & $-$01:34:35.1\\
Redshift                            & $0.375$\\
Velocity dispersion (km~s$^{-1}$)   & $1520\pm 93$\\
X-ray temperature  (keV)            & $8.77^{+0.33}_{-0.34}$\\
Einstein radius ($\arcsec$)         & $33.9\pm 1.1$ for $\zs=2$
\enddata
\tablecomments{
The optical cluster center is at the midpoint of the two BCGs \citep{Lotz2017hff,Buffalo}. Units of right ascension are hours, minutes, and seconds, and units of declination are degrees, arcminutes, and arcseconds. The cluster velocity dispersion is derived from spectroscopic observations of \citet{Lagattuta2022} in the core region of the cluster \citep[see also][]{Lagattuta2019a370}. The X-ray emission centroid is determined from a 2D $\beta$-model fit to \Chandra X-ray observations (see Section~\ref{sec:xray}). The average temperature of the cluster is measured from the \Chandra X-ray spectrum in the radial range $\in [50,500]\kpch$ centered on the X-ray centroid. The Einstein radius is constrained by detailed strong lens modeling by \citet{Kawamata2018}.
}
\end{deluxetable}


\section{Weak-lensing Methodology}
\label{sec:method}

\subsection{Basics of Galaxy--Cluster Weak Lensing}
\label{subsec:basics}

The effects of weak gravitational lensing on background galaxies are characterized by the convergence, $\kappa$, and the shear with spin 2 rotational symmetry, $\gamma=|\gamma|e^{2i\phi_\gamma}$ \citep[for reviews, see][]{2001PhR...340..291B,Umetsu2020rev}. In this work, we closely follow the notation of \citet{Umetsu2020rev}.

The lensing convergence $\kappa$ alone causes an isotropic magnification of galaxy images and it is defined as the surface mass density $\Sigma$ of a lens in units of the critical surface density for gravitational lensing, $\kappa=\Sigma/\SigmaCrit$, where
\begin{equation}
 \begin{aligned}
  \SigmaCrit(\zl, \zs) &= \frac{c^2}{4\pi G D_l(\zl)}\frac{1}{\beta(\zl,\zs)},\\
  \beta(\zl,\zs) &= 
  \begin{cases} 
   D_{ls}(\zl,\zs)/D_s(\zs) & \mathrm{for}~\zs > \zl\\
   0 & \mathrm{for}~\zs \le \zl
  \end{cases},
 \end{aligned}
\end{equation}
with $c$ the speed of light, $G$ the gravitational constant, and $D_l(\zl)$, $D_s(\zs)$, and $D_{ls}(\zl,\zs)$ the observer--lens, observer--source, and lens--source angular diameter distances, respectively. The dimensionless factor $\beta(\zl,\zs)$ describes the geometric lensing efficiency as a function of lens redshift $\zl$ and source redshift $\zs$. The shear and convergence thus depend on $(\zl, \zs)$ as well as on the image position $\btheta$.
 
The gravitational shear field $\gamma(\btheta)$ is directly observable from image ellipticities of background galaxies in the weak-lensing regime, $|\kappa|\ll 1, |\gamma|\ll 1$. The shear and convergence fields are related by 
\begin{equation}
\label{eq:kappa2gamma}
\gamma(\btheta) =
\int\!d^2\theta^\prime\,D(\btheta-\btheta^\prime)\kappa(\btheta^\prime)
\end{equation}
with $D(\btheta)$ the complex kernel $D(\btheta)=(\theta_2^2-\theta_1^2-2i\theta_1\theta_2)/(\pi|\btheta|^4)$.
The key observable for weak shear lensing in the subcritical regime is the complex reduced shear,
\begin{equation}
\label{eq:redshear}
g(\btheta) =\frac{\gamma(\btheta)}{1-\kappa(\btheta)},
\end{equation}
which remains invariant under the global transformation $\kappa(\btheta)\to \lambda \kappa(\btheta) + 1-\lambda$ and $\gamma(\btheta)\to \lambda \gamma(\btheta)$ with an arbitrary constant $\lambda\ne 0$ (for a fixed source redshift $\zs$). This is referred to as the mass-sheet degeneracy \citep[][]{Schneider+Seitz1995}. This degeneracy can be broken or alleviated, for example, by measuring the magnification factor $\mu$ in the subcritical regime,
\begin{equation}
\label{eq:mu}
\mu(\btheta) = \frac{1}{[1-\kappa(\btheta)]^2-|\gamma(\btheta)|^2}
\equiv \frac{1}{\Delta_\mu(\btheta)}.
\end{equation}
We note that in practical applications to magnification bias measurements, this degeneracy can be lifted only if the unlensed mean source background density is known or can be estimated from the data (see Section~\ref{subsec:magbias}). The magnification factor $\mu$ transforms as $\mu(\btheta)\to \lambda^{-2}\mu(\btheta)$. For simplicity of notation, we often use the inverse magnification $\Delta_\mu=\mu^{-1}$.

The reduced shear $g_{1,2}$ can be decomposed into the tangential component $g_+=\gamma_+/(1-\kappa)$ and the $45^\circ$-rotated cross-shear component $g_\times=\gamma_\times/(1-\kappa)$ with respect to a given reference point. The tangential shear $\gamma_+(\theta)$ averaged around a circle of projected radius $\theta$ is related to the excess surface mass density $\Delta\Sigma(\theta)$ through the following identity:
\begin{equation}
 \begin{aligned}
  \SigmaCrit \gamma_+(\theta) &= \Sigma(<\theta)-\Sigma(\theta) \equiv \Delta\Sigma(\theta),
 \end{aligned}
\end{equation}
where $\Sigma(\theta)$ is the azimuthally averaged surface mass density at radius $\theta$ and $\Sigma(<\theta)$ is the average surface mass density interior to $\theta$. The azimuthally averaged cross-shear $\gamma_\times(\theta)$ is expected to vanish if the signal is due to weak lensing.

\subsection{Source Redshift Distribution}
\label{subsec:Nz}

We consider a population of source galaxies characterized by their mean (unlensed) redshift distribution, $\overline{N}(z)$. In general, we use different magnitude, color, size, and quality cuts in background selection for measuring the shear and magnification effects. This results in different $\overline{N}(z)$ for shear and magnification. 
The source-averaged mean lensing depth $\langle\beta^n\rangle_X$ ($n=1,2,\dots$) for a given population ($X=g,\mu$) is
\begin{equation}
\label{eq:depth}
\langle\beta^n\rangle_X =\left[
\int_0^\infty\!dz\, \overline{N}_X(z) \beta^n(\zl,z)\right]
\left[
\int_0^\infty\!dz\, \overline{N}_X(z)
\right]^{-1}.
\end{equation} 
In general, $\overline{N}(z)$ for a given lens can include foreground galaxies. The contribution from unlensed objects with $\beta=0$ is thus taken into account in the calculation of $\langle\beta^n\rangle_X$.

We introduce the relative lensing strength of a given source population $\langle W\rangle_X = \langle\beta\rangle_X  / \beta_\infty$ with $\beta_\infty\equiv \beta(\zl, z_{s,\infty})$ defined relative to a reference source in the far background at redshift $z_{s,\infty}$ \citep{2001PhR...340..291B}. We use a reference redshift of $z_{s,\infty}= 20000$, which was adopted in the CLASH program \citep{Umetsu2014clash,Merten2015clash}. The associated critical surface density is $\Sigma_{\mathrm{cr},\infty}(\zl)=c^2/(4\pi G D_l)\beta_{\infty}^{-1}$. Hereafter, we use the far-background fields $\kappa_\infty(\btheta)$ and  $\gamma_\infty(\btheta)$ to describe the projected mass distribution of the cluster.

\subsection{Pixelized Mass Distribution}
\label{subsec:massmodel}

We pixelize the convergence field, $\kappa_\infty(\btheta)=\Sigma_{\mathrm{cr},\infty}^{-1}\Sigma(\btheta)$, into a regular grid of pixels and describe $\kappa_\infty(\btheta)$ by a linear combination of basis functions $B(\btheta-\btheta^\prime)$ as
\begin{equation}
\label{eq:basis}
\kappa_\infty(\btheta) =\Sigma_{\mathrm{cr},\infty}^{-1} \sum_{n=1}^{\Npix} B(\btheta-\btheta_n)\, \Sigma_n.
\end{equation}
To avoid the loss of information due to oversmoothing, we choose the basis function to be the Dirac delta function, $B(\btheta-\btheta^\prime)=(\Delta\theta)^2\delta^2_\mathrm{D}(\btheta-\btheta^\prime)$, with $\Delta\theta$ a constant grid spacing. The 2D cluster lensing signal is specified by a vector of model parameters containing cell-averaged surface mass densities \citep{Umetsu2015,Umetsu2018clump3d},
\begin{equation}
 \bm = \left\{\Sigma_n\right\}_{n=1}^{\Npix}
\end{equation}
with $\Sigma_n=\Sigma(\btheta_n)$ ($n=1,2,\dots,\Npix$). The complex shear field is then expressed as 
\begin{equation}
\label{eq:shear2m}
\gamma_\infty(\btheta)= \Sigma_{\mathrm{cr},\infty}^{-1}\sum_{n=1}^{\Npix} (D\otimes B)(\btheta-\btheta_n)\,\Sigma_n
\end{equation}
with $D\otimes B = \pi^{-1} (\Delta\theta)^2D$ an effective complex kernel (Equation~(\ref{eq:kappa2gamma})).
Hence, both $\kappa_\infty(\btheta)$ and $\gamma_\infty(\btheta)$ can be expressed as linear combinations of mass coefficients.\footnote{Because of the choice of the basis function, an unbiased extraction of the mass coefficients $\{\Sigma_n\}_{n=1}^{\Npix}$ is possible by performing a spatial integral of $\kappa_\infty(\btheta)$ over a certain area. Such operations include spatial smoothing, azimuthal averaging for the radial profile extraction, and fitting with smooth parametric functions.}

\subsection{Reduced Shear Field}
\label{subsec:shear}

We use the reduced shear field as the primary constraint from our weak-lensing observations. The source-averaged reduced shear $g_n=g(\btheta_n)$ is measured from shape measurements of background galaxies onto a regular grid of $\Npix$ pixels ($n=1,2, \dots, \Npix$) as
\begin{equation}
\label{eq:bin_shear} 
g_n
=
\left[
\displaystyle\sum_k
S(\btheta_{(k)},\btheta_n)
w_{g (k)}g_{(k)}
\right]
\left[
\displaystyle\sum_{k} 
S(\btheta_{(k)},\btheta_n)w_{g(k)}
\right]^{-1} 
\end{equation}
where $S(\btheta,\btheta^\prime)$ is a spatial window function, $g_{(k)}$ is an estimate of $g(\btheta)$ for the $k$th galaxy at $\btheta_{(k)}$, and $w_{g(k)}$ is its statistical weight, $w_{g(k)} = 1/(\sigma^2_{g(k)}+\alpha^2_g)$, with $\sigma^2_{g(k)}$ the error variance of $g_{(k)}$. The $\alpha_g$ parameter is set to a typical value of the shear dispersion $\sigma_g = 0.4$ found in Subaru weak-lensing observations \citep[e.g.,][]{Umetsu+2009,Umetsu2014clash}. 

The source-averaged expectation (denoted by a hat symbol) for the observable $g_n$ (Equation~(\ref{eq:bin_shear})) is given by \citep{Seitz+Schneider1997,Umetsu2015}
\begin{equation}
\label{eq:g_ave}
\widehat{g}(\btheta_n) \simeq \frac{\langle W\rangle_g
 \gamma_\infty(\btheta_n)}{1-f_{W,g} \langle W \rangle_g \kappa_\infty(\btheta_n)},
\end{equation}
where $\langle W\rangle_g$ is the source-averaged relative lensing strength (see Section~\ref{subsec:Nz}) and $f_{W,g}\equiv \langle W^2\rangle_g/\langle W\rangle_g^2 = \langle\beta^2\rangle_g/\langle\beta\rangle_g^2$ is a dimensionless correction factor of the order unity. The error variance  $\sigma_{g,n}^2$ for $g_n$ is expressed as
\begin{equation}
 \begin{aligned}
\label{eq:bin_shearvar}
\sigma^2_{g,n}=
 \frac{ \sum_k S^2(\btheta_{(k)},\btheta_n)w_{g(k)}^2 \sigma^2_{g(k)} }
{ \left[
 {\sum_k S(\btheta_{(k)},\btheta_n)w_{g(k)}}
  \right]^{2}}.
  \end{aligned}
\end{equation}
We adopt the top-hat window of radius $\theta_\mathrm{f}$ \citep{Merten+2009,Umetsu2015,Umetsu2018clump3d}, $S(\btheta,\btheta^\prime)=H(\theta_\mathrm{f}-|\btheta-\btheta^\prime|)$, with $H(x)$ the Heaviside function defined such that $H(x)=1$ if $x\ge 0$ and $H(x)=0$ otherwise. The shape-noise covariance matrix for $g_{\alpha,n}=g_\alpha(\btheta_n)$ is then given as \citep{Oguri2010LoCuSS}
\begin{equation}
 \label{eq:Cshape}
 \left(C_g\right)_{\alpha\beta,mn}
 = 
  \frac{1}{2}\delta_{\alpha\beta} \sigma_{g,m} \sigma_{g,n}
  \xi_{H}(|\btheta_m-\btheta_n|),
\end{equation}
where the indices $\alpha$ and $\beta$ run over the two components of the reduced shear ($\alpha,\beta=1,2$),
$\delta_{\alpha\beta}$ denotes the Kronecker delta, and $\xi_H(x)$ is the autocorrelation of a pillbox of  radius 
$\theta_\mathrm{f}$ \citep{1999ApJ...514...12W,Park+2003,Umetsu2015}, 
\begin{equation}
\label{eq:xi_H}
 \xi_{H}(x)=\frac{2}{\pi}\left[\cos^{-1}\left(\frac{x}{2\theta_{\rm
  f}}\right)-\left(\frac{x}{2\theta_{\rm
  f}}\right)\sqrt{1-\left(\frac{x}{2\theta_\mathrm{f}}\right)^2}\right]
\end{equation}
for $|x|\le 2\theta_\mathrm{f}$
and $\xi_{H}(x)=0$ for $|x|>2\theta_\mathrm{f}$.

\subsection{Flux Magnification Bias}
\label{subsec:magbias}

Lensing magnification influences the observed surface number density of background sources behind lenses, enhancing the apparent source fluxes and expanding the area of sky. The former effect increases the source counts above the limiting flux, whereas the latter reduces the effective observing area in the source plane, thus decreasing the observed source counts per unit solid angle. The net effect, known as magnification bias \citep{BTP1995}, depends on the intrinsic slope of the source luminosity function. 

Deep multi-band photometry can be used to sample the faint end of the luminosity function for quiescent galaxies at $z\sim 1$ \citep{Ilbert2010}. The effect of magnification bias for such a population is dominated by the geometric area distortion, because there are relatively few fainter objects that can be magnified into the flux-limited sample. This effect results in a net depletion of source counts \citep[e.g.,][]{BTU+05,Ford2012,Coe2012a2261,Radovich2015planck,Ziparo2016locuss}. The key advantage in the regime of density depletion, at the expense of deep multi-band imaging, is that the effect is not sensitive to the exact form of the source luminosity function \citep{Umetsu2014clash}.
 
In cluster--galaxy weak lensing, the change in magnitude $\delta m=2.5\log_{10}\mu$ due to magnification is small compared to the range over which the slope of the luminosity function varies. The source counts can thus be locally approximated by a power law at a given cutoff magnitude $\mcut$ \citep{Umetsu2020rev}.
Following \citet{Umetsu2014clash,Umetsu2016clash}, we interpret the source-averaged magnification bias as (see Appendix~\ref{appendix:magslope})
\begin{equation}
 \label{eq:magbias}
   \begin{aligned}
    \widehat{b}_\mu(\btheta) &\equiv
    \frac{\widehat{N}_\mu(\btheta| < \mcut)}{\overline{N}_\mu(< \mcut)}
    \simeq \Delta_\mu(\btheta)^{1-2.5\slim},\\
    \Delta_\mu(\btheta) 
    &= \left[1-\langle W\rangle_\mu
    \kappa_\infty(\btheta)\right]^2- \langle W\rangle_\mu^2|\gamma_\infty(\btheta)|^2,
   \end{aligned}
\end{equation}
where the expected value of a weak-lensing observable is denoted by a hat symbol, $\overline{N}_\mu(< \mcut)=\int_0^\infty\!dz\,\overline{N}_{\mu}(z| < \mcut)$ is the unlensed mean counts per cell, $\langle W\rangle_\mu$ is the source-averaged relative lensing strength (Section~\ref{subsec:Nz}), and $\slim$ is the logarithmic count slope evaluated at the cutoff magnitude $\mcut$,
\begin{equation}
 \label{eq:slope}
 \slim(\mcut) = \frac{d\log_{10}\overline{N}(<m)}{dm}\Bigg|_{\mcut}.
\end{equation}

Since a given magnitude cut corresponds to different luminosities at different source redshifts, different source populations probe different regimes of magnification bias \citep{Umetsu2013}. A net depletion (or enhancement) of source counts results when $s < 0.4$ (or $> 0.4$). In this study, we measure the density depletion signal using a source population with $s<0.4$. For simplicity, we write $N_\mu(\btheta) = N_\mu(\btheta|< \mcut)$ and $\overline{N}_\mu = \overline{N}_\mu(<\mcut)$. In the weak-lensing limit, $\widehat{b}_\mu -1\simeq (5\slim-2)\langle W\rangle_\mu \kappa_\infty$.

The covariance matrix $\mathrm{Cov}[N(\btheta_m),N(\btheta_n)] \equiv (C_N)_{mn}$ of the counts in cell includes the clustering and Poisson contributions, $(C_N)_{mn}=(\overline{N}_\mu)^2\omega_{mn} +\delta_{mn}N_\mu(\btheta_m)$ \citep{Hu+Kravtsov2003} with $\omega_{mn}$ the cell-averaged angular correlation function of source galaxies. As discussed in detail by \citet{Umetsu2015}, $C_N$ can be approximated as
\begin{equation}
\label{eq:covN}
\left(C_N\right)_{mn} \simeq
\left[
\langle \delta N_\mu^2(\btheta_m)\rangle +  
N_\mu(\btheta_m)
\right]\delta_{mn},
\end{equation}
with $\langle \delta N_\mu^2(\btheta_m)\rangle$ the variance of the $m$th counts.

To overcome this noise, we azimuthally average the observed counts $N_\mu(\btheta)$ in a set of clustercentric annuli and calculate the surface number density profile $\{n_{\mu,i}\}_{i=1}^{\Nbin}$ of background galaxies as
\citep{Umetsu2015,Umetsu2016clash}
\begin{equation}
 \label{eq:nb}
  n_{\mu,i} =
  \frac{1}{(1-f_{\mathrm{mask},i})\Omega_\mathrm{cell}}
  \sum_{m} {\cal P}_{im}N_\mu(\btheta_m),
\end{equation}
where $\Omega_\mathrm{cell}$ is the solid angle of each cell and ${\cal P}_{im}=(\sum_m A_{mi})^{-1}A_{mi}$ is the projection matrix normalized by $\sum_m {\cal P}_{im}=1$; $A_{mi}$ denotes the area fraction of the $m$th cell lying within the $i$th radial bin and $f_{\mathrm{mask},i}$ is the mask correction factor for the $i$th bin,  $(1-f_{\mathrm{mask},i})^{-1} \equiv \left[ \sum_{m} (1-f_m)A_{mi}\right]^{-1} \sum_{m} A_{mi}$, with $f_m$ the masked area fraction in the $m$th cell due to saturated objects, foreground galaxies, and cluster members. We use Monte Carlo integration to calculate the area fractions $A_{mi}$ for individual cells \citep{UB2008}. The Poisson and clustering contributions to the uncertainty in $n_{\mu,i}$ are 
\begin{equation} 
\label{eq:Covn}
\sigma_{\mu,i}^2
=
\frac{1}{(1-f_{\mathrm{mask},i})^2\Omega_\mathrm{cell}^2}
\sum_{m}
{\cal P}_{im}^2
\left(C_N\right)_{mm}.
\end{equation}
Additionally, we account for systematic uncertainties in the magnification analysis. In Appendix~\ref{appendix:magbiasdata}, we describe the procedure used to estimate the uncertainties $\sigma_{\mu,i}$ in $n_{\mu,i}$.

The expectation for the observable $n_{\mu,i}$ (Equation~(\ref{eq:nb})) is 
\begin{equation}
 \label{eq:nb_th}
\widehat{n}_{\mu,i} =
\overline{n}_\mu
\sum_m {\cal P}_{im}\Delta_\mu(\btheta_m)^{1-2.5\slim}
\end{equation}
with $\overline{n}_\mu =\overline{N}_\mu/\Omega_\mathrm{cell}$.

\subsection{Mass Reconstruction Algorithm}
\label{subsec:massrec}

A practical limitation of the shear-only lensing analysis is the inherent mass-sheet degeneracy, which can be alleviated by using the complementary combination of shear and magnification \citep{Schneider+2000,UB2008,Rozo+Schmidt2010}. Measuring the two complementary effects also enables us to check the internal consistency of weak-lensing measurements \citep{Umetsu2014clash}. Moreover, obtaining accurate mass maps has the important advantage of being able to identify local mass structures and to directly compare them with multiwavelength observations.

In this work, we use the mass inversion algorithm developed by \citet{Umetsu2015}, who generalized the cluster lensing mass inversion (\textsc{clumi}) code of \citet{Umetsu2013} into a 2D description of the pixelized mass distribution. This free-form algorithm combines a 2D shear pattern ($g_1(\btheta), g_2(\btheta)$) with azimuthally averaged measurements of magnification bias $\{n_{\mu,i}\}_{i=1}^{\Nbin}$. The latter imposes a set of azimuthally averaged constraints on $\Sigma(\btheta)$ to effectively break the mass-sheet degeneracy. The \textsc{clumi}-2D algorithm takes full account of the nonlinear subcritical regime of lensing. 

Given a model $\blambda$ and observed data $\bd$, the Bayes' theorem states that the joint posterior probability $P(\blambda|\bd)$ is proportional to the product of the likelihood ${\cal L}(\blambda)\equiv P(\bd|\blambda)$ and the prior probability $P(\blambda)$. In our inversion problem, $\blambda$ is a signal vector containing the pixelized mass coefficients $\bm=\{\Sigma_n\}_{n=1}^{\Npix}$ (Section~\ref{subsec:massmodel}) and calibration nuisance parameters $\bc$ (see Section~\ref{subsubsec:calib}),  so that $\blambda\equiv (\bm,\bc)$.

We express the joint likelihood function ${\cal L}$ for combined weak-lensing data $\bd$ as a product of the two separate likelihood functions, ${\cal L}={\cal L}_{g} \times {\cal L}_\mu$ with ${\cal L}_{g}$ and ${\cal L}_\mu$ the likelihood functions for shear and magnification, respectively. 
We assume that the observational errors follow a Gaussian distribution, so that ${\cal L}\propto \exp(-\chi^2/2)$, with $\chi^2$ the standard misfit statistic.

\subsubsection{Shear Log-likelihood Function}
\label{subsubsec:lg}

The log-likelihood function $l_{g}\equiv -\ln{\cal L}_{g}$ for 2D shear data is written as
\citep{Oguri2010LoCuSS,Umetsu2015,Umetsu2018clump3d} 
\begin{equation}
 \begin{aligned}
l_{g}(\blambda) =& \frac{1}{2}
 \sum_{m,n=1}^{\Npix}
 \sum_{\alpha,\beta=1}^{2}
[g_{\alpha,m}-\widehat{g}_{\alpha,m}(\blambda)]
\left({\cal W}_g\right)_{\alpha\beta,mn}\\
  &\times [g_{\beta,n}-\widehat{g}_{\beta,n}(\blambda)] + \mathrm{const.},
  \end{aligned}
\end{equation}
where $\widehat{g}_{\alpha,m}(\blambda)$ is the theoretical expectation for $g_{\alpha,m}=g_\alpha(\btheta_m)$ and $({\cal W}_g)_{\alpha\beta,mn}$ is the shear weight matrix,
\begin{equation}
  \left({\cal W}_g\right)_{\alpha\beta,mn} = M_m M_n \left(C_g^{-1}\right)_{\alpha\beta,mn}.
\end{equation}
Here, $M_m$ is a mask weight, defined such that $M_m=0$ if the $m$th cell is masked out and $M_m=1$ otherwise, and $C_g$ is the shear covariance matrix given by Equation~(\ref{eq:Cshape}).

\subsubsection{Magnification Log-likelihood Function}
\label{subsubsec:lmu}

The log-likelihood function for magnification bias data  $l_{\mu}\equiv -\ln{\cal L}_\mu$ is written as \citep{Umetsu2015,Umetsu2018clump3d}
\begin{equation}
 \begin{aligned}
  l_\mu(\blambda) =& \frac{1}{2}\sum_{i=1}^{\Nbin}
  [n_{\mu,i}-\widehat{n}_{\mu,i}(\blambda)]
  \left({\cal W}_\mu\right)_{ij}\\
  &\times [n_{\mu,j}-\widehat{n}_{\mu,j}(\blambda)] + \mathrm{const.},
 \end{aligned}
\end{equation}
where $\widehat{n}_{\mu,i}(\blambda)$ is the theoretical expectation for $n_{\mu,i}$ and $({\cal W}_\mu)_{ij}$ is the magnification weight matrix, ${\cal W}_\mu = C^{-1}_\mu$, with $C_\mu$ the corresponding covariance matrix,
\begin{equation}
 \label{eq:C_mu}
  (C_\mu)_{ij} = \sigma_{\mu,i}^2\delta_{ij},
\end{equation}
where the diagonal errors $\sigma_{\mu,i}$ ($i=1,2,\dots,\Nbin$) are given by Equation~(\ref{eq:sigma_mu}). 

The $l_\mu$ function sets azimuthally integrated constraints on $\Sigma(\btheta)$, providing the otherwise unconstrained normalization of $\Sigma(\btheta)$ over a set of concentric annuli where magnification measurements are obtained. No assumption is made about the azimuthal symmetry of $\Sigma(\btheta)$ in our analysis. We use Monte Carlo integration to compute the projection matrix ${\cal P}_{im}$ (Equation~(\ref{eq:nb})) of size $\Nbin \times \Npix$, which is necessary to predict  $\{\widehat{n}_{\mu,i}(\blambda)\}_{i=1}^{\Nbin}$ for a given model $\blambda=(\bm,\bc)$.

\subsubsection{Calibration Parameters}
\label{subsubsec:calib}

In our joint likelihood analysis, we account for the uncertainty in the observational calibration parameters,
\begin{equation}
\label{eq:calib}
\bc = \{\langle W\rangle_g, f_{W,g}, \langle W\rangle_\mu, \overline{n}_\mu, \slim\},
\end{equation}
with $\langle W\rangle_g=\langle \beta\rangle_g/\beta_{\infty}$, $\langle W\rangle_\mu=\langle \beta\rangle_\mu/\beta_{\infty}$, and $f_{W,g}=\langle \beta^2\rangle_g/\langle \beta\rangle_g^2$ (Section~\ref{subsec:Nz}). To this end, we include Gaussian priors on $\bc$ defined with mean values and uncertainties directly estimated from data. Specifically, we use for each parameter the mean and uncertainty estimated from the Suprime-Cam data (Tables~\ref{tab:gsample} and \ref{tab:musample}) as the center and dispersion of the prior distribution, respectively.

\subsection{Best-fit Solution and Covariance Matrix}
\label{subsec:cmat}

The log-posterior function $F(\blambda) =-\ln{P(\blambda|\bd)}$ is written as a linear sum of the log-likelihood and log-prior (or quadratic penalty) terms. The global maximum of the joint posterior probability distribution function (PDF) over $\blambda$ is found by minimizing $F(\blambda)$ with respect to $\blambda$. We use the conjugate-gradient algorithm \citep[see][]{1992nrfa.book.....P} to find the global solution $\widehat{\blambda}$. We employ an analytic expression for the gradient function $\bnabla F$ obtained in the nonlinear, subcritical regime (see Appendix~B of \citealt{Umetsu2018clump3d}). 

The reconstructed mass pixels are correlated primarily because the relation between the shear and convergence is nonlocal (Equation~(\ref{eq:kappa2gamma})). Additionally, the effects of spatial averaging (Equation~(\ref{eq:xi_H})) and cosmic noise due to projected uncorrelated large scale structure can produce a covariance between different pixels. In our analysis, the effects of correlated errors are modeled analytically. Specifically, we take into account the statistical and cosmic-noise contributions to the total covariance matrix $C_{mn}=\mathrm{Cov}(\Sigma_m,\Sigma_n)$ ($m,n=1,2,\dots,\Npix$) as
\begin{equation}
 \label{eq:ctot}
  C = C_\mathrm{stat} + C_\mathrm{lss},
\end{equation}
where $C_\mathrm{stat}$ is given by $(C_\mathrm{stat})_{mn} =\left({\cal F}^{-1}\right)_{mn}$ with ${\cal F}$ the Fisher matrix evaluated at the best-fit solution $\blambda=\widehat{\blambda}$ (see Appendix~B of \citealt{Umetsu2018clump3d}),
\begin{equation}
{\cal F}_{mn} = \left\langle \frac{\partial^2 F(\blambda)}{\partial \lambda_m \partial \lambda_{n}} \right\rangle\Bigg|_{\widehat{\blambda}},
\end{equation}
and $(C_\mathrm{lss})_{mn}=\SigmaCrit^{2}\xi_\mathrm{lss}(|\btheta_m-\btheta_n|)$ with $\xi_\mathrm{lss}(|\btheta|)$ the cell-averaged two-point angular correlation function for the cosmic convergence field $\kappa_\mathrm{lss}(\btheta)$ \citep{Kaiser1992}. In this work, we approximate the pixel window function \citep[e.g.,][]{Hu+White2001} by a Dirac delta function centered at each pixel and compute the elements of the $C_\mathrm{lss}$ matrix for a given source population (see Section~\ref{subsec:back}), using the nonlinear matter power spectrum of \citet{Smith+2003halofit} for the base-\LCDM model from \Planck 2018 cosmic microwave background (CMB) anisotropy data in combination with CMB lensing \citep[][see their Table~2]{Planck2018VI}.


\section{Subaru Data and Weak-lensing Analysis}
\label{sec:data}

In this section, we describe our new weak-lensing analysis of A370 based on deep Suprime-Cam \BRz imaging. In this study, we analyze the Suprime-Cam data using our reduction and analysis pipelines presented in \citet[][see also \citealt{Umetsu2015}]{Umetsu2014clash}, who performed a homogeneous weak-lensing analysis of 20 high-mass clusters targeted by the CLASH program. As detailed in Section~\ref{subsec:photometry}, the present analysis further implements an improved astrometry based on the Gaia mission \citep[][]{2021brown}.

\subsection{Data and Photometry}
\label{subsec:photometry}

\begin{figure*}[tbp] 
 \begin{center}
  \includegraphics[width=0.9\textwidth,angle=0,clip]{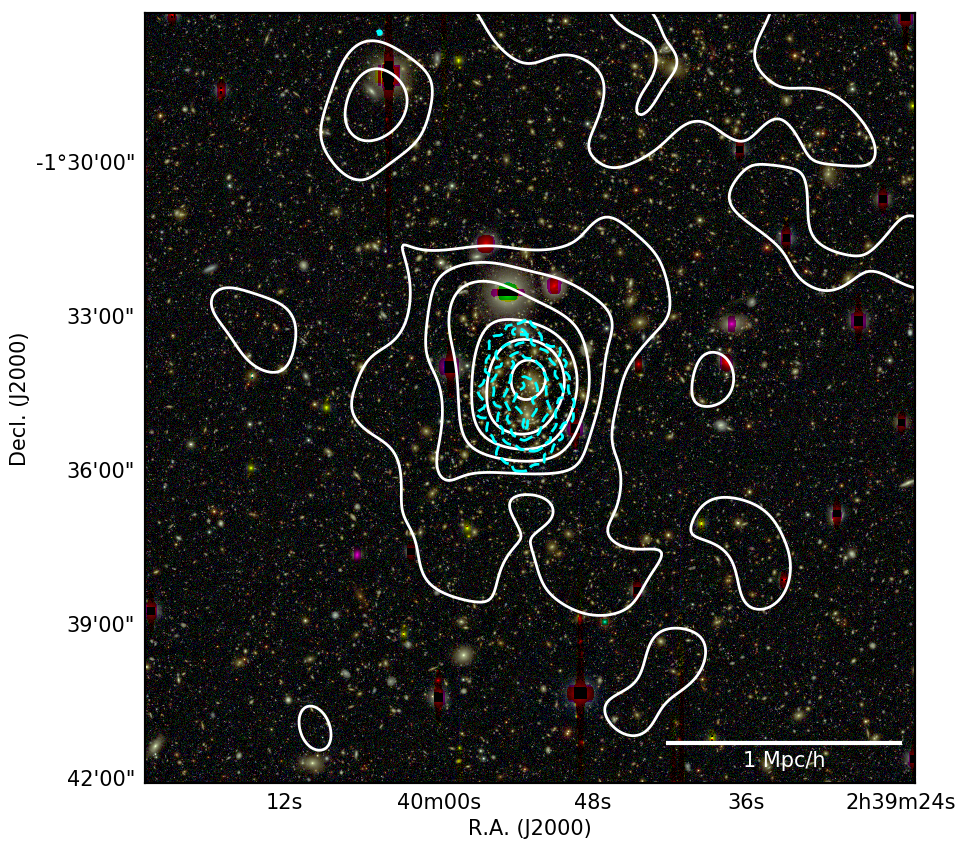}
 \end{center}
\caption{
\label{fig:rgb}
Subaru/Suprime-Cam \BRz composite color image centered on A370, overlaid with mass contours from our weak-lensing analysis (Section~\ref{sec:mrec}). The image is $15\arcmin\times 15\arcmin$ in size. The mass map is smoothed with a Gaussian of $1.2\arcmin$~FWHM. The lowest contour level and the contour interval are $\Delta\kappa=0.08$. Also overlaid are logarithmically spaced X-ray brightness contours (cyan dashed) from \Chandra observations in the $0.5$--$7$~keV energy band (Section~\ref{sec:xray}). The horizontal bar represents $1\Mpch$ at the cluster redshift. North is up and east is to the left.}
\end{figure*}

\begin{deluxetable}{cccc}[tbp]
\tabletypesize{\footnotesize}
\centering
\tablecaption{Subaru Suprime-Cam data} 
\label{tab:subaru}
\tablehead{ 
 \multicolumn{1}{c}{Filter} &
 \multicolumn{1}{c}{Total exposure time} &
 \multicolumn{1}{c}{Seeing\tablenotemark{a}} &
 \multicolumn{1}{c}{$m_\mathrm{lim}$\tablenotemark{b}}  
\\
 \colhead{} &
 \multicolumn{1}{c}{(ks)} &
 \multicolumn{1}{c}{(arcsec)} &
 \multicolumn{1}{c}{(AB~mag)} 
} 
\startdata
 $B$            & $3.0$ & $0.74$  & $27.3$\\
 $R_\mathrm{C}$ & $3.6$ & $0.63$  & $26.3$\\ 
 $z^\prime$     & $8.4$ & $0.72$  & $25.8$ 
\enddata
\tablenotetext{a}{Seeing FWHM from the full stack of images.}
\tablenotetext{b}{Limiting magnitude for a $3\sigma$ detection in a $2\arcsec$ diameter aperture.}
\end{deluxetable}

We analyze deep \BRz images centered on A370 observed with the wide-field camera Suprime-Cam \citep[$34\arcmin\times 27\arcmin$;][]{2002PASJ...54..833M} mounted at the prime focus of the 8.2~m Subaru Telescope.  Details of the Subaru/Suprime-Cam observations are summarized in Table~\ref{tab:subaru}. We use existing archival data taken from SMOKA.\footnote{\href{http://smoka.nao.ac.jp}{http://smoka.nao.ac.jp}}  
The $R_\mathrm{C}$-band images used in this work were taken in excellent seeing conditions on the night of 2005 December 4 (Proposal~ID: o05319). The $R_\mathrm{C}$ images were obtained at two different camera orientations separated by 90~degrees. The $B$ images were taken on the night of 2010 October 12 (Proposal~ID: o10314). For the $z^\prime$ band, we use data taken on the nights of 2009 September 17 and 2010 October 12 (Proposal~ID: o10314) after the Suprime-Cam CCD upgrade in 2008. For the weak-lensing shape measurements (Section~\ref{subsec:shape}), we use the $R_\mathrm{C}$-band data, which have the best image quality in our data sets. 

Figure~\ref{fig:rgb} shows a Suprime-Cam \BRz composite color image of the cluster field, produced using the \textsc{trilogy} software \citep{Coe2012a2261}. The image is overlaid by mass contours from our weak-lensing analysis (Section~\ref{sec:mrec}) and X-ray brightness contours from our \Chandra analysis (Section~\ref{sec:xray}).

The image reduction pipeline used in this study derives from \citet{Nonino+2009}. Several modifications and improvements have been applied to the original pipeline \citep[e.g.,][]{Umetsu+2012,Umetsu2014clash,Umetsu2015,Medezinski+2013,Medezinski2016}. In particular, it has been optimized separately for accurate photometry and shape measurements. For multi-band photometry, standard reduction steps include bias subtraction, super-flat-field correction, and masking of saturated star trails and other artifacts. Photometric catalogs are created using \textsc{sextractor} \citep{SExtractor} in the dual-image mode on PSF-matched images, with the Suprime-Cam $z^\prime$ band image as the detection image.

An accurate astrometric solution was derived with the \textsc{scamp} software \citep{SCAMP} using Gaia Data Release 2  \citep[DR2;][]{2016prusti,2018brown} as an external reference catalog. An astrometric solution has been obtained at the camera level using Gaia DR2 sources extracted from individual exposures for each CCD chip. This astrometric solution does not account for the proper motions of Gaia DR2 sources since the epoch of the Suprime-Cam observations. Comparing with the astrometric solution obtained from proper-motion-corrected source positions based on Gaia Early Data Release 3 \citep[EDR3;][]{2021brown}, we find a mean positional offset of $\approx 2$--$8$~mas and an rms of $\approx 20$--$30$~mas for the $R_\mathrm{C}$-band astrometry. 

The \textsc{swarp} software \citep{Bertin+2002Swarp} is used to stack individual exposures on a common World Coordinate System (WCS) grid with pixel scale of $0.2\arcsec$. No point spread function (PSF) matching is applied. For each passband, we create a full stack of co-added images from which to measure source photometry. For the weak-lensing band ($R_\mathrm{C}$), we additionally create two separate co-added images, each from different camera rotation angles (see Section~\ref{subsec:shape}). 

Once the Subaru images had been combined into a full stack of co-added images, a catalog was then produced and matched directly to the corresponding Gaia DR2 sources to validate the astrometric properties of the full stack. A total of 428 sources from the full stack catalog were matched directly to Gaia DR2 sources, and the positional offsets were measured between each of these catalog sources and their matched Gaia DR2 counterparts. The resulting distribution of positional offsets displays well-behaved symmetry, with an rms uncertainty of 35~mas in R.A. and 33~mas in decl., demonstrating that the positional accuracy of the full stack is in good agreement with the accuracy of the Gaia DR2 alignment carried out on each of the individual single-exposure frames that were used to construct the full stack, as previously described.

Finally, after having verified that the positional uncertainties of the full stack catalog sources were comparable to those of all the individual single-exposure frames, this full stack catalog was then aligned to Gaia EDR3, to ensure that the absolute astrometry could be as up-to-date as possible. The astrometric difference between sources from the full-stack catalog (which was still on Gaia DR2) and the matched sources from Gaia EDR3 show excellent agreement, with only a small difference needing to be applied to place these sources onto Gaia EDR3, namely 2.1~mas in R.A. and 1.4~mas in decl., perhaps due to slight residual differences in proper motion corrections, and not significant compared to the rms uncertainties of $33$--$35$~mas in the catalog source positions.

The photometric zero point for the Suprime-Cam $z^\prime$ filter was calibrated against stars from the Pan-STARRS data release 1 (DR1) catalog \citep{PS1DR1}. The zero points for the Suprime-Cam $B$ and $R_\mathrm{C}$ filters were derived by matching the stellar locus in the $B-R_\mathrm{C}$ vs. $R_\mathrm{C}-z^\prime$ diagram to the COSMOS2020 photometry \citep{COSMOS2020}. These zero points were further refined by matching the color distributions in the $B-R_\mathrm{C}$ vs. $R_\mathrm{C}-z^\prime$ diagram between our Suprime-Cam data and the COSMOS2020 data. The magnitudes for galaxies were corrected for foreground Galactic extinction according to \citet{1998ApJ...500..525S}.  Full details of our photometric calibration are described in Appendix~\ref{appendix:CCcalib}.

\subsection{Shape Measurement}
\label{subsec:shape}

We use our shape measurement pipeline based in part on the \textsc{imcat} package \citep[][KSB]{1995ApJ...449..460K}, with modifications incorporating several key improvements developed by \citet{Umetsu+2010CL0024,Umetsu2014clash}. In this work, we perform a weak shear analysis of A370 following the procedure of \citet{Umetsu2014clash}. 

Here we briefly summarize some of the main features and refer to \citet{Umetsu2014clash} for details of the analysis pipeline. We select isolated galaxy images for the shape measurement, reducing the impact of crowding and blending. After the rejection of close pairs, objects detected with low significance $\nu_g<10$ are excluded from our analysis. Here $\nu_g$ is the peak detection significance given by \textsc{imcat}'s peak-finding algorithm. We select galaxies detected with high significance $\nu_g\ge 30$ as a sample of \textit{shape calibrators}, which is a subset of the target galaxy sample with $\nu_g\ge 10$. The key feature of our analysis pipeline is that only those galaxies detected with sufficiently high significance, $\nu_g\ge 30$, are used to model the isotropic PSF correction as a function of object size and magnitude \citep{Umetsu+2010CL0024}. This calibration method is designed to minimize the inherent noise bias and was employed by the CLASH and LoCuSS collaborations in their cluster weak-lensing studies based on Subaru/Suprime-Cam data \citep{Umetsu2014clash,Okabe+Smith2016}.

For the shape measurement, we separately stack $R_\mathrm{C}$-band images collected at two different camera rotation angles (Section~\ref{subsec:photometry}). In this way, we do not smear individual exposures before stacking, so as not to degrade the weak-lensing signal derived from the shapes of galaxies \citep[][]{Umetsu2014clash,Umetsu2015}. A shape catalog is created for each camera rotation separately. The two subcatalogs are combined by properly weighting and stacking the calibrated distortion measurements for galaxies in the overlapping region \citep[][see their Section~4.3]{Umetsu2014clash}. All galaxies with usable shape measurements are matched to those in our \BRz-selected background samples (see Section~\ref{subsec:back}). 

Our KSB+ implementation has been extensively tested and applied to ground-based observations of a large number of massive clusters including 20 CLASH clusters \citep{Umetsu2014clash,Merten2015clash}. Full details of our shear recovery test based on simulated Subaru/Suprime-Cam images are found in \citet[][see their Appendix A]{Umetsu2018clump3d}. They found that the reduced shear signal $g_\alpha$ ($\alpha=1,2$) can be recovered with $m_\alpha\approx -0.05$ of the multiplicative calibration bias and $|c_\alpha|\sim 10^{-4}$ of the additive shear bias. Here the observed and true values of the reduced shear ($g^\mathrm{obs}, g^\mathrm{true}$) are related by \citep[]{2006MNRAS.368.1323H,2007MNRAS.376...13M},
\begin{equation}
g_\alpha^\mathrm{obs}=(1+m_\alpha) g_\alpha^\mathrm{true} + c_\alpha
\end{equation}
Accordingly, we include for each galaxy a shear calibration factor of $1/0.95$ ($g\to g/0.95$) to account for the residual multiplicative bias. 

\subsection{Background Galaxy Selection}
 \label{subsec:back}


\begin{figure*}[tbp]
 \begin{center}
 $
 \begin{array}
  {c@{\hspace{0.3in}}c}
  \includegraphics[width=0.45\textwidth,angle=0,clip]{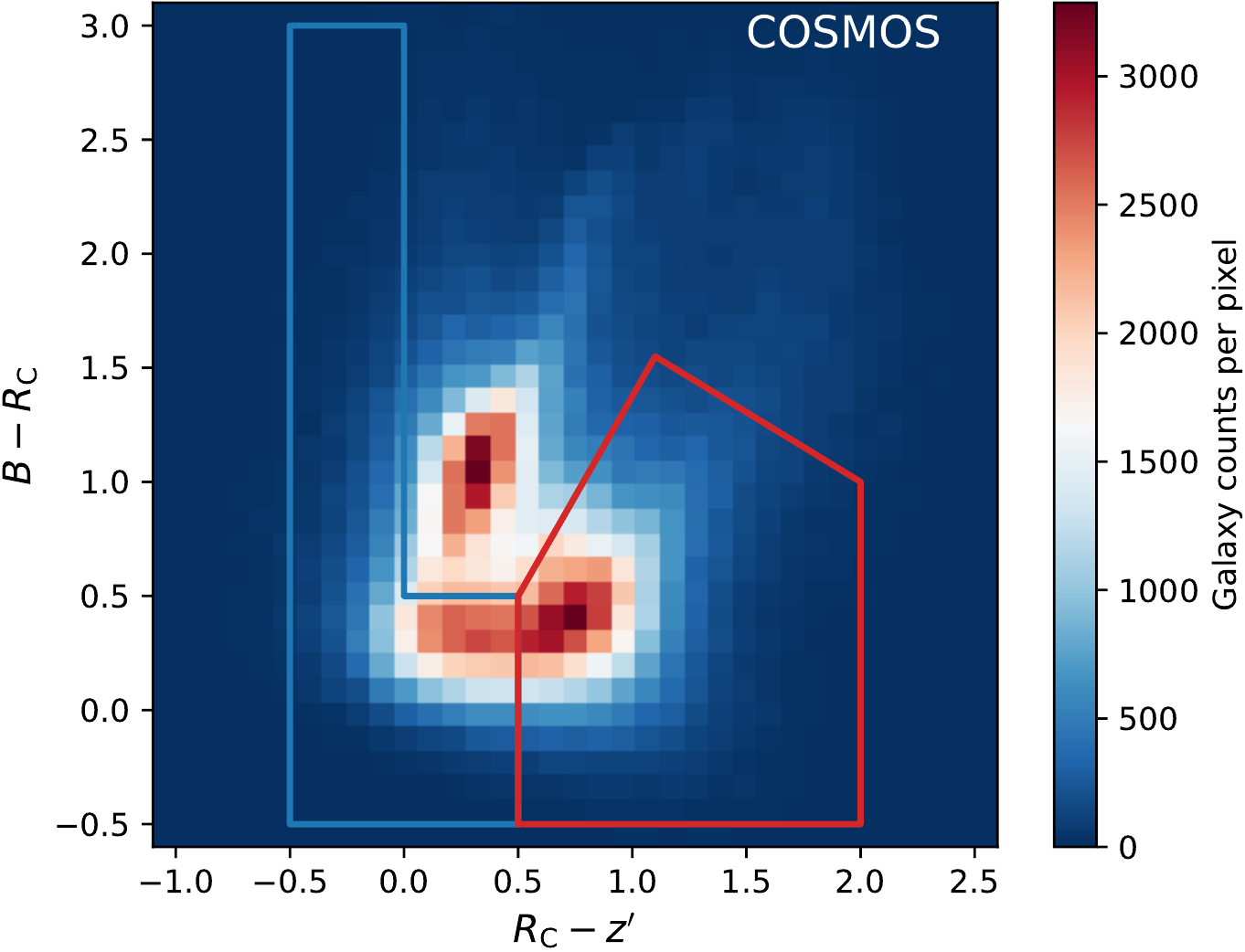} &
  \includegraphics[width=0.45\textwidth,angle=0,clip]{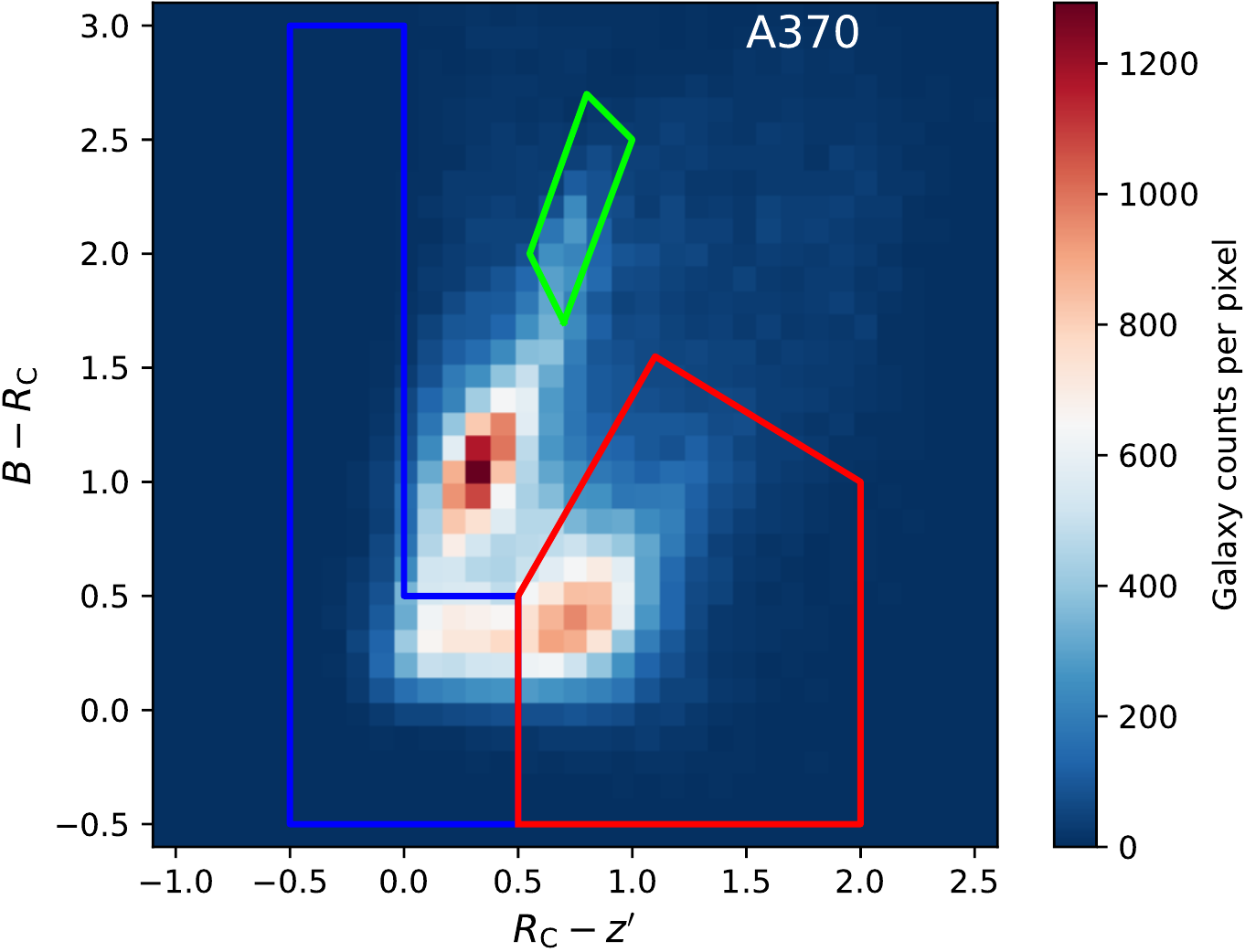} \\
 \end{array}
 $
 \end{center}
\caption{Binned distribution of galaxies in color--color space for the COSMOS field  (left) and A370 (right). Color boundaries of the blue and red background samples  (left blue and lower-right red regions, respectively) selected on the basis of Subaru \BRz photometry are indicated in each panel. In the right panel, the green polygon marks the boundaries of our green sample dominated by red-sequence galaxies of A370 at $z=0.375$. The middle peak with colors bluer than the cluster sequence shows the overdensity of foreground galaxies (see also Figure~\ref{fig:cc_pz}). The plots in both panels are limited to $z^\prime < 26$~mag, which is close to our detection limit (Table~\ref{tab:subaru}). 
 \label{fig:cc_ng} }
\end{figure*}


\begin{figure}[tbp]
  \begin{center}
   \includegraphics[width=0.45\textwidth,angle=0,clip]{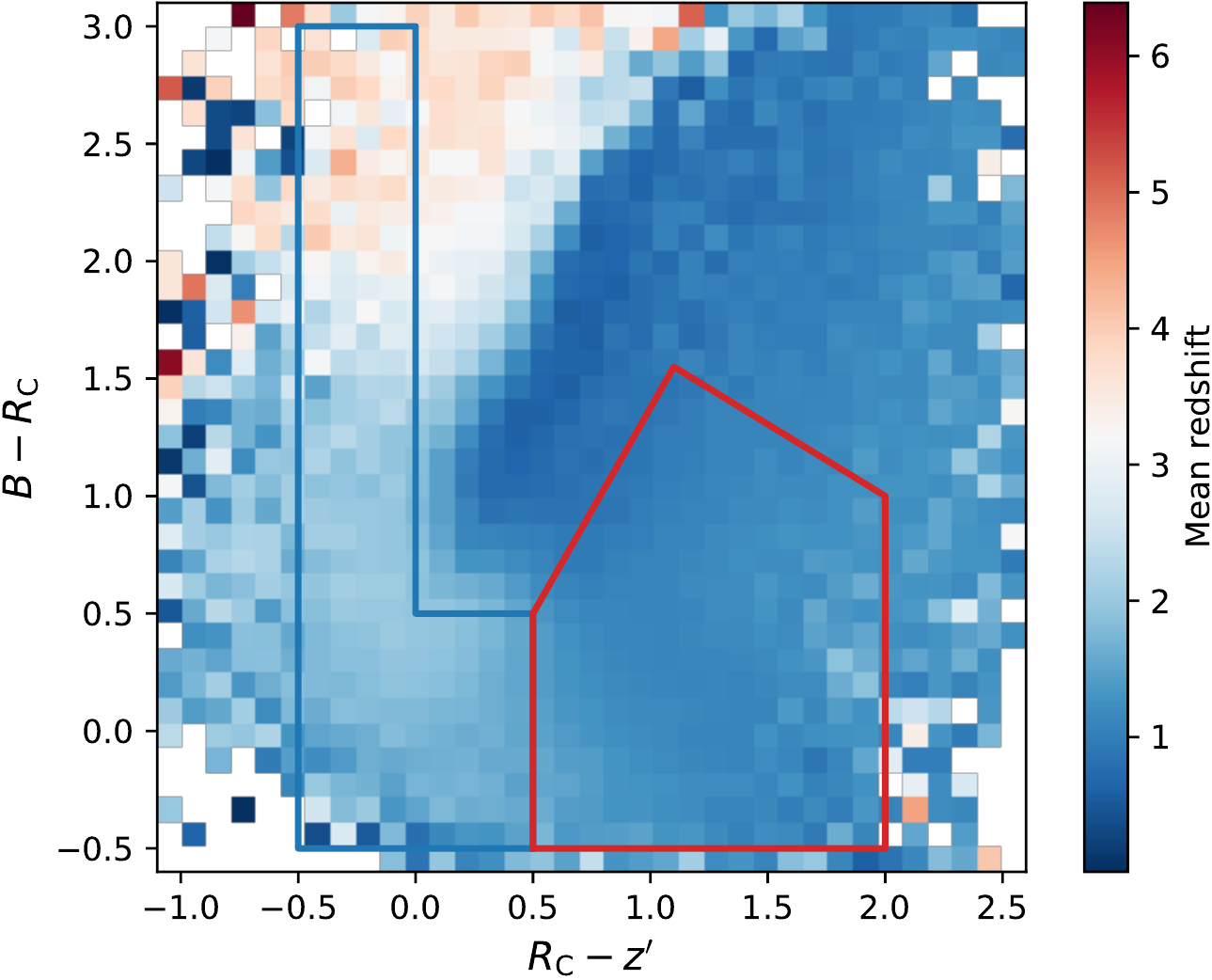}
  \end{center}
 \caption{Binned photometric redshift of COSMOS field galaxies displayed in color--color space. Blue and red polygons mark the boundaries of the blue and red background samples, respectively.}
 \label{fig:cc_pz}
\end{figure}

Contamination of background galaxy samples by unlensed objects, when not accounted for, leads to a systematic underestimation of the true lensing signal. Inclusion of foreground galaxies produces a dilution of the lensing signal that is independent of the cluster radius. In contrast, the inclusion of cluster members dilutes the lensing signal more strongly at smaller cluster radii \citep{BTU+05}. A secure selection of background galaxies is thus essential for obtaining accurate cluster mass estimates from weak lensing \citep[e.g.,][]{Medezinski+2010,Okabe+2013,Gruen2014}. 

In this study, we employ the color--color (CC) selection method of \citet{Medezinski+2010} \citep[see also][]{Medezinski2018src} to define background galaxy samples for measuring both shear and magnification effects. We use \BRz photometry from Subaru/Suprime-Cam, which spans the full optical wavelength range. The CC-cut selection method has been calibrated with evolutionary color tracks of galaxies \citep{Kotulla2009,Medezinski+2010,Medezinski+2011} as well as with photometric-redshift (photo-$z$) catalogs from deep multiwavelength surveys such as COSMOS \citep{Ilbert+2009COSMOS,Laigle2016cosmos,COSMOS2020}. For this purpose, we use the photometric properties and redshifts derived from the COSMOS2020 catalog \citep{COSMOS2020} based on the \textsc{farmer} photometry using the \LePhare code \citep{Ilbert2006}.

In Figure~\ref{fig:cc_ng}, we show the distribution of galaxies in the $B-R_\mathrm{C}$ vs. $R_\mathrm{C}-z^\prime$ plane obtained for the COSMOS field (left panel) and A370 (right panel). Similarly, Figure~\ref{fig:cc_pz} shows the binned average photo-$z$ distribution of COSMOS field galaxies in CC space. As demonstrated by \citet{Medezinski+2010,Medezinski+2011}, the color region dominated by the foreground population is well defined in CC space as a clear overdensity (at $B-R_\mathrm{C}\sim 1$ and $R_\mathrm{C}-z^\prime\sim 0.3$ with $\langle z\rangle\sim 0.5$). Following \citet{Medezinski+2010,Medezinski+2011}, we select two distinct populations that encompass the ``red'' and ``blue'' branches of background galaxies in CC space, each with typical redshift distributions $N(z)$ peaked around $z \sim 1$ and $\sim 2$, respectively  \citep[see][]{Medezinski+2011,Lilly+2007}. 

The color boundaries of our CC-cut samples are shown in Figure~\ref{fig:cc_ng}. The green polygon shown in the right panel marks the boundaries of the ``green'' sample comprising mostly the red-sequence galaxies of A370. We see in Figure~\ref{fig:cc_ng} that the foreground peak for A370 is more pronounced compared to the COSMOS field. This enhancement is likely due to the contribution from bluer cluster members and galaxies in the surrounding regions \citep[see][]{Umetsu+2012,Umetsu2015}.

\begin{deluxetable}{ccccccc}
\centering
\tabletypesize{\scriptsize}
\tablecaption{Background Galaxy Samples for Weak-lensing Shape Measurements}  
\label{tab:gsample}
\tablehead{ 
 \multicolumn{1}{c}{Sample} &
 \multicolumn{1}{c}{$N_g$} &
 \multicolumn{1}{c}{$\overline{n}_g$} & 
 \multicolumn{1}{c}{$\langle\beta\rangle$} &
 \multicolumn{1}{c}{$f_W$} &
 \multicolumn{1}{c}{$\zeff$} &
 \multicolumn{1}{c}{$\mathrm{S/N}$}
 \\
 \colhead{} & 
 \colhead{} &
 \multicolumn{1}{c}{(arcmin$^{-2}$)} &
 \colhead{} &
 \colhead{} &
 \colhead{} &
 \colhead{}
} 
\startdata  
 Red      &  9988& $12.6$  & $0.56\pm 0.03$ & $1.05$ & $1.02$ & $13.5$\\
 Blue     &  6679&  $8.4$  & $0.64\pm 0.03$ & $1.10$ & $1.33$ & $10.9$\\
 Blue+red & 16667& $21.0$  & $0.59\pm 0.03$ & $1.08$ & $1.12$ & $16.8$
\enddata 
\tablecomments{Subaru \BRz\-selected samples of background galaxies. We use the composite blue+red background sample for our weak-lensing shear analysis. The mean lensing depth $\langle\beta\rangle$ and the spread parameter $f_W=\langle\beta^2\rangle/\langle\beta\rangle^2$ for each source population are estimated using photometric redshifts from the COSMOS2020 \textsc{farmer} catalog. The quantity $\zeff$ represents the effective source redshift of each sample, defined as $\beta(\zeff)=\langle \beta\rangle$. The $\mathrm{S/N}$ is the detection significance for the tangential distortion profile $g_+(\theta)$.} 
\end{deluxetable}

\begin{deluxetable*}{cccccccc}[tbp]
\centering
\tabletypesize{\footnotesize}
\tablecaption{Background Galaxy Samples for Magnification-bias Measurements}
\label{tab:musample}
\tablehead{ 
 \multicolumn{1}{c}{Sample} &
 \multicolumn{1}{c}{Magnitude limits} &
 \multicolumn{1}{c}{$N_\mu$} &
 \multicolumn{1}{c}{$\overline{n}_\mu$} & 
 \multicolumn{1}{c}{$\slim$} & 
 \multicolumn{1}{c}{$\langle\beta\rangle$} &
 \multicolumn{1}{c}{$\zeff$} &
 \multicolumn{1}{c}{$\mathrm{S/N}$}
 \\
 \colhead{} & 
 \multicolumn{1}{c}{(AB mag)} &
 \colhead{} &
 \multicolumn{1}{c}{(arcmin$^{-2}$)} &
 \colhead{} &
 \colhead{} & 
 \colhead{} &
 \colhead{}
} 
\startdata  
 Lensing cut & $21.0 <z^\prime< 25.6$ & 22142 & $21.1\pm 0.6$  & $0.168\pm 0.037$ & $0.57\pm 0.03$ & $1.04$ & $5.7$\\
 Null-test   & $21.0 <z^\prime< 23.6$ &  6344 & $ 5.6\pm 0.3$  & $0.404\pm 0.069$ & $0.57\pm 0.03$ & $1.04$ & $2.7$
\enddata 
\tablecomments{Lensing-cut and null-test samples of CC-red background galaxies selected for our weak-lensing magnification analysis. Apparent magnitude cuts are applied in the reddest CC-selection band available ($z^\prime$) to avoid incompleteness near the detection limit (Table~\ref{tab:subaru}). The mean lensing depth $\langle\beta\rangle$ for each source population is estimated using photometric redshifts from the COSMOS2020 \textsc{farmer} catalog. The quantity $\zeff$ represents the effective source redshift corresponding to the mean lensing depth $\langle \beta\rangle$ of each sample, defined as $\beta(\zeff)=\langle \beta\rangle$. The $\mathrm{S/N}$ is the detection significance for the magnification bias profile $b_\mu(\theta)=n_\mu(\theta_)/\overline{n}_\mu$.}
\end{deluxetable*}

To further reduce residual contamination by bright foreground objects, we apply bright magnitude cuts of $z^\prime > 21$ and $22$~mag for the red and blue photometry samples, respectively \citep{Medezinski+2010,Medezinski2018src}. These selection criteria yield a total of $31306$ and $14954$ galaxies for the red and blue photometry samples, respectively. For our shear analysis, we use the weak-lensing-matched, blue and red composite sample containing $16667$ galaxies with usable $R_\mathrm{C}$ shape measurements, corresponding to a mean surface number density of $\overline{n}_g\approx 21$~galaxies~arcmin$^{-2}$ (Table~\ref{tab:gsample}).

To measure the magnification bias, we use magnitude-limited samples of CC-red galaxies. For the measurement of density depletion (Section~\ref{subsec:magbias}), we define a ``lensing-cut'' sample by applying a faint magnitude cut of $z^\prime=25.6$~mag to the red photometry sample (Table~\ref{tab:musample}).\footnote{Our CC-cut selection is not expected to cause incompleteness at the faint end in the bluer filters (see \citealt{Hildebrandt+2012cfhtlens} for a general discussion) because we have deeper photometry in the bluer bands (\citealt{BTP1995}) and our CC-red galaxies are relatively blue in $B-R_\mathrm{C}$ (Figure~\ref{fig:cc_ng}).}  On the other hand, since the net effect of magnification bias is expected to vanish at $s=0.4$ (Section~\ref{subsec:magbias}), lensing magnification also provides a null test, which allows us to assess the level of residual bias that could be present in the measurement for the lensing-cut sample \citep[see][]{Chiu2020hscmag,Umetsu2020rev}. To this end, we define a ``null-test'' sample with a faint magnitude cut of $z^\prime=23.6$~mag, at which the count slope is found to be $s=0.404\pm 0.069$ (Table~\ref{tab:musample}).

\subsection{Lensing Depth Estimation}
\label{subsec:depth}

To assess the mean lensing depth ($\langle\beta\rangle,\langle\beta^2\rangle$; see Equation~(\ref{eq:depth})) for our CC-cut samples, we use the COSMOS2020 \textsc{farmer} catalog with robust photometry and photo-$z$ measurements. For each background sample, we apply the same cuts to the COSMOS multi-band photometry and obtain the redshift distribution $N(z)$ of the selected galaxies. The lensing weight $w_g$ (see Section~\ref{subsec:shear}) is not taken into account in the depth estimation, because there are no photo-$z$ estimates available for our background sample in the A370 field.\footnote{The effect of neglecting the lensing weight $w_g$ was checked using photo-$z$ and shape catalogs based on Suprime-Cam 5-band imaging available for CLASH clusters at similar redshifts, $\zl\in [0.35,0.40]$ \citep{Umetsu2014clash}. The fractional differences in the estimated $\langle\beta\rangle$ values are found to be $<1\percent$, which is not significant compared to the total fractional uncertainty of $5\percent$ adopted in this study.} The resulting depth estimates for our shear and magnification analyses are summarized in Tables~\ref{tab:gsample} and \ref{tab:musample}, respectively.\footnote{The expected contribution of foreground galaxies with $\beta(\zl,\zs)=0$ is accounted for in our lensing depth estimation (see Equation~(\ref{eq:depth})).}

For a consistency check, we also make use of photo-$z$ estimates from alternative aperture-based COSMOS2020 photometry, \textsc{Classic} \citep{COSMOS2020}. For each sample, we obtain consistent depth estimates (to within $1\percent$) from the \textsc{farmer} and \textsc{classic} catalogs. Taking into account the field-to-field variance in $N(z)$ \citep[][see their Section~4.4]{Umetsu2014clash}, we assume a fractional uncertainty of $5\%$ in the COSMOS-based estimates of $\langle\beta\rangle$. We marginalize over this uncertainty in our mass reconstruction (Section~\ref{sec:mrec}).

The level of residual cluster contamination for the CC-cut method has been assessed by \citet{Umetsu2016clash} using large spectroscopic samples from the CLASH-VLT program \citep{Rosati2014VLT}. Combining VLT spectroscopic redshifts and Subaru multi-band photometry available for 10 southern CLASH clusters with a mean redshift of $z\approx 0.37$, \citet{Umetsu2016clash} found a mean contamination fraction of $(2.4\pm 0.7)\percent$ in the blue+red CC-cut sample. This level of residual contamination is subdominant compared to other uncertainties in our lensing analysis.

\subsection{Null Tests}
\label{subsec:null}


\begin{figure}[tbp]
  \begin{center}
   \includegraphics[width=0.4\textwidth,angle=0,clip]{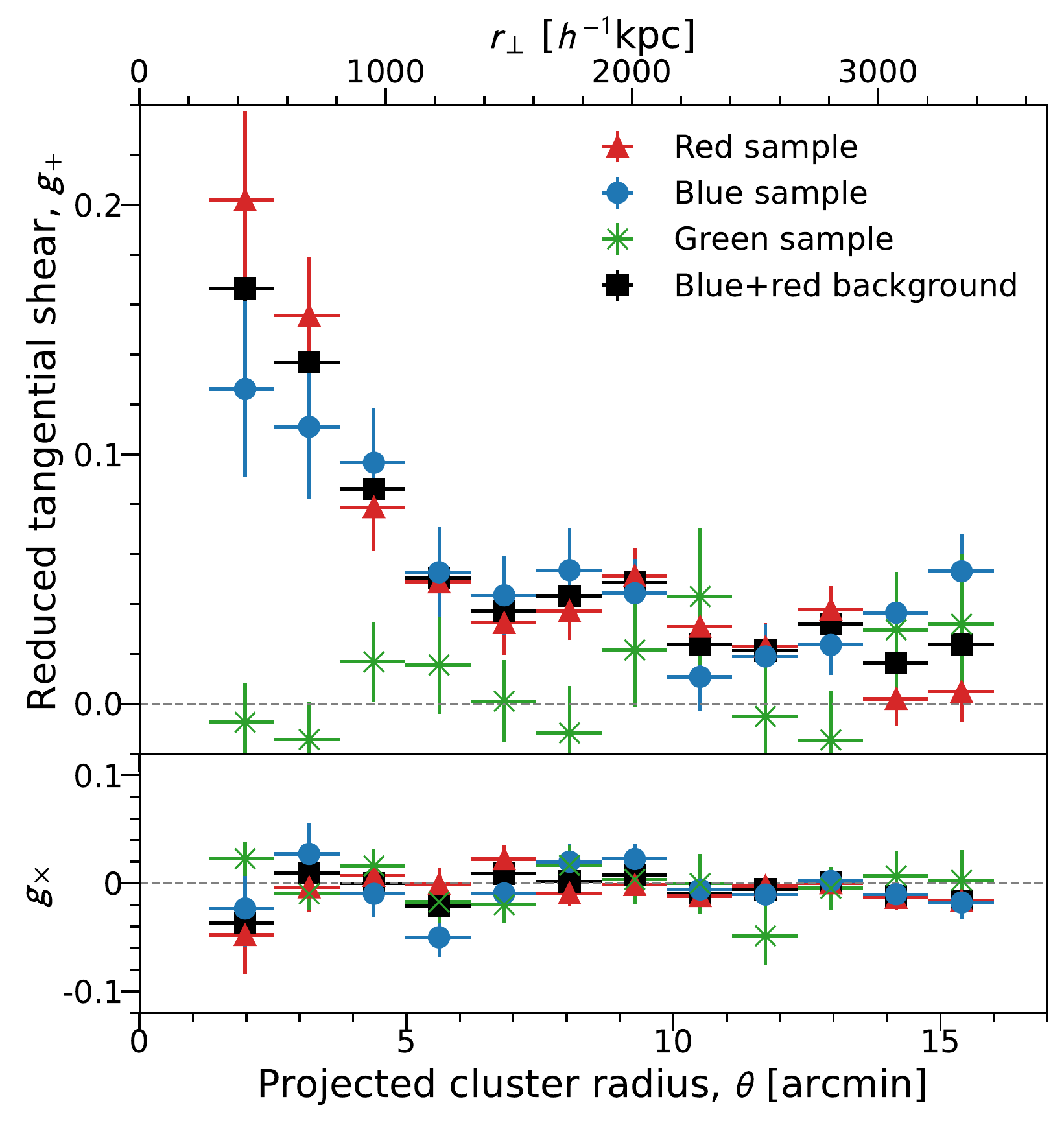}
  \end{center}
\caption{Azimuthally averaged radial profiles of the tangential and cross shear components, $g_+$ (upper panel) and $g_\times$ (lower panel), respectively, for our red (triangles), blue (circles), green (crosses), and blue+red (squares) galaxy samples.}
\label{fig:gt}
\end{figure}

Figure~\ref{fig:gt} shows the azimuthally averaged tangential ($g_+$) and cross ($g_\times$) components of the reduced shear as a function of projected cluster radius. We find a rising $g_+(\theta)$ profile toward the cluster center from both blue and red background samples. In contrast, the $g_+(\theta)$ signal for the green sample is suppressed by the inclusion of cluster members and consistent with zero at $R\simlt 2\Mpch$, while it becomes comparable to the pure background signal outside the cluster region.  

In the absence of higher-order effects, weak lensing only produces tangential shape distortions (Section~\ref{subsec:basics}). The presence of $\times$ distortions can thus be used to check for systematic errors. Here we use a $\chi^2$ test to assess the statistical significance of the measured $\times$-mode signal against the null hypothesis. We find $\chi^2$ values of the null hypothesis to be $10.6, 16.4, 9.8$, and $14.9$ for $\Nbin=12$ degrees of freedom, for the red, blue, green, and blue+red samples, respectively. For all the cases tested, the $\times$-component signal is statistically consistent with a null detection.


\begin{figure}[htbp]
  \begin{center}
   \includegraphics[width=0.4\textwidth, angle=0, clip]{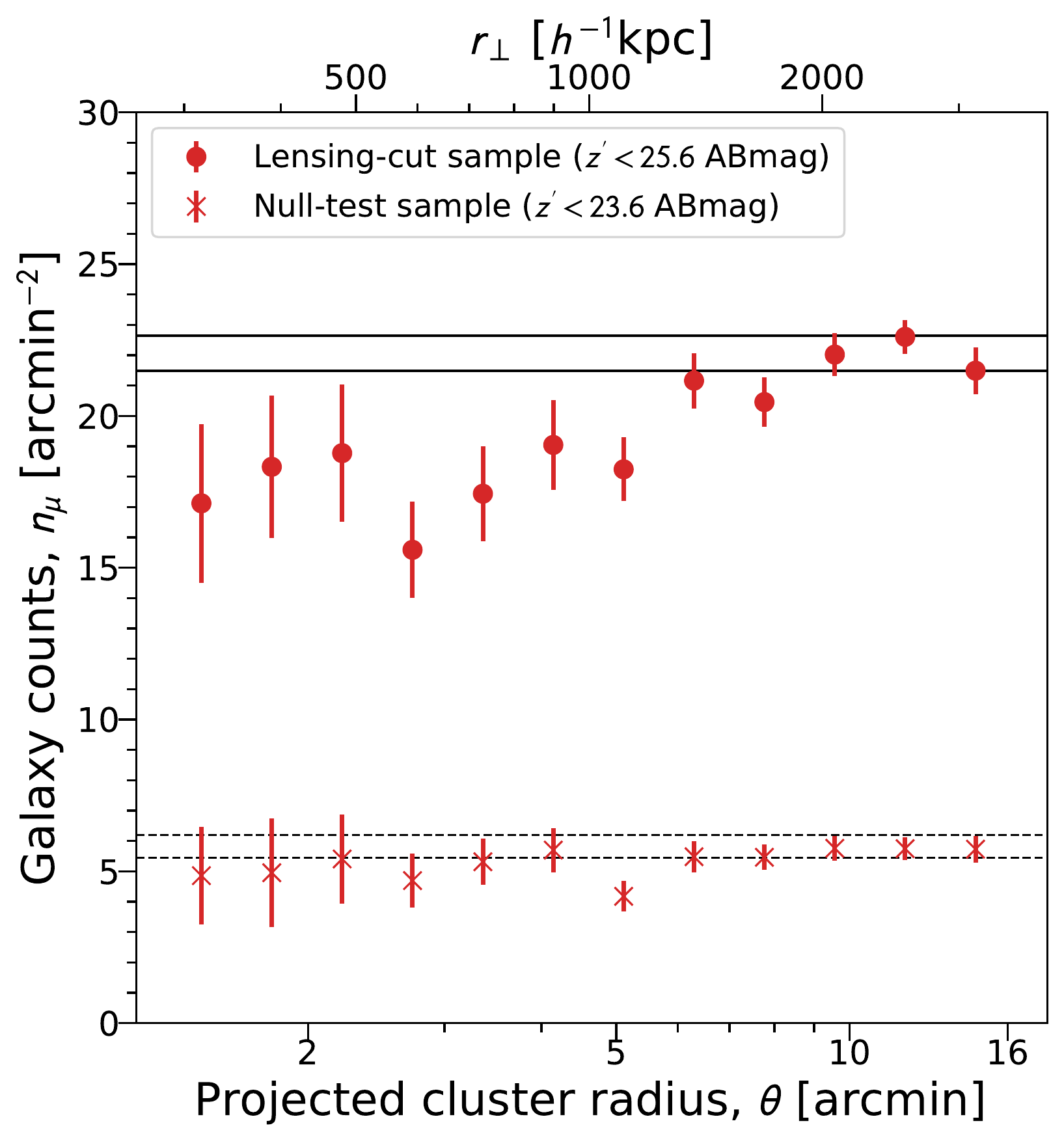}
  \end{center}
\caption{Coverage- and mask-corrected surface number density profiles $n_\mu$ of \BRz-selected red background samples. The results are shown for our lensing-cut (circles) and null-test (crosses) samples. The error bars include both Poisson and clustering contributions estimated from the data. For the lensing-cut sample, a radial count depletion due to magnification of the sky area is seen toward the cluster center. For the null-test sample with $s\approx 0.4$, the net effect of magnification bias is expected to vanish. The mean background levels estimated for the lensing-cut and null-test samples are marked with solid and dashed horizontal lines, respectively.
\label{fig:null}}
\end{figure}

Figure \ref{fig:null} shows the coverage- and mask-corrected surface number density of background galaxies as a function of projected cluster radius, for the lensing-cut and null-test samples. In both cases, no clustering is observed toward the center, demonstrating that there is no detectable contamination by cluster members. The lensing-cut sample reveals a systematic decrease in their counts toward the cluster center, caused by magnification of the sky area. In contrast, the null-test sample shows no significant evidence for radial count variations with $\chi^2=10.0$ for $12$ degrees of freedom, as expected by their count slope. A more quantitative magnification analysis will be discussed in Section~\ref{sec:mrec}.


\section{Weak-lensing Mass Reconstruction}
\label{sec:mrec}

\subsection{Mass Profile Reconstruction (WL-1D)}
\label{subsec:wl1d}


\begin{figure}[tbp]
  \begin{center}
   \includegraphics[width=0.4\textwidth, angle=0, clip]{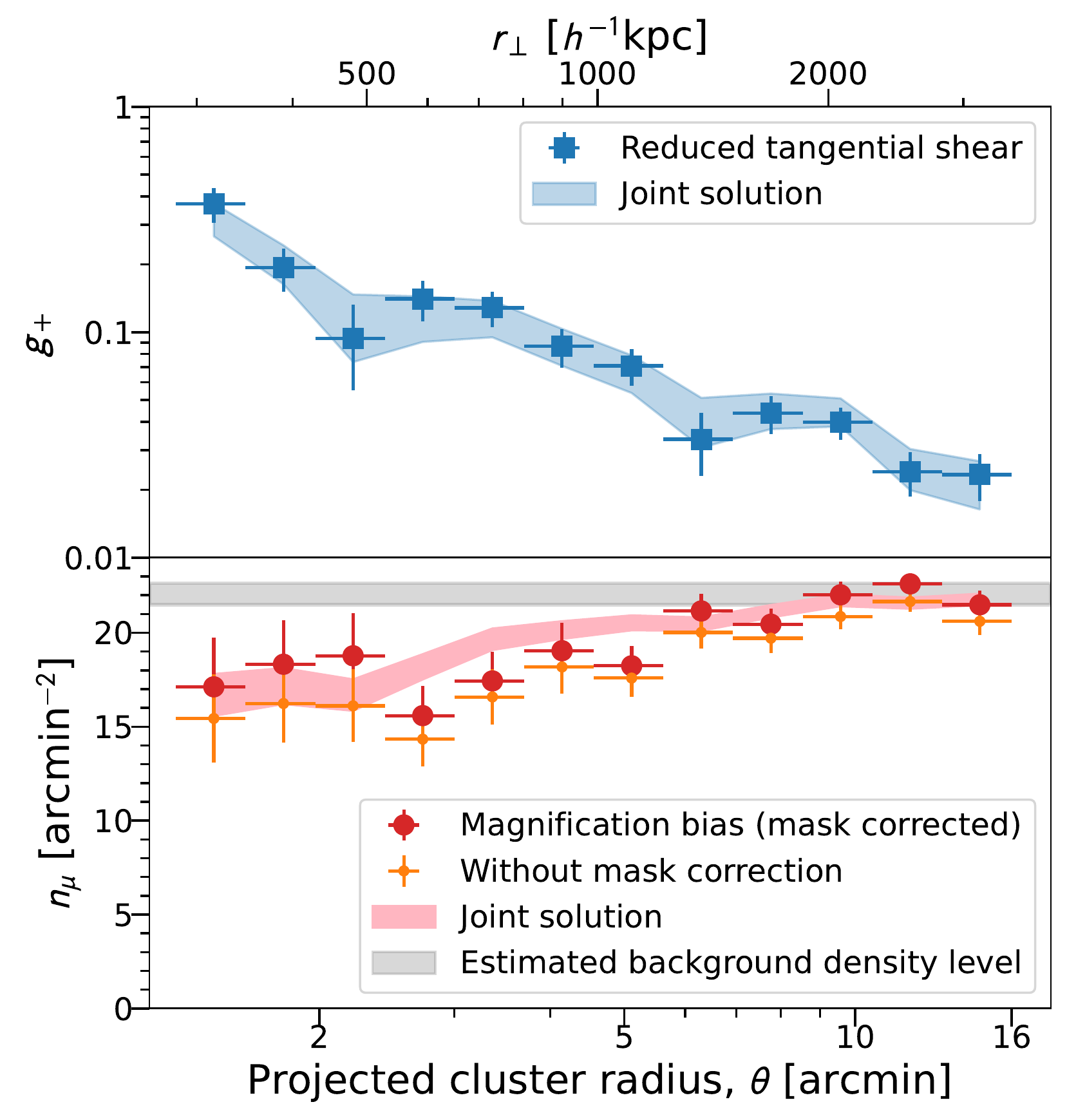}
  \end{center}
\caption{
Azimuthally averaged cluster lensing profiles of A370 obtained from Subaru observations. The upper panel shows the reduced tangential shear profile $g_+$ (blue squares) based on the blue+red background sample. The lower panel shows the magnification bias profile $n_\mu$ measured from our $z^\prime$-limited sample of red background galaxies, with (red circles) and without (orange dots) the mask correction. For each observed profile, the shaded area represents the $1\sigma$ confidence region of the joint reconstruction from the shear and magnification profile measurements. The horizontal bar (gray shaded region) shows the estimated mean background level.}
\label{fig:wl}
\end{figure}


\begin{figure}[tbp]
  \begin{center}
   \includegraphics[width=0.45\textwidth, angle=0, clip]{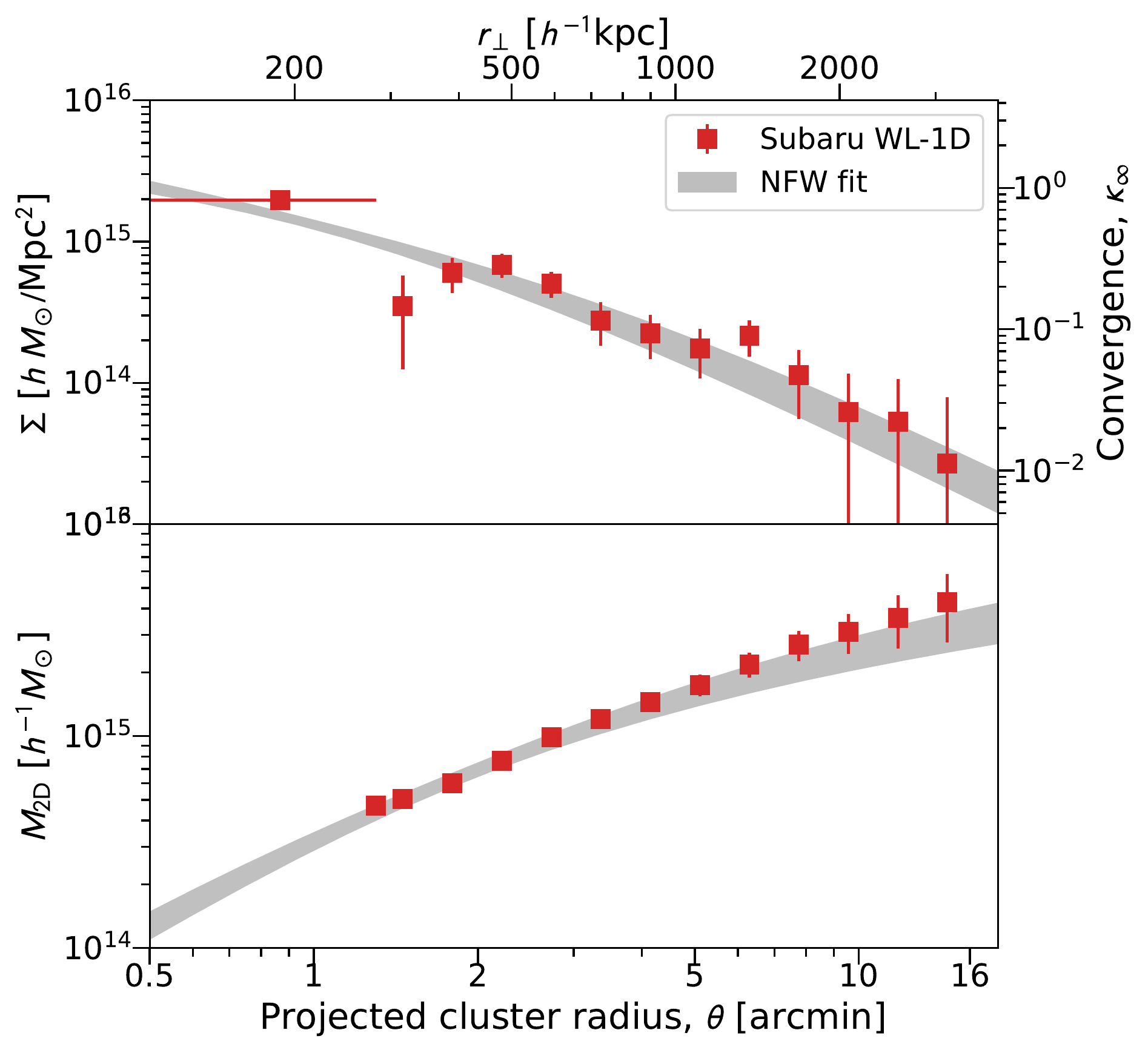}
  \end{center}
 \caption{Surface mass density profile $\Sigma(\theta)$ (upper panel, red squares) derived from a joint analysis of azimuthally averaged shear and magnification measurements shown in Figure~\ref{fig:wl}. The lower panel shows the corresponding cumulative mass profile $\Mproj(<\theta)$ (red squares). The innermost bin is an integrated average inside $\theta_\mathrm{min}=1.3\arcmin$. The gray shaded area in each panel represents the marginalized $1\sigma$ confidence interval of the NFW fit to the $\Sigma$ profile.}
 \label{fig:mplot} 
\end{figure}

Before carrying out a 2D mass reconstruction, we first perform a weak-lensing 1D radial profile analysis (WL-1D) of our Subaru observations (Section~\ref{sec:data}). A370 has two central BCGs separated by $\approx 37\arcsec$ (about $140\kpch$ at $z=0.375$) along the north--south direction (Figure~\ref{fig:rgb}). In this work, we adopt the optical center, or the midpoint of the two BCGs (see Table~\ref{tab:cluster}), as the cluster center for our radial profile analysis.

We derive azimuthally averaged radial profiles of tangential reduced shear ($g_+$) and magnification bias ($n_\mu$) from our Subaru/Suprime-Cam data. We calculate the binned lensing profiles, $\{g_{+,i}\}_{i=1}^{\Nbin}$ and $\{n_{\mu,i}\}_{i=1}^{\Nbin}$, in $\Nbin=12$ logarithmically spaced radial bins centered on the cluster, spanning the range from $\theta_\mathrm{min}=1.3\arcmin$ to $\theta_\mathrm{max}=16\arcmin$, with a logarithmic spacing of $\Delta\ln\theta\approx 0.21$. Our radial profile analysis begins at $\theta_\mathrm{min}=1.3\arcmin$, which is sufficiently large compared to twice the effective Einstein radius, $2\Rein\sim 1.1\arcmin$ (for $\zs=2$; Table~\ref{tab:cluster}), determined from strong-lens modeling by \citet{Kawamata2018}. Hence, our analysis does not include outer multiple images of strongly lensed galaxies lying at $\theta\sim [\Rein, 2\Rein]$, and our data do not resolve the central substructures. The outer boundary $\theta_\mathrm{max}=16\arcmin$ ($\approx 3.5\Mpch$) is large enough to encompass the entire cluster region with $r_\mathrm{vir}\sim 2\Mpch$ \citep{Umetsu+2011}, but sufficiently small compared to the size of the Suprime-Cam field of view so as to ensure accurate PSF correction. 

For the magnification analysis, the count normalization parameter $\overline{n}_\mu$ is estimated in the reference background region at $\theta\in [12\arcmin,16\arcmin]$.\footnote{The 2-halo term ($\kappa\simlt 10^{-2}$) does not cause bias in the mass reconstruction, because the range of the prior on $\overline{n}_\mu$ is sufficiently wide. See \citet[][their Section 7.4.2]{Umetsu2014clash} for detailed discussion.}  The estimated values and errors for $\overline{n}_\mu$ and $\slim$ are summarized in Table~\ref{tab:musample}. Details of the error analysis and the mask correction procedure are described in Appendix~\ref{appendix:magbiasdata}. 

We reconstruct the radial mass profile of A370 from a joint likelihood analysis of azimuthally averaged shear and magnification constraints, using the \textsc{clumi} code of \citet[][]{Umetsu2013}. We have a total of 24 constraints $\{g_{+,i}, n_{\mu,i}\}_{i=1}^{\Nbin}$ in 12 radial  bins. The model is described by $\Nbin+1=13$ parameters, $\bmprof=\{\Sigma_\mathrm{min},\Sigma_i\}_{i=1}^{\Nbin}$, where $\Sigma_\mathrm{min}\equiv \Sigma(<\theta_\mathrm{min})$ is the average surface mass density interior to $\theta_\mathrm{min}$\footnote{The central surface mass density $\Sigma(<\theta_\mathrm{min})$ can be fully determined by the combination of tangential-shear and magnification measurements outside $\theta_\mathrm{min}$ \citep[see][]{Umetsu2013}.} and $\Sigma_i$ is the surface mass density averaged in the $i$th bin. In addition, we account for the calibration uncertainty in the observational parameters $\bc$ (Equation~(\ref{eq:calib}); see Tables~\ref{tab:gsample} and \ref{tab:musample}).\footnote{The \textsc{clumi} algorithm employs uniform priors on the calibration parameters $\bc$ and explores parameter space with a Markov chain sampling method \citep{Umetsu2013}, whereas \textsc{clumi}-2D uses Gaussian priors on $\bc$ to find the global solution $\widehat{\blambda}$ with the conjugate-gradient method (Section~\ref{subsec:cmat}).} Following \citet{Umetsu2014clash}, we fix $f_{W,g}$ to the estimated value.

Figure~\ref{fig:wl} compares the observed lensing profiles $\{g_{+,i}, n_{\mu,i}\}_{i=1}^{\Nbin}$ with the respective joint reconstructions. The joint solution has a $\chi^2$ value of $18.7$ for $11$ degrees of freedom, indicating a slight (but statistically not significant) discrepancy between the two data sets. We see from the lower panel of Figure~\ref{fig:wl} that the measured $n_\mu$ value at $\theta\sim 5\arcmin$ is $\sim 2\sigma$ lower than expected from the joint reconstruction. This is consistent with the result for the null-test sample, which exhibits a similar local deficit of the galaxy counts in the same radial bin (see Figure~\ref{fig:null}). For the other bins, we find a good agreement between the shear and magnification data. The reconstructed $\Sigma(\theta)$ profile is shown in the upper panel of Figure~\ref{fig:mplot}, along with the $1\sigma$ confidence interval of the spherical Navarro--Frenk--White \citep[][hereafter NFW]{NFW1996,NFW1997} model (see Section~\ref{sec:model} for details of the modeling). The corresponding cumulative mass profile $\Mproj(<\theta)=\pi (D_l\theta)^2\Sigma(<\theta)$ is shown in the lower panel of Figure~\ref{fig:mplot}.

\subsection{Two-dimensional Map Making (WL-2D)}
\label{subsec:wl2d}


\begin{figure}[tbp] 
  \begin{center}
   \includegraphics[width=0.4\textwidth, angle=0, clip]{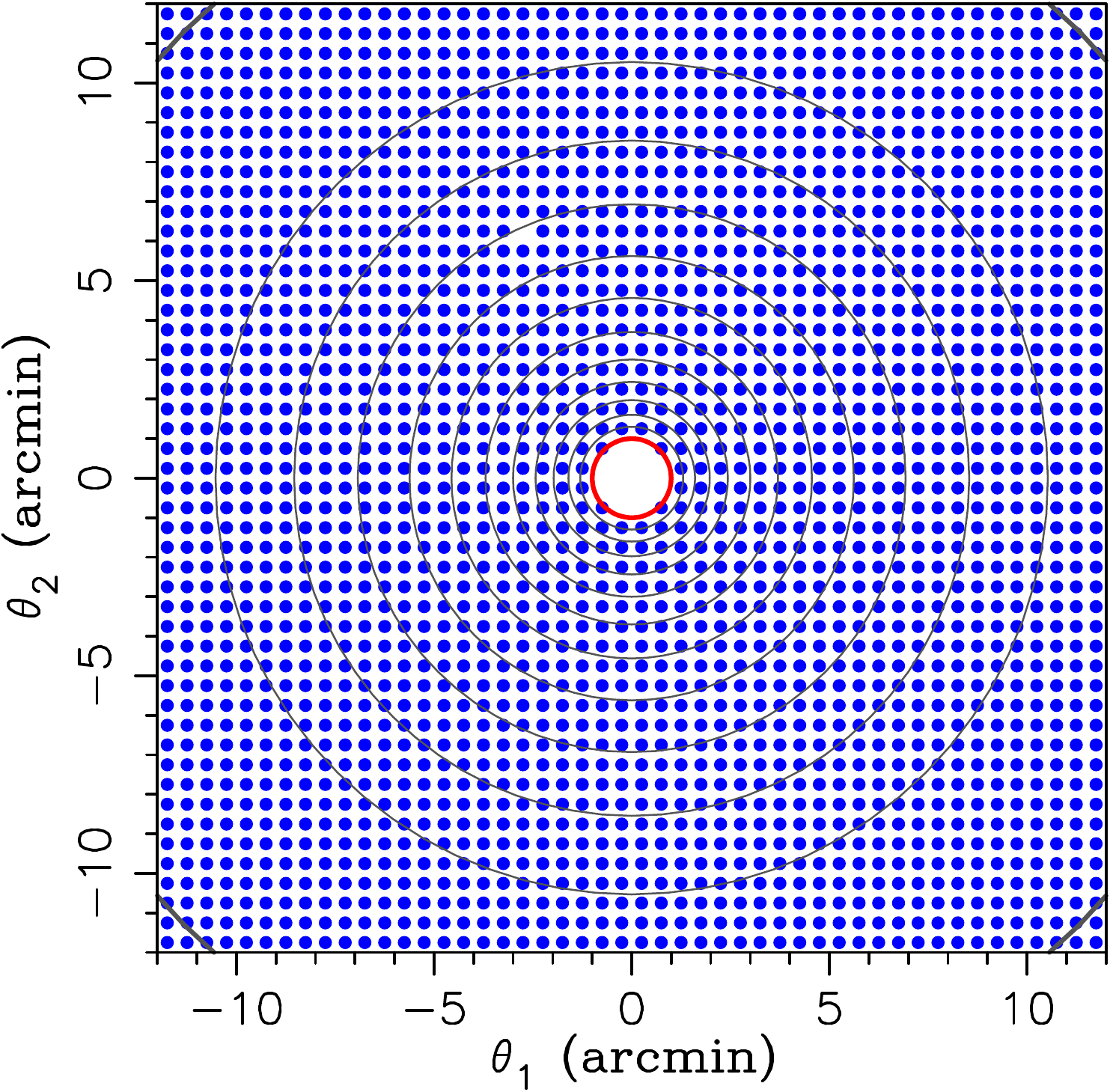}
  \end{center}
\caption{
Spatial distribution of reduced shear constraints  ($g_1,g_2$) averaged onto a mass grid of $48\times 48$ pixels, covering a field of $24\arcmin \times 24\arcmin$ centered on A370. Each point represents a single pixel with measured ($g_1,g_2$) averaged within a top-hat region with radius $\theta_\mathrm{f}=0.4\arcmin$. We exclude from our analysis those pixels lying within the central $1\arcmin$ region (red circle) and those having no background galaxies with usable shape measurements. Azimuthally averaged magnification constraints are obtained in $12$ logarithmically spaced annuli centered on the cluster spanning the range $\theta\in[1.3\arcmin,16\arcmin]$. 
 \label{fig:grid} }
\end{figure}


\begin{figure}[tbp] 
  \begin{center}
   \includegraphics[width=0.45\textwidth, angle=0, clip]{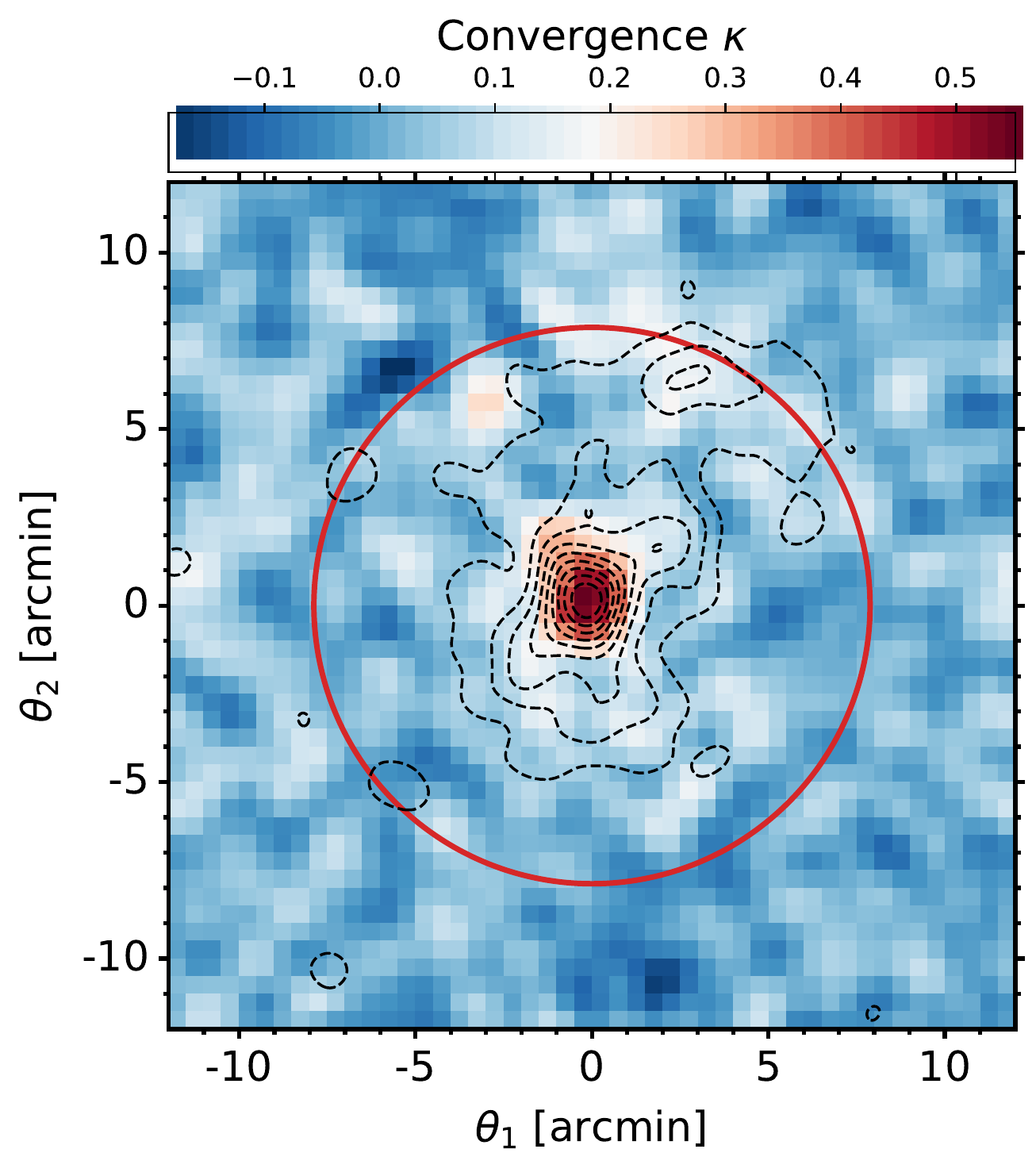} 
  \end{center}
 \caption{Projected mass distribution $\Sigma(\btheta)$ of A370 reconstructed from a joint analysis of 2D shear and azimuthally averaged magnification data (Figure~\ref{fig:grid}). The map is $24\arcmin\times 24\arcmin$ in size ($5.2\Mpch$ on a side at $z=0.375$) and centered on the cluster. The color bar indicates $\kappa=\SigmaCrit^{-1}\Sigma$ scaled to the mean lensing depth, $1/\langle\SigmaCrit^{-1}\rangle = 3.77\times 10^{15}~h\Msun\Mpc^{-2}$. For visualization purposes, the mass map is smoothed with a $3\times 3$~pixel boxcar kernel. The dashed contours show the surface density distribution of red-sequence cluster galaxies, smoothed with a Gaussian of $1.2\arcmin$~FWHM. The lowest contour level and the contour interval are both $10\percent$ of the peak density $n\approx 47$~galaxies~arcmin$^{-2}$. The red circle indicates the cluster radius of $r_{200}\approx 1.7\Mpch$. North is to the top, east to the left.}
 \label{fig:kmap}
\end{figure}

We apply our \textsc{clumi}-2D method (Section~\ref{subsec:massrec}) to our Subaru/Suprime-Cam data (Section~\ref{sec:data}) for obtaining an unbiased recovery of the projected mass distribution $\Sigma(\btheta)$ in A370. In this analysis (WL-2D), we combine the observed shear field $(g_1(\btheta),g_2(\btheta))$ with the azimuthally averaged magnification data $\{n_{\mu,i}\}_{i=1}^{\Nbin}$ (Section~\ref{subsec:wl1d}), which impose a set of azimuthally integrated constraints on the underlying $\Sigma(\btheta)$ field. \textsc{clumi}-2D takes into account the nonlinear subcritical regime of the lensing properties.

For mapmaking, we pixelize the $\gamma_\infty(\btheta)$ and $\kappa_\infty(\btheta)$ fields into a uniform grid of $48\times 48$ pixels with $\Delta\theta=0.5\arcmin$ spacing, covering the central $24\arcmin\times 24\arcmin$ field. The model $\blambda=(\bm,\bc)$ is specified by $\Npix=48^2$ parameters, $\bm=\{\Sigma(\btheta_n)\}_{n=1}^{\Npix}$, and a set of calibration parameters $\bc$ to marginalize over (Tables~\ref{tab:gsample} and \ref{tab:musample}). We utilize the \textsc{fftw} implementation of fast Fourier transforms to compute $\gamma_\infty(\btheta)$ from $\kappa_\infty(\btheta)$ using Equation~(\ref{eq:shear2m}). To avoid spurious aliasing effects from the periodic boundary condition, the maps are zero padded to twice the data length in each spatial dimension \cite[e.g.,][]{1998ApJ...506...64S,UB2008}. 

We use a top-hat window of $\theta_\mathrm{f}=0.4\arcmin$ (Section~\ref{subsec:shear}) to average over a local ensemble of galaxy image ellipticities at each grid point, accounting for the intrinsic ellipticity distribution of background galaxies. To avoid potential systematic errors (see Section~\ref{subsubsec:lg}), we exclude from our analysis 12 pixels lying within the central $\theta_\mathrm{cut}=1\arcmin$ and one pixel containing no background galaxies. For distortion measurements ($g_1(\btheta),g_2(\btheta)$), this leaves us with a total of $2291$ usable measurement pixels (blue points in Figure~\ref{fig:grid}), corresponding to $4582$ constraints. For magnification measurements, we have $12$ azimuthally averaged constraints $\{n_{\mu,i}\}_{i=1}^{\Nbin}$ (Figure~\ref{fig:grid}). The total number of constraints is thus $N_\mathrm{data}=4594$, yielding $N_\mathrm{data}-\Npix=2290$ degrees of freedom.

In Figure~\ref{fig:kmap}, we show the $\Sigma(\btheta)$ field reconstructed from the joint analysis of the 2D shear and azimuthally averaged magnification data. The $\chi^2$ value for the global maximum posterior solution is $\chi^2(\widehat{\blambda})=2871$ for $2290$ degrees of freedom. For comparison, we plot in Figure~\ref{fig:kmap} the surface density distribution of the green sample (dashed contours) composed mostly of cluster members. The projected mass distribution is elongated in the north--south direction and similar to that of cluster member galaxies (Figure~\ref{fig:kmap}). Our mass reconstruction barely resolves substructure features (e.g., a north--south mass extension located about $1\arcmin$ north and south of the cluster center) revealed by the free-form mass inversion of \citet{Ghosh2021} based on BUFFALO strong-lensing data. We defer a more detailed investigation of weak-lensing substructures in the A370 field to a forthcoming paper (S.-I. Tam et~al. 2022, in preparation).

\begin{deluxetable}{ccc}[tbp]
\tabletypesize{\footnotesize}
\tablecaption{Projected total mass estimates for A370 from the WL-2D analysis}
\label{tab:m2d}
\tablehead{ 
 \multicolumn{1}{c}{Aperture radius, $\theta$} &
 \multicolumn{1}{c}{$r_\perp$\tablenotemark{a}}&
 \multicolumn{1}{c}{$M_\mathrm{2D}(<\theta)$}\\
 \multicolumn{1}{c}{(arcmin)} &
 \multicolumn{1}{c}{($\Mpch$)} &
 \multicolumn{1}{c}{($10^{15}\Msunh$)}
 }
\startdata
$1.3$ &  $0.28$ & $0.51 \pm 0.05$\\
$1.6$ &  $0.35$ & $0.52 \pm 0.05$\\
$2.0$ &  $0.43$ & $0.58 \pm 0.06$\\
$2.4$ &  $0.53$ & $0.82 \pm 0.07$\\
$3.0$ &  $0.65$ & $0.99 \pm 0.09$\\
$3.7$ &  $0.80$ & $1.16 \pm 0.11$\\
$4.6$ &  $0.99$ & $1.37 \pm 0.14$\\
$5.6$ &  $1.22$ & $1.61 \pm 0.18$\\
$6.9$ &  $1.50$ & $2.09 \pm 0.24$\\
$8.5$ &  $1.85$ & $2.61 \pm 0.33$\\
$10.5$ &  $2.28$ & $3.11 \pm 0.47$\\
$13.0$ &  $2.81$ & $3.69 \pm 0.68$\\
$16.0$ &  $3.47$ & $4.30 \pm 1.02$
\enddata
\tablenotetext{a}{Clustercentric radius in physical units, $r_\perp=D_l\theta$.}
\end{deluxetable}

We construct the binned radial profiles $\Sigma(\theta)$ and $\Sigma(<\theta)$ and their associated covariance matrices from an optimally weighted projection of the $\Sigma(\btheta)$ map using the method described in Appendix~\ref{appendix:2dto1d}. We thus obtain model-independent constraints on the projected total mass $\Mproj(<\theta)=\pi (D_l\theta)^2\Sigma(<\theta)$ from our WL-2D analysis. The resulting projected mass estimates are listed in Table~\ref{tab:m2d}.

\subsection{Radial Mass Profiles}
\label{subsec:kcomp}


\begin{figure}[tbp]
 \begin{center}
  \includegraphics[width=0.45\textwidth, angle=0, clip]{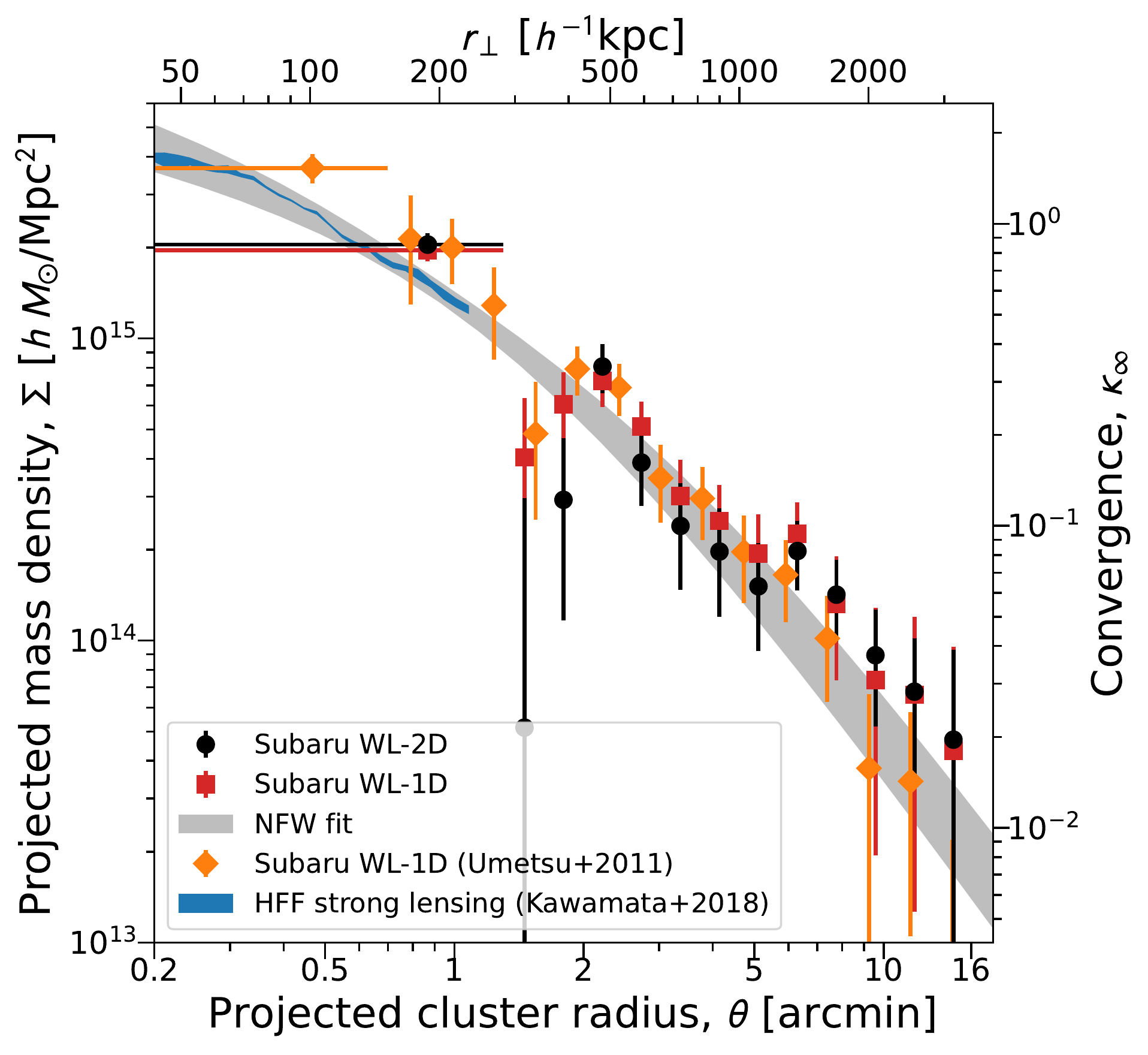}
 \end{center}
\caption{
Comparison of reconstructed surface mass density profiles $\Sigma(\theta)$ of A370. The black circles (red squares) with error bars show the results from our 2D (1D) weak-lensing reconstruction. The blue shaded area represents the $\Sigma$ profile with $2\sigma$ uncertainty derived from strong lens modeling of Hubble Frontier Fields data \citep[][]{Kawamata2018}. The orange diamonds with error bars show the results from our earlier Subaru weak-lensing analysis \citep{Umetsu+2011}. For each case, the innermost central bin $\Sigma(<\theta_\mathrm{min})$ is marked with a horizontal bar. The gray shaded area represents the marginalized $1\sigma$ confidence interval of the spherical NFW fit to the $\Sigma$ profile from our 1D reconstruction.
\label{fig:kcomp} }
\end{figure}

In Figure~\ref{fig:kcomp}, we compare the surface mass density profiles $\Sigma(\theta)$ of A370 obtained from our WL-1D (Section~\ref{subsec:wl1d}) and WL-2D (Section~\ref{subsec:wl2d}) analyses. Our 1D- and 2D-based $\Sigma$ profiles are consistent within the errors in each radial bin. The gray shaded area in the figure represents the $1\sigma$ confidence region of the spherical NFW fit to the 1D-based $\Sigma$ profile (see Section~\ref{sec:model} for details of the modeling).

For comparison, we overplot in Figure~\ref{fig:kcomp} the azimuthally averaged $\Sigma$ profile (shown out to $2\Rein\approx 1.1\arcmin$ for $\zs = 2$) based on strong lens modeling of Hubble Frontier Fields data performed by \citet[][see also \citealt{Oguri2010,Kawamata2016}]{Kawamata2018}, obtained using the technique detailed in \citet{Oguri2021enfw} to speed up lensing calculations. The inner $\Sigma$ profile derived from \HST strong lensing is in excellent agreement with our WL-1D constraints on the NFW profile.

In Figure~\ref{fig:kcomp}, our $\Sigma$ profiles are also compared with the 1D results of \citet{Umetsu+2011} based on their joint shear and magnification analysis of Suprime-Cam data. In the \citet{Umetsu+2011} analysis, the innermost measurement radius was taken to be $\theta_\mathrm{min}=0.7\arcmin$ ($\approx 1.2\Rein$ for $\zs=2$), in contrast to the conservative choice adopted in this work ($\theta_\mathrm{min}=1.3\arcmin$). The shear and magnification measurements of \citet{Umetsu+2011} are based on Suprime-Cam images reduced by \citet{Medezinski+2010}, who used the \textsc{sdfred} package \citep{Yagi2002,Ouchi2004} for flat-fielding, instrumental distortion correction, PSF matching, sky subtraction, and image stacking. Moreover, implementation details of the 1D reconstruction of \citet{Umetsu+2011} (e.g., the choice of summary statistics) are different from those of the \textsc{clumi} code that has been tested and calibrated with simulations \citep[see][]{Umetsu2013}. Nevertheless, our 1D and 2D results are both in agreement with the $\Sigma$ profile of \citet{Umetsu+2011} in the region of overlap.


\section{Mass Modeling of A370}
\label{sec:model}

In this section, we present mass modeling of A370. With ground-based Subaru weak-lensing observations alone, we cannot spatially resolve the bimodal structure of the cluster in the supercritical region (Sections~\ref{subsec:wl1d} and \ref{subsec:wl2d}). In this study, we thus restrict ourselves to single-component mass models of a spherical or ellipsoidal halo. We forward model projected cluster lensing observations by projecting a triaxial or spherical NFW halo model along the line of sight \citep[e.g.,][]{Corless2009triaxial,Sereno+Umetsu2011,Umetsu2015,Chiu2018clump3d}.

\subsection{Dynamical State of A370}
\label{subsec:dynamical}


\citet{Molnar2020} studied the dynamics of A370 using dedicated $N$-body hydrodynamical simulations of binary cluster mergers constrained by multi-probe observations. In their simulations, the initial virial masses of two progenitors were fixed to $1.2\times 10^{15}\Msunh$ and $1.1\times 10^{15}\Msunh$ with a sum of $2.3\times 10^{15}\Msunh$, according to the result of \citet{Umetsu+2011} who conducted a weak-lensing shear and magnification analysis (see Section~\ref{subsec:kcomp}) of five superlens clusters with Suprime-Cam data. In combination with mass profile constraints from \HST strong-lensing data, \citet{Umetsu+2011} obtained $M_\mathrm{vir}=2.28^{+0.26}_{-0.22}\times 10^{15}\Msunh$ for A370 assuming a generalized form of the NFW profile. 

\citet{Molnar2020} found that initial conditions of the two progenitors with an infall velocity of $3500$~km~s$^{-1}$ and an impact parameter of $70\kpch$ can reproduce the positions and the offsets between the peaks of the X-ray emission and the total mass surface density, the amplitude of the integrated SZE signal \citep{Czakon2015}, and the relative line-of-sight velocity between the two BCGs ($V\approx 1024$~km~s$^{-1}$). Moreover, the best-matching simulation reproduces well the velocity dispersion and the line-of-sight velocity distribution of cluster member galaxies \citep{Lagattuta2019a370,Molnar2020}. These simulation results support the large total mass of A370 derived from lensing \citep{Umetsu+2011}.  

The binary merger simulations of \citet{Molnar2020} suggest that A370 is a post-major merger of two similar-mass clusters, viewed after the second core passage in the infalling phase, just before the third core passage. These results also suggest that the mass distribution of A370 is highly elongated along the current direction of the collision axis, which is oriented close to the line of sight in their best simulation, with a viewing angle of $\vartheta=17.6^\circ\pm 3.5^\circ$, or $\cos\vartheta=0.95\pm 0.02$.

\subsection{Triaxial NFW Model}
\label{subsec:triNFW}

Triaxial modeling of density profiles gives an improved description of simulated \LCDM halos over the conventional spherical model \citep{2002ApJ...574..538J,Kasun+Evrard2005}. In this work, we model the cluster mass distribution with a triaxial NFW density profile. The radial dependence of the spherical NFW profile is given by \citep[][]{NFW1996,NFW1997}
\begin{equation}
\label{eq:triNFW}
  \rho(r) = \frac{\rho_\mathrm{s}}{(r/r_\mathrm{s})(1+r/r_\mathrm{s})^2},
\end{equation}
where $\rho_\mathrm{s}$ is the scale density and $r_\mathrm{s}$ is the characteristic scale radius at which the logarithmic slope of the density profile equals $-2$. We generalize the NFW density profile $\rho(r)$ to obtain its triaxial expression by replacing $r$ and $r_\mathrm{s}$ with their respective ellipsoidal radii $R$ and $R_\mathrm{s}$ as
\begin{equation}
 \label{eq:ellipsoid}
  R^2 = \frac{X^2}{q_a^2} + \frac{Y^2}{q_b^2} + Z^2,
\end{equation}
with $q_a$ and $q_b$ the minor--major and intermediate--major axis ratios, respectively.  By definition, we have $0<q_a\le q_b\le 1$.

The degree of triaxiality is defined as \citep{Sereno2013glszx}
\begin{equation}
\label{eq:Tri}
{\cal T} = \frac{1-q_b^2}{1-q_a^2},
\end{equation}
where $0\le {\cal T}\le 1$ by construction. The value of ${\cal T}$ approaches unity at $q_a=q_b$ (or zero at $q_b=1$), if the halo shape is maximally prolate (or oblate). For $q_a=q_b=1$, Equation~(\ref{eq:triNFW}) reduces to the spherical NFW profile $\rho(r)$ with $r=\sqrt{X^2+Y^2+Z^2}$.
 
We define an ellipsoidal overdensity radius $R_{\Delta}$ \citep[][]{Corless2009triaxial,Sereno+Umetsu2011,Buote+Humphrey2012p2} such that the mean interior density contained within an ellipsoidal volume of semimajor axis $R_{\Delta}$ is $\Delta\times \rho_\mathrm{c}(\zl)$. The total mass enclosed within $R_{\Delta}$ is expressed as 
\begin{equation}
\label{eq:Mdelta}
M_{\Delta}=\frac{4\pi \Delta}{3} q_a q_b \rho_\mathrm{c}(\zl) R_{\Delta}^3. 
\end{equation}
Spherical-equivalent overdensity radii $r_\Delta$ are defined by
\begin{equation}
 r_\Delta = (q_a q_b)^{1/3} R_\Delta.
\end{equation}
Similarly, we define $r_\mathrm{s} = (q_aq _b)^{1/3}R_\mathrm{s}$.

The triaxial concentration parameter is defined as the ratio of the ellipsoidal overdensity radius $R_\Delta$ to the scale radius $R_\mathrm{s}$ along the major axis,
\begin{equation}
c_{\Delta} := \frac{R_{\Delta}}{R_\mathrm{s}} = \frac{r_\Delta}{r_\mathrm{s}}.
\end{equation}
The characteristic density is then expressed as $\rho_\mathrm{s}=(\Delta\rho_\mathrm{c}/3)\times c_\Delta^3/[\ln(1+c_\Delta)-c_\Delta/(1+c_\Delta)]$.  In this study, we use $\Delta=200$ to define the halo mass, $M_{200}$, and the concentration parameter, $c_{200}$. 

A triaxial halo is projected onto the lens plane as elliptical isodensity contours, which can be expressed as a function of the intrinsic halo axis ratios ($q_a,q_b$) and orientation angles ($\vartheta,\phi,\psi$) with respect to the observer's line of sight. Following \citet{Umetsu2015}, we adopt the $z$-$x$-$z$ convention of Euler angles ($\vartheta,\phi,\psi$) to be consistent with \citet[][see also \citealt{Sereno+Ettori+Baldi2012}]{Stark1977}. The angle $\vartheta$ represents the inclination of the major axis ($Z$) with respect to the line of sight.

After a rotation by the first two Euler angles ($\vartheta,\phi$), elliptical isodensity contours of the projected ellipsoid can be described as a function of the elliptical radius $\zeta$,
expressed in terms of projected Cartesian coordinates $(x^\prime, y^\prime)$ as
\begin{equation}
\label{eq:rellip}
\zeta^2 = 
\frac{1}{f}\left(
j x^{\prime 2} + 2k x^\prime y^\prime + l y^{\prime 2}
\right),
\end{equation}
where
\begin{equation}
 \label{eq:ABC}
 \begin{aligned}
j &=\cos^2\vartheta\left(\frac{1}{q_a^2}\cos^2\phi + \frac{1}{q_b^2}\sin^2\phi\right) + \frac{1}{q_a^2 q_b^2}\sin^2\vartheta,\\
k &=  \sin\phi \cos\phi \cos\vartheta \left(\frac{1}{q_a^2}-\frac{1}{q_b^2}\right),\\
l &= \frac{1}{q_a^2}\sin^2\phi + \frac{1}{q_b^2}\cos^2\phi,\\
f &= \sin^2\vartheta\left(
\frac{1}{q_a^2} \sin^2\phi 
+ 
\frac{1}{q_b^2} \cos^2\phi
\right) + \cos^2\vartheta.
 \end{aligned}
\end{equation}
The minor--major axis ratio ($\le 1$) of the elliptical isodensities is expressed as
$q_\perp(q_a, q_b, \vartheta,\phi) = \sqrt{\frac{j+l-\sqrt{(j-l)^2+4k^2}}{j+l+\sqrt{(j-l)^2+4k^2}}}.$
Finally, the third Euler angle $\psi$ describes the additional rotational degree of freedom in the sky plane to specify the observer's coordinate system $(x,y)$, defined such that $x^\prime=x\cos\psi - y\sin\psi$ and $y^\prime=x\sin\psi + y\cos\psi$.

For a self-similar mass model expressed as $\rho(R) = \rho_\mathrm{s} f_\mathrm{3D}(R/R_\mathrm{s})$, the projected mass density $\Sigma(\zeta)$ is related to $\rho(R)$ (see Equation~(\ref{eq:triNFW})) as \citep{Umetsu2015}
\begin{equation}
  \Sigma(\zeta) = \frac{2R_\mathrm{s}\rho_\mathrm{s}}{\sqrt{f}} \int_{\zeta/R_\mathrm{s}}^{\infty}\frac{f_\mathrm{3D}(u)udu}{\sqrt{u^2-(\xi/\xi_\mathrm{s})^2}} \equiv \Sigma_\mathrm{s} f_\mathrm{2D}(\xi/\xi_\mathrm{s}),
\end{equation}
where $\Sigma_\mathrm{s}$ is the scale surface mass density defined by
\begin{equation}
 \Sigma_\mathrm{s} = 2 R_\mathrm{s} \rho_\mathrm{s}/\sqrt{f},
\end{equation}
$\xi = \sqrt{x^{\prime\prime 2}+y^{\prime\prime 2}/q_\perp^2}$, and $\xi_\mathrm{s}$ is the semi-major scale length of the projected halo. Here we have chosen the new coordinate system $(x^{\prime\prime},y^{\prime\prime})$ such that the $x^{\prime\prime}$ axis is aligned with the major axis of the projected ellipse. In this study, we employ the radial dependence of the projected NFW profile $f_\mathrm{2D}(u)$ as given by \citet{2000ApJ...534...34W}.


To summarize, our mass model is specified by a total of seven parameters describing the total matter ellipsoid, namely, halo mass and concentration ($M_{200}, c_{200}$), intrinsic axis ratios ($q_a,q_b$), and three Euler angles ($\vartheta,\phi,\psi$):
\begin{equation}
 \bp = \{M_{200}, c_{200}, q_a, q_b, \vartheta, \phi, \psi\}.
\end{equation}
In this way, for a given set of the model parameters, we can project a triaxial (or spherical) NFW halo onto the lens plane and compute the surface mass density $\Sigma(x,y)$ at each angular position. As discussed in \citet{Umetsu2015} \citep[see also][]{Sereno+Umetsu2011}, however, it should be noted that 2D lensing observations can effectively constrain only four observationally accessible parameters, namely, $\Sigma_\mathrm{s}$, $\xi_\mathrm{s}$, $q_\perp$, and the position angle of the projected major axis \citep{Gavazzi2005}. That is, the deprojection of triaxial systems is intrinsically underconstrained \citep[][]{Limousin2013}. On the other hand, the spherical NFW model ($q_a=q_b=1$) is specified by two parameters, $(M_{200},c_{200})$, which can be constrained by data in principle.

\subsection{Bayesian Inference Procedure}
\label{subsec:trinfwprior}

The likelihood function ${\cal L}$ of the 2D mass distribution data $\bm=\{\Sigma(\btheta_n)\}_{n=1}^{\Npix}$ given a set of model parameters $\bp$ is expressed as \citep{Oguri2005,Umetsu2015}
\begin{equation}
 \begin{aligned}
 -2\ln{\cal L}(\bp) =& \sum_{m,n=1}^{\Npix}
 \left[\Sigma-\widehat\Sigma(\bp)\right]_m
 \left(C^{-1}\right)_{mn}
 \left[\Sigma-\widehat\Sigma(\bp)\right]_n\\
  &+ \ln\left[(2\pi)^{\Npix}\mathrm{det(C)}\right]
 \end{aligned}
\end{equation}
where $\widehat{\Sigma}_n(\bp)=\Sigma(\btheta_n|\bp)$ is the surface mass density at the grid position $\btheta_n$ predicted by the model $\bp$ and $C=C_\mathrm{stat}+C_\mathrm{lss}$ is the total covariance matrix (Equation~(\ref{eq:ctot})).

We use a Bayesian Markov Chain Monte Carlo (MCMC) algorithm to obtain a well-characterized inference of the model $\bp$. We consider the following three different modeling approaches: (1) spherical modeling with uninformative uniform priors on $\log{M_{200}}$ and $\log{c_{200}}$, (2) fiducial triaxial modeling with uninformative uniform priors on all parameters, and (3) triaxial modeling incorporating an informative line-of-sight (LOS) prior from \citet{Molnar2020}. For simplicity, we refer to these three modeling approaches as Spherical, Triaxial, and Triaxial+LOS modeling, respectively. 

Here we briefly summarize the assumed priors for each case.
\begin{enumerate}
\item Spherical modeling: We float only two parameters ($M_{200},c_{200}$) and fix the remaining parameters ($q_a=q_b=1$ and $\vartheta=\phi=\psi=0$). We employ uninformative log-uniform priors for $M_{200}$ and $c_{200}$ in the range $M_{200}/(\Msunh)\in [10^{14}, 10^{16}]$ and $c_{200}\in [1,10]$.\footnote{It is appropriate to assume a log-uniform prior, instead of a uniform prior, for a positive-definite quantity such as $M_{200}$ and $c_{200}$, especially if the quantity spans a wide dynamic range \citep{Umetsu2020rev}. Since the corresponding prior distributions in $M_{200}$ and $c_{200}$ scale as $1/M_{200}$ and $1/c_{200}$, the choice of their lower bounds is relatively important.}

\item Fiducial triaxial modeling: We use uniform priors on the intrinsic shapes ($q_a,q_b$) and orientation angles ($\cos\vartheta,\phi,\psi$), while keeping the same log-uniform priors on $M_{200}$ and $c_{200}$ as in the spherical case. We assume the following form of the prior PDF for the intrinsic axis ratios:
\begin{equation}
 P(q_a,q_b) = P(q_b|q_a) P(q_a),
\end{equation}
where
\begin{equation}
 \begin{aligned}
  &P(q_a) =
  \begin{cases}
   1/(1-q_\mathrm{min}) &  \mathrm{for}~q_\mathrm{min}<q_a\le 1\\
   0                    &  \mathrm{for}~q_a\le q_\mathrm{min}
  \end{cases},\\
  &P(q_b|q_a) =
  \begin{cases}
   1/(1-q_a) & \mathrm{for}~q_b\ge q_a\\
   0         & \mathrm{for}~q_b < q_a
  \end{cases},
 \end{aligned}
\end{equation}
and $q_\mathrm{min}=0.1$ is the lower bound of the minor-to-major axis ratio $q_a$ \citep[e.g.,][]{Oguri2005,Chiu2018clump3d}, which is introduced to exclude unstable configurations that are not expected for cluster halos. For the orientation angles, we consider a population of randomly oriented halos with $P(\cos\vartheta)=1$ for $0\le \cos\vartheta\le 1$, $P(\phi)=1/\pi$ for $-\pi/2\le \phi\le \pi/2$, and $P(\psi)=1/\pi$ for $-\pi/2\le \psi\le \pi/2$.

\item Triaxial+LOS modeling: We adopt an informative prior on $\cos\vartheta$ based on the binary merger simulations of \citet{Molnar2020}. For the other parameters, we use the same priors as for the fiducial triaxial modeling. Specifically, we employ a Gaussian prior on $\cos\vartheta$ of $0.95\pm 0.02$ (Section~\ref{subsec:dynamical}) truncated in the range $0\le\cos\vartheta\le 1$.
\end{enumerate}

For comparison purposes, we perform spherical NFW modeling with the surface mass density profile $\bmprof=\{\Sigma_\mathrm{min},\Sigma_i\}_{i=1}^{\Nbin}$ derived from the WL-1D analysis (Section~\ref{subsec:wl1d}). The likelihood function ${\cal L}(\bp)$ for the WL-1D analysis is defined as in Equation~(26) of \citet{Umetsu2014clash}. In the covariance matrix of WL-1D, we account for systematic effects due to the residual mass-sheet degeneracy, in addition to the measurement error and cosmic noise contributions (see Section~\ref{subsec:cmat}). This residual uncertainty is estimated in each $\Sigma$ bin as a difference between the joint and marginal posterior solutions \citep[see][]{Umetsu2014clash}.\footnote{Because of the large number of parameters involved, we do not explore the whole likelihood surface in the \textsc{clumi}-2D code, and thus we are not able to include the systematic term in the WL-2D analysis. As we have seen in Figure~\ref{fig:kcomp}, our WL-1D and WL-2D results are consistent with each other, with no significant evidence for a systematic offset.}

Similarly, we also perform spherical NFW modeling with the reduced tangential shear profile $\{g_{+,i}\}_{i=1}^{\Nbin}$ obtained in our WL-1D analysis, because this tangential shear fitting is the standard approach to infer cluster masses from weak-lensing data \citep[e.g.,][]{Okabe+2013,WtG3,Hoekstra2015CCCP,Schrabback2018spt}. Here we account for the measurement error and cosmic noise contributions in the covariance matrix \citep[see Section~4.4 of][]{Umetsu2020rev}.

\subsection{Posterior Parameter Constraints}
\label{subsec:posterior}


\begin{figure}[tbp]
  \begin{center}
   \includegraphics[width=0.4\textwidth, angle=0, clip]{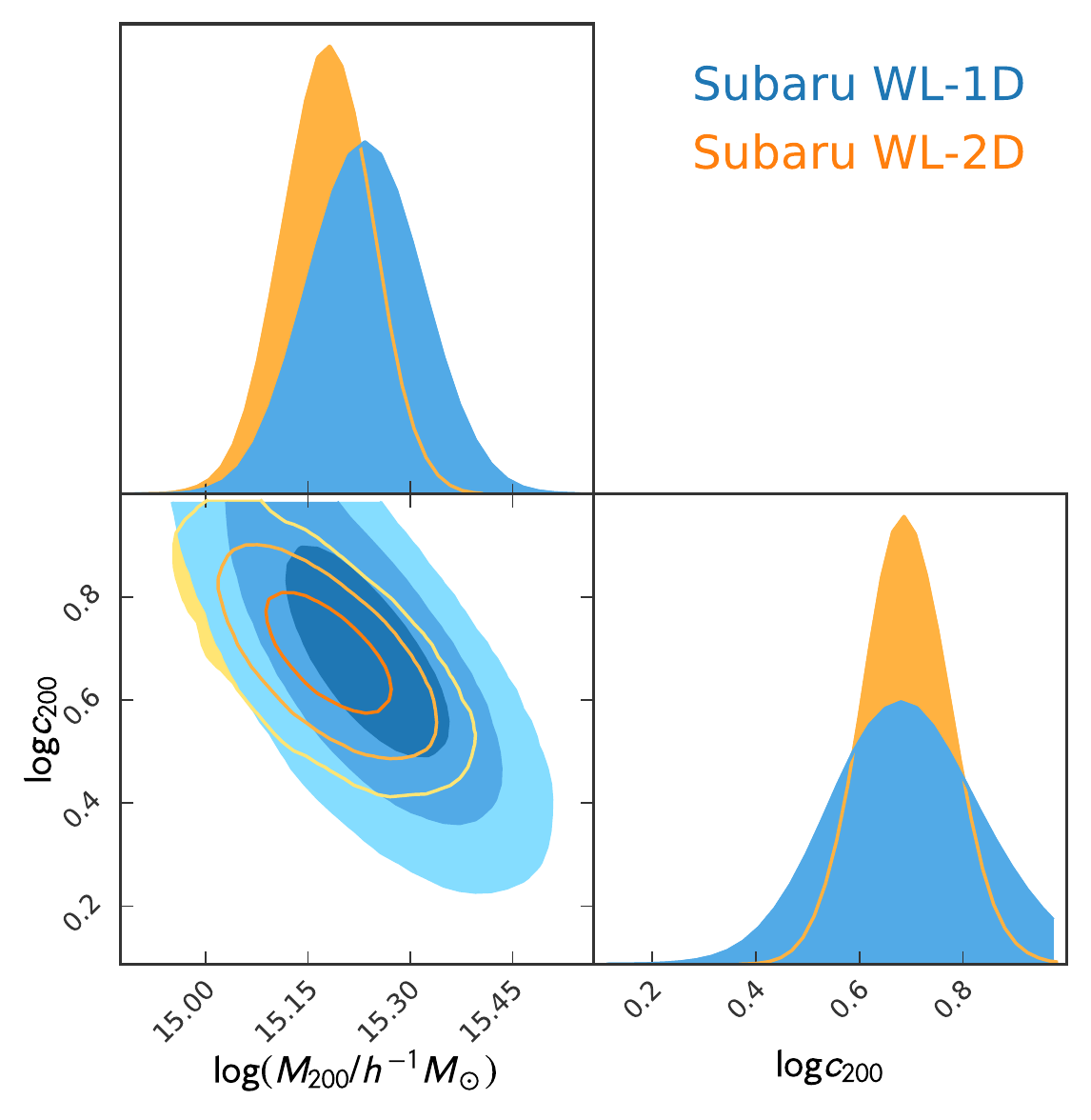}
  \end{center}
 \caption{Marginalized 1D and 2D (68\percent, 95\percent, and 99.7\percent\ confidence level contours) posterior distributions for the NFW model parameters ($\log_{10}{M_\mathrm{200}}, \log_{10}{c_\mathrm{200}}$) obtained using uniform priors assuming spherical symmetry. Blue shaded contours show the constraints obtained from the 1D mass reconstruction (WL-1D; see Figure~\ref{fig:mplot}). Orange contours show the constraints from the 2D reconstruction (WL-2D; see Figure~\ref{fig:kmap}).} 
 \label{fig:NFW} 
\end{figure}


\begin{figure*}[tbp]
  \begin{center}
   \includegraphics[width=0.5\textwidth, angle=0, clip]{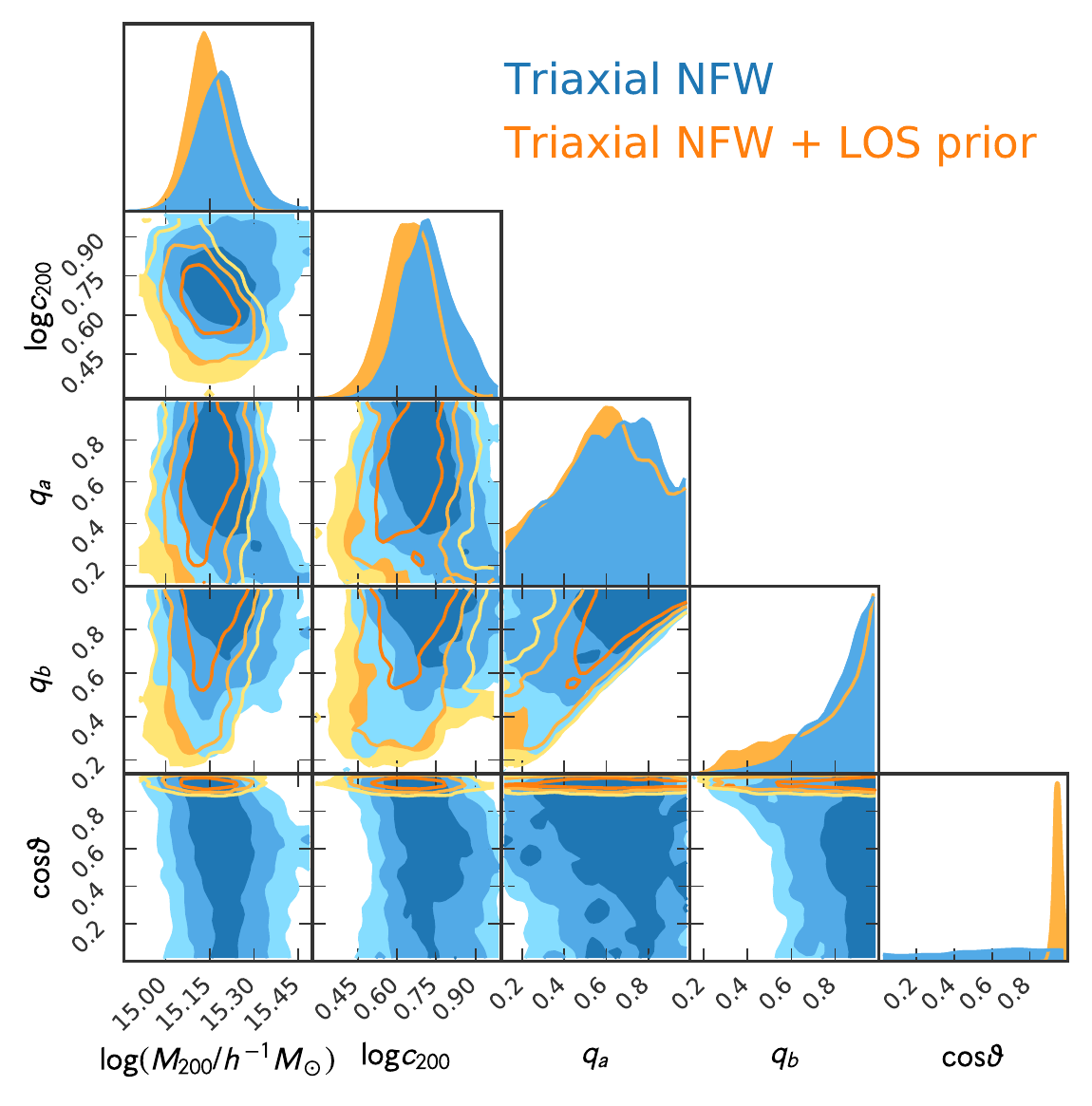} 
  \end{center}
\caption{Marginalized 1D and 2D (68\percent, 95\percent, and 99.7\percent\ confidence level contours) posterior distributions for the triaxial NFW parameters of interest $(\log_{10}{M_{200}}, \log_{10}{c_{200}}, q_a, q_b, \cos\vartheta)$. Orange and blue shaded contours show the results obtained with and without the line-of-sight prior on $\cos\vartheta$ based on binary merger simulations of \citet{Molnar2020}. For each parameter, the dashed line shows the biweight central location (\CBI) of the marginalized 1D distribution.}
 \label{fig:triNFW} 
\end{figure*}

\begin{deluxetable*}{ccccccccc}[ht]
\tabletypesize{\footnotesize}
\tablecaption{Posterior summary of the NFW halo parameters for A370}
\label{tab:nfw}
\tablehead{
\multicolumn{1}{c}{Method} &
\multicolumn{1}{c}{Data} &
\multicolumn{1}{c}{Prior} &
\multicolumn{1}{c}{$M_{200}$} &
\multicolumn{1}{c}{$c_{200}$} &
\multicolumn{1}{c}{$q_a$} &
\multicolumn{1}{c}{$q_b$} &
\multicolumn{1}{c}{$\cos\vartheta$} &
\multicolumn{1}{c}{${\cal T}$}\\
\colhead{} &
\colhead{} &
\colhead{} &
\multicolumn{1}{c}{$(10^{15}\Msunh)$} &
\colhead{} &
\colhead{} &
\colhead{} &
\colhead{} &
\colhead{} }
\startdata
    Tangential shear & $g_{+}$~profile &  Spherical & $1.55 \pm 0.26$ & $5.61 \pm 1.68$ & --- & --- & ---\\
               WL-1D & $\kappa$~profile &  Spherical & $1.72 \pm 0.32$ & $4.90 \pm 1.58$ & --- & --- & ---\\
               WL-2D &    $\kappa$~map &  Spherical & $1.51 \pm 0.22$ & $4.91 \pm 0.93$ & --- & --- & --- & ---\\
               WL-2D &    $\kappa$~map & Triaxial (fiducial) & $1.54 \pm 0.29$ & $5.27 \pm 1.28$ & $0.62 \pm 0.23$ & $0.86 \pm 0.13$ & $0.58 \pm 0.29$ & $0.51 \pm 0.29$\\
               WL-2D &    $\kappa$~map & Triaxial+LOS & $1.38 \pm 0.20$ & $4.45 \pm 0.93$ & $0.59 \pm 0.23$ & $0.80 \pm 0.21$ & $0.95 \pm 0.02$ & $0.58 \pm 0.30$
\enddata
\tablecomments{Cluster halo parameters derived from a spherical or triaxial NFW fit to Subaru weak-lensing data. We adopt a concordance cosmology of $h=0.7$, $\Omega_\mathrm{m}=0.3$, and $\Omega_\Lambda=0.7$. We note that the degree of triaxiality ${\cal T}$ is a derived parameter that depends on $q_a$ and $q_b$ (see Equation~(\ref{eq:Tri})). As posterior summary statistics, we use the biweight estimator of \citet{1990AJ....100...32B} to represent the center location (\CBI) and the spread (\SBI) of marginalized 1D posterior distributions. For each parameter, symmetrized biweight statistics $C_\mathrm{BI}\pm S_\mathrm{BI}$ are shown. The $\kappa(\theta)$ profile is reconstructed from the WL-1D analysis of the $\{g_+(\theta),n_\mu(\theta)\}$ data set (Section~\ref{subsec:wl1d}), while the $\kappa(\btheta)$ map from the WL-2D analysis of the $\{g_1(\btheta),g_2(\btheta),n_\mu(\theta)\}$ data set (Section~\ref{subsec:wl2d}).}
\end{deluxetable*}

\begin{deluxetable*}{ccccccccc}[ht]
\tabletypesize{\scriptsize}
\tablecaption{Weak-lensing mass estimates for A370}
\label{tab:mass}
\tablehead{
\multicolumn{1}{c}{Method} &
\multicolumn{1}{c}{Data} &
\multicolumn{1}{c}{Prior} &
\multicolumn{1}{c}{$M_{2500}$} &
\multicolumn{1}{c}{$M_{1000}$} &
\multicolumn{1}{c}{$M_{500}$} &
\multicolumn{1}{c}{$M_\mathrm{vir}$} &
\multicolumn{1}{c}{$M_\mathrm{200m}$} &
\multicolumn{1}{c}{$M_\mathrm{s}$\tablenotemark{a}} \\
\colhead{} &
\colhead{} &
\colhead{} &
\multicolumn{1}{c}{$(10^{15}\Msunh)$} &
\multicolumn{1}{c}{$(10^{15}\Msunh)$} &
\multicolumn{1}{c}{$(10^{15}\Msunh)$} &
\multicolumn{1}{c}{$(10^{15}h^{-1}\Msunh)$} &
\multicolumn{1}{c}{$(10^{15}\Msunh)$} &
\multicolumn{1}{c}{$(10^{15}\Msunh)$}  }
\startdata
    Tangential shear & $g_{+}$~profile &  Spherical & $0.53 \pm 0.07$ & $0.85 \pm 0.10$ & $1.13 \pm 0.15$ & $1.76 \pm 0.32$ & $1.86 \pm 0.35$ & $0.29 \pm 0.11$ \\
               WL-1D & $\kappa$~profile &  Spherical & $0.54 \pm 0.08$ & $0.90 \pm 0.11$ & $1.23 \pm 0.18$ & $1.97 \pm 0.40$ & $2.09 \pm 0.44$ & $0.36 \pm 0.14$ \\
               WL-2D &    $\kappa$~map &  Spherical & $0.48 \pm 0.05$ & $0.80 \pm 0.09$ & $1.09 \pm 0.13$ & $1.73 \pm 0.27$ & $1.84 \pm 0.29$ & $0.31 \pm 0.08$ \\
               WL-2D &    $\kappa$~map & Triaxial (fiducial) & $0.50 \pm 0.11$ & $0.82 \pm 0.16$ & $1.11 \pm 0.20$ & $1.76 \pm 0.33$ & $1.86 \pm 0.35$ & $0.31 \pm 0.08$ \\
               WL-2D &    $\kappa$~map & Triaxial+LOS & $0.42 \pm 0.07$ & $0.71 \pm 0.09$ & $0.97 \pm 0.13$ & $1.58 \pm 0.24$ & $1.68 \pm 0.26$ & $0.31 \pm 0.08$ 
\enddata
\tablenotetext{a}{Total mass enclosed within the NFW scale radius $r_\mathrm{s}$, $M(<r_\mathrm{s})$}
\end{deluxetable*}

The main results from our Bayesian inference of the spherical and triaxial NFW models are summarized in Table~\ref{tab:nfw}. As summary statistics, we employ the biweight estimator of \citet{1990AJ....100...32B} to represent the center location (\CBI) and the scale or spread (\SBI) of marginalized 1D posterior PDFs \citep[e.g.,][]{Umetsu2020xxl}. For a lognormally distributed quantity, \CBI approximates the median of the distribution. 

Triaxial modeling allows for a more general description of the intrinsic shape of cluster halos, leading to broader posterior distributions than the spherical case \citep{Oguri2005,Sereno+Umetsu2011}. The parameter constraints become more degenerate because of the lack of information of the halo elongation along the line of sight. These trends are found in the posterior distributions from our data. 

Our spherical modeling of the WL-2D data yields $M_{200}=(1.51\pm 0.22)\times 10^{15}\Msunh$ (or $r_{200}=(1.64\pm 0.08)\Mpch$) and $c_{200}=4.91\pm 0.93$, which are consistent with results from the tangential-shear and WL-1D methods. From triaxial modeling, we obtain $M_{200}=(1.54\pm 0.29)\times 10^{15}\Msunh$ (or $r_{200}=(1.65\pm 0.10)\Mpch$), $c_{200}=5.27\pm 1.28$, and a degree of triaxiality ${\cal T}=0.51\pm 0.29$ with the fiducial priors and $M_{200}=(1.38\pm 0.20)\times 10^{15}\Msunh$ (or $r_{200}=(1.59\pm 0.08)\Mpch$), $c_{200}=4.45\pm 0.93$, and ${\cal T}=0.58\pm 0.30$ when the LOS prior is employed. Thus, the level of mass bias due to the LOS elongation is found to be $\sim 10\percent$ for A370.

We now turn to the impact of the prior on the inferred uncertainty in the mass determination. Spherical (or Triaxial+LOS) modeling of A370 yields a fractional uncertainty in $M_{200}$ of $S_\mathrm{BI}(M_{200})/C_\mathrm{BI}(M_{200})\approx 15\percent$. In contrast, the uncertainty in $M_{200}$ from Triaxial modeling is $\approx 19\percent$, which is a factor of $\sim 1.3$ larger than that of Spherical modeling. It is insightful to compare our results with those of \citet{Umetsu2015}, who performed a WL-2D analysis of the superlens cluster Abell~1689 based on deeper Suprime-Cam observations. Analyzing their WL-2D data, \citet{Umetsu2015} obtained fractional uncertainties in $M_{200}$ of $\approx 8\percent$ and $20\percent$ for their spherical and full-triaxial NFW models, respectively (see the first and second rows of Table~7 in \citealt{Umetsu2015}). For both clusters, the fractional mass uncertainty in full triaxial modeling is $\sim 20\percent$, suggesting that the mass accuracy in deep weak-lensing observations is essentially limited by the uncertainty in the intrinsic shape and orientation of the cluster. Similar trends are also found for the concentration parameter. To accurately infer the cluster mass and concentration from lensing, it is thus necessary to directly model or marginalize over the 3D shape of clusters; when spherical symmetry is assumed, the effect of the intrinsic shape of the cluster should be accounted for in the error analysis \citep[e.g.,][]{Gruen2015,Umetsu2016clash}.

We also derive summary statistics on the total mass $M_\Delta$ evaluated at several characteristic interior overdensities $\Delta$. Table~\ref{tab:mass} lists the results of our cluster mass estimates. Our estimates of $M_\mathrm{vir}$ obtained without the LOS information are consistent within the errors with $M_\mathrm{vir}=2.28^{+0.26}_{-0.22}\times 10^{15}\Msunh$ from the combined weak- and strong-lensing analysis of \citet[][see Section~\ref{subsec:dynamical}]{Umetsu+2011}. In particular, our WL-1D analysis yields $M_\mathrm{vir}=(1.97\pm 0.40)\times 10^{15}\Msunh$ and $c_\mathrm{vir}=5.91\pm 1.87$, in agreement with the results of \citet{Umetsu+2011}. We find that spherical mass estimates from the WL-2D analysis are slightly lower than but consistent within the errors with the WL-1D results.

It should be noted that the halo mass $M_\Delta$ constrained using the LOS prior is likely to be considerably lower than the sum of the initial bound masses of the two progenitors, because A370 is expected to be in a highly disturbed dynamical state (see Section~\ref{subsec:dynamical}). Our estimates of $M_\mathrm{vir}$ obtained without the LOS prior are consistent to better than $2\sigma$ with the total mass of the system $M_\mathrm{vir}=2.3\times 10^{15}\Msunh$ adopted in the binary merger simulations of \citet{Molnar2020}.

Overall, our results agree well with weak-lensing mass estimates of \citet[][]{Hoekstra2015CCCP}, who obtained $M_\mathrm{vir}=2.13^{+0.39}_{-0.37}\times 10^{15}\Msunh$ and $M_{500}=(1.23\pm 0.22)\times 10^{15}\Msunh$ for A370. Our estimates of $M_{500}$ are also in agreement with the recent caustic mass estimate by \citet{Lagattuta2022},  $M_{500}=(0.92\pm 0.11)\times 10^{15}\Msunh$, obtained from a detailed phase-space analysis of \HST BUFFALO imaging and VLT Multi-Unit Spectroscopic Explorer (MUSE) spectroscopic observations.

In Figure~\ref{fig:NFW}, we show posterior constraints on the NFW parameters $(\log{M_{200}},\log{c_{200}})$ inferred from spherical modeling of both WL-1D and WL-2D data. The blue and orange contours in the lower-left panel represent the joint posterior PDFs for WL-1D and WL-2D, respectively, showing good agreement between the two methods. In both cases, the marginalized posterior PDFs for $(\log{M_{200}},\log{c_{200}})$ are unimodal and symmetric. 

Figure~\ref{fig:triNFW} displays the marginalized posterior PDFs for the triaxial NFW parameters of interest ($\log{M_{200}}, \log{c_{200}}, q_a, q_b, \cos{\vartheta}$) based on our WL-2D analysis. The results with and without employing the LOS prior are compared in the figure. The posterior PDFs for $\log{M_{200}}$ and $\log{c_{200}}$ are clearly unimodal and fairly symmetric. For both parameters, there is no significant shift in the PDF with respect to the spherical case (Figure~\ref{fig:NFW}). The posterior PDFs for the shape and orientation parameters ($q_a,q_b,\cos{\vartheta}$) from our fiducial modeling are very broad, reflecting the fact that the deprojection of triaxial halos is intrinsically underconstrained. In contrast, the axis ratio of the projected mass distribution, $q_\perp(q_a,q_b,\vartheta,\phi)$ (Section~\ref{subsec:triNFW}), can be directly constrained by the WL-2D data. Our posterior inference of the projected axis ratio is $q_\perp = 0.78 \pm 0.13$ and $0.79\pm 0.13$ with and without using the LOS prior, respectively. This is slightly larger than, but consistent with, the median axis ratio $\overline{q}_\perp\sim 0.6$ expected for randomly oriented cluster-scale CDM halos \citep[][see also \citealt{Bonamigo2015,Suto2016}]{Umetsu2018clump3d}.

Compared to the fiducial results obtained with uniform priors, our inference with the informative Gaussian prior on $\cos{\vartheta}$ (Triaxial+LOS) prefers a more prolate geometry with lower mass and lower concentration. In fact, there is a slight increase in the posterior probability for a prolate configuration ($q_a\sim q_b\simlt 0.6$) with lower mass and lower concentration. This can be understood as a consequence of the boosted surface mass density of the cluster lens due to the strongly aligned configuration.


\section{X-ray Data and Analysis}
\label{sec:xray}


\begin{figure}[tbp]
 \begin{center}
  \includegraphics[width=0.4\textwidth, angle=0, clip]{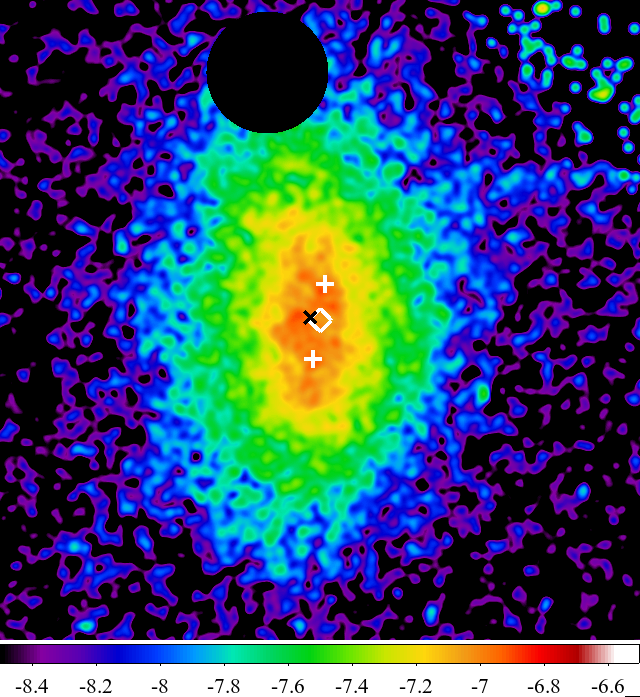}
 \end{center}
\caption{Logarithmically scaled, exposure-corrected, and background-subtracted \Chandra ACIS image of A370 in the $0.5$--$7$~keV band, smoothed with a Gaussian of $\mathrm{FWHM}=4.6\arcsec$. The image is $5.2\arcmin\times 5.2\arcmin$ ($1.1\Mpch$ on a side at $z=0.375$) in size and centered on the optical cluster center (open diamond symbol). The positions of the two BCGs are marked with $+$ symbols. The X-ray centroid position is marked with a $\times$ symbol.  A bright X-ray source in the north corresponding to a foreground elliptical galaxy is masked by a black solid circle. North is up and east is to the left.}
 \label{fig:chandra} 
\end{figure}

Here we describe our analysis of archival \Chandra X-ray data (Section~\ref{subsec:chandra}). We use two complementary approaches to determining the 3D gas density and temperature profiles of A370 under the assumption of spherical symmetry. First, we derive gas densities and temperatures of the cluster in concentric spherical shells from a spectral deprojection analysis (Section~\ref{subsec:deproj}). Second, we perform forward modeling to simultaneously fit X-ray surface brightness profiles binned in multiple energy bands to infer the 3D gas density and temperature profiles in a parametric form (Section~\ref{subsec:forward}). With the forward-fitting method, we will also derive the total mass profile assuming hydrostatic equilibrium.

\subsection{Chandra Data Reduction}
\label{subsec:chandra}

We analyze archival X-ray data of A370 taken with the Advanced CCD Imaging Spectrometer \citep[ACIS;][]{Garmire03} on board the \Chandra X-ray Observatory. The observation identification (ObsID) numbers of \Chandra observations analyzed in this study are 515 and 7715. Our analysis uses the \Chandra Interactive Analysis of Observations software \citep[\textsc{ciao}, version~4.13;][]{Fruscione06} and the \Chandra Calibration Database (\textsc{caldb}, version~4.9.5). We checked the light curve of each data set using the \texttt{lc\_clean} task in \textsc{ciao}, filtering flare data. The net exposure time of each data set is 62.9~ks and 7.1~ks. Point sources were identified using the \texttt{wavdetect} task in \textsc{ciao} and excluded from the analysis. 

In our spectral analysis, we use the X-ray Spectral Fitting Package \citep[][\textsc{xspec} version~12.11.1]{Arnaud96} and the \textsc{atomdb} code \citep[][version~3.0.9]{atomdb2021} for plasma emission modeling, assuming that the ICM is in collisional ionization equilibrium \citep{Smith01}. The abundance table of \cite{Anders89} is used in \textsc{xspec}. Here, the abundance of a given element is defined as $Z_{i} = (n_{i, \mathrm{obs}}/n_\mathrm{H, obs}) / (n_{i, \odot} / n_\mathrm{H, \odot})$, where $n_{i}$ and $n_\mathrm{H}$ are the number densities of the $i$th element and hydrogen, respectively.  We use the iron abundance to represent the ICM metal abundance, such that the abundance of other elements is tied to the iron abundance as $Z_{i} = Z_\mathrm{Fe}$ \citep{Ueda2021}. The Galactic absorption column density is estimated at $N_\mathrm{H}=2.89 \times 10^{20}$~cm$^{-2}$ according to \cite{HI4PI16} and fixed in our X-ray spectral analysis. The blank-sky data included in \textsc{caldb} are used to determine the background contribution.

To determine the centroid of X-ray emission in A370, we fit the surface brightness distribution with a 2D $\beta$-model using the \textsc{sherpa} fitting package in \textsc{ciao} \citep{Freeman01, Doe07, sherpa_2021_5554957}. The surface brightness map was extracted from the ACIS S3 chip in the data set of ObsID~515 to reduce the uncertainty in the background determination. A bright foreground galaxy lying about $2\arcmin$ north of the cluster center was masked with a circle of radius $30\arcsec$ from its X-ray peak. 
From the best-fit model, we find the X-ray centroid of $\mathrm{R.A.} = \mathrm{2:39:53.2}$ and $\mathrm{decl.}=-\mathrm{1:34:35.1}$ (Table~\ref{tab:cluster}), with positional uncertainties of $(\Delta\mathrm{R.A.}, \Delta\mathrm{decl.}) = (0.33\arcsec, 0.39\arcsec)$.

Figure~\ref{fig:chandra} shows the exposure-corrected and background-subtracted \Chandra ACIS image of A370 in the $0.5$--$7$~keV band, smoothed with $4.6\arcsec$ FWHM Gaussian. The X-ray emission centroid determined from \Chandra observations is $4.9\arcsec$ ($\approx 18\kpch$) away from the optical center defined as the midpoint of the two BCGs.


\subsection{Spectral Deprojection Analysis}
\label{subsec:deproj}


\begin{figure}[tbp]
  \begin{center}
   \includegraphics[width=0.45\textwidth, angle=0, clip]{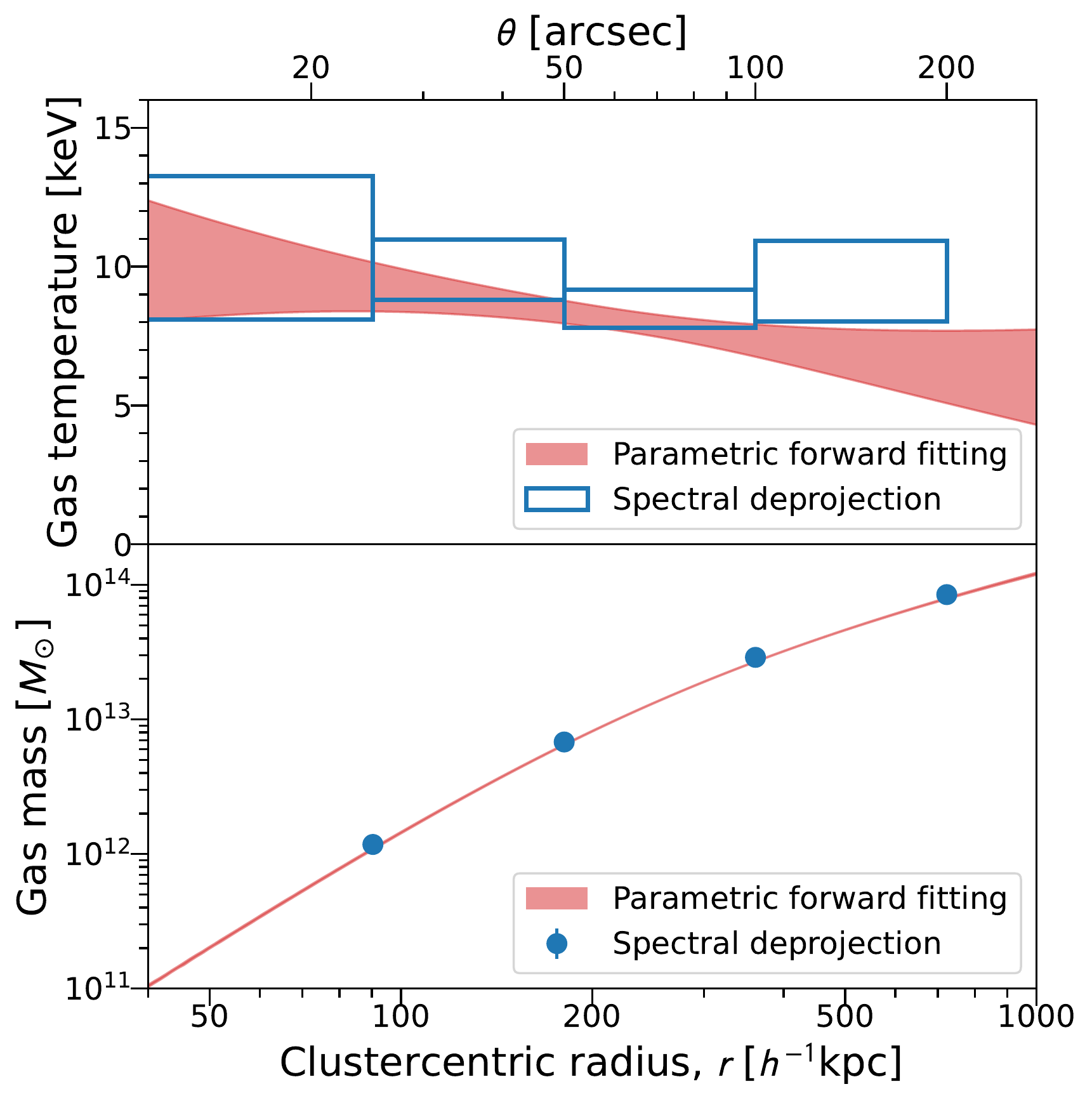}
  \end{center}
\caption{
Three-dimensional gas temperature $T_\mathrm{gas}(r)$ (upper panel) and enclosed gas mass $M_\mathrm{gas}(<r)$ (lower panel) profiles of A370 derived from \Chandra X-ray observations. The red shaded region in each panel shows the marginalized $1\sigma$ confidence region of the respective profile obtained from forward modeling of \Chandra X-ray data. Blue open boxes show the gas temperatures with $1\sigma$ uncertainties from spectral deprojection in concentric spherical shells. Blue filled circles with error bars in the lower panel show the results obtained from spectral deprojection.
\label{fig:xmodel} }
\end{figure}

\begin{deluxetable}{ccccc}[tbp]
\centering
\tabletypesize{\footnotesize}
\tablecaption{Best-fit parameters from the \Chandra spectral deprojection analysis}
\label{tab:spec} 
\tablehead{
 \multicolumn{1}{c}{$\theta_\mathrm{in}$} &
 \multicolumn{1}{c}{$\theta_\mathrm{out}$} &
 \multicolumn{1}{c}{$T_\mathrm{gas}$} &
 \multicolumn{1}{c}{$Z$} &
 \multicolumn{1}{c}{$n_\mathrm{e}$} \\
 \multicolumn{1}{c}{($\arcsec$)} &
 \multicolumn{1}{c}{($\arcsec$)} &
 \multicolumn{1}{c}{(keV)} &
 \multicolumn{1}{c}{($Z_\odot$)} &
 \multicolumn{1}{c}{($10^{-3}$~cm$^{-3}$)} }
\startdata
$0$	& $25$	& $10.16_{-2.06}^{+3.11}$ & $0.58_{-0.41}^{+0.45}$ & $4.754 \pm 0.191$\\
$25$ 	& $50$ 	& $9.77_{-0.97}^{+1.20}$  & $0.71_{-0.21}^{+0.23}$ & $3.241 \pm 0.067$\\
$50$ 	& $100$ & $8.40_{-0.61}^{+0.77}$  & $0.35 \pm 0.11$	   & $1.595 \pm 0.020$\\
$100$ 	& $200$ & $9.29_{-1.25}^{+1.64}$  & $0.3$ (fixed)	   & $0.502 \pm 0.005$\\
\enddata 
\end{deluxetable}

We jointly fit background-subtracted \Chandra spectra in the $0.4$--$7.0$~keV band extracted from four concentric annular regions ($\theta_\mathrm{in},\theta_\mathrm{out}$) of $0\arcsec$--$25\arcsec$, $25\arcsec$--$50\arcsec$, $50\arcsec$--$100\arcsec$, and $100\arcsec$--$200\arcsec$ centered on the X-ray centroid using the \texttt{projct} routine implemented in \textsc{xspec} \citep[][]{Smith02}. Here \texttt{projct} allows us to fit spectra extracted from a series of concentric annuli simultaneously, assuming spherical symmetry to calculate suitable geometric factors \citep[e.g.,][]{Fabian80, Fabian81, Kriss83, Arabadjis02}. In this analysis, the cluster redshift and $N_\mathrm{H}$ are fixed at $0.375$ and $2.89 \times 10^{20}$~cm$^{-2}$, respectively. The metal abundance of the ICM in the $100\arcsec$--$200\arcsec$ region is assumed to be $0.3Z_\odot$ \citep{Fujita2008,Werner2013,Urban2017,Ghizzardi2021}. 

To set the outer boundary conditions, we fit the background-subtracted X-ray spectrum in the $0.4$--$2.0$~keV band extracted from the outermost annular region of $200\arcsec$--$400\arcsec$, ignoring the emission from gas outside the outermost shell \citep[see][]{Humphrey06} and fixing the metal abundance to $0.3Z_\odot$. The best-fit parameters for the outermost shell, $T_\mathrm{3D}=9.7^{+10.2}_{-3.3}$~keV and $n_\mathrm{e}=(1.76\pm 0.08)\times 10^{-4}$~cm$^{-3}$, are included and fixed in our deprojection analysis of the inner concentric regions.

The resulting best-fit parameters for each spherical shell are summarized in Table~\ref{tab:spec} (see also Figure~\ref{fig:xmodel}). The gas density $\rho_\mathrm{gas}$ is related to the electron number density $n_\mathrm{e}$ as $\rho_\mathrm{gas} = \mu_\mathrm{e} m_\mathrm{p} n_\mathrm{e}$, with $\mu_\mathrm{e}\approx 1.11$ the mean mass per electron in units of proton mass $m_\mathrm{p}$.

\subsection{Parametric Forward Fitting}
\label{subsec:forward}

We perform a forward model fitting of the \Chandra observations for A370. The \textsc{mbproj2} algorithm developed by \citet[][see also \citealt{Sanders18}]{Sanders17} is capable of modeling radial X-ray surface brightness profiles in multiple energy bins, with or without assuming hydrostatic equilibrium. Motivated by their work, we have implemented a forward-modeling algorithm to simultaneously fit the X-ray brightness profiles binned in multiple energy bands to infer the 3D gas density and temperature profiles in a parametric form, without assuming hydrostatic equilibrium. Both algorithms assume spherical symmetry.

We model the 3D gas density profile $n_\mathrm{e}(r)$ as a $\beta$-profile and the 3D temperature profile $T_\mathrm{3D}(r)$ as a universal temperature profile of \cite{Vikhlinin06}:
\begin{equation}
\label{eq:3dmodel}
 \begin{aligned}
 n_\mathrm{e}(r) &= n_\mathrm{e0} \left[1 + (r/r_\mathrm{c})^2 \right]^{-3\beta/2},\\
 T_\mathrm{3D}(r) &= T_{0} \frac{(r/r_\mathrm{t})^{a}}{\left[1 + (r/r_\mathrm{t})^{2}\right]^{c/2}},
 \end{aligned}
\end{equation}
where $n_\mathrm{e0}$ is the central electron number density, $\beta$ is the slope parameter, and $r_\mathrm{c}$ is the core radius of the $\beta$ profile; $T_{0}$ is the central gas temperature, $a$ and $c$ are the temperature slope parameters, and $r_\mathrm{t}$ is the temperature scale radius.

In this analysis, we fix the cluster redshift to $z=0.375$, the Galactic absorption column density to $N_\mathrm{H}=2.89 \times 10^{20}$~cm$^{-2}$, and the metal abundance of the ICM to $Z=0.3Z_\odot$. We use the spectroscopic-like temperature $T_\mathrm{2D}(r)$ of \citet{Mazzotta04} to approximate spectroscopic temperatures extracted from \Chandra X-ray observations:
\begin{equation}
T_\mathrm{2D} = \frac{\int\! w T_\mathrm{3D}  dV}{\int\! w dV},
\end{equation}
with $w = n_\mathrm{e}^2(r) T_\mathrm{3D}^{-3/4}(r)$. The X-ray surface brightness $S_\mathrm{X}(r_\perp)$ as a function of projected cluster radius $r_\perp=D_l\theta$ is modeled by the following equation \citep{Ettori00}:
\begin{equation}
S_\mathrm{X}(r_\perp) = n_\mathrm{e0} n_\mathrm{p0} r_\mathrm{c} \Lambda_\mathrm{X} 
 B\left(3\beta - \frac{1}{2}, \frac{1}{2}\right)
 \left[ 1 + \left(\frac{r_\perp}{r_\mathrm{c}}\right)^2 \right]^{\frac{1}{2} - 3\beta}, \label{eq:Sx}
\end{equation}
where $\Lambda_\mathrm{X}(T_\mathrm{2D},Z)$ is the cooling function, $n_\mathrm{p0}\approx n_\mathrm{e0}/1.17$ is the central proton number density, and $B(x,y)$ is the beta function. We use the \textsc{pyatomdb} python package \citep{Foster20} to evaluate the cooling function $\Lambda_\mathrm{X}(T_\mathrm{2D},Z)$ in each energy band for a given value of the spectroscopic-like temperature $T_\mathrm{2D}(r_\perp)$. 

We have extracted the radial profiles of X-ray surface brightness in $N_\mathrm{spec}=10$ energy bands between neighbouring energies of 0.5, 0.75, 1, 1.25, 1.5, 2, 3, 4, 5, 6, and 7~keV.  In each energy band, the X-ray surface brightness is sampled in $49$ linearly spaced radial bins in the range $\theta\in [10\arcsec,200\arcsec]$ centered on the X-ray centroid. Following \citet{Sanders18}, we have chosen these bands so as to capture most of the spectral information without overly increasing the computational time. We estimate in each radial bin the pixel-to-pixel variance of X-ray brightness over the $0.4$--$7$~keV energy band, finding that the standard errors of the mean based on the estimated variance are highly consistent with the errors determined based on the photon counts. In this work, we use the standard error based on the estimated variance to characterize the uncertainty in the mean X-ray surface brightness in each bin. 

The background contribution in each energy band is determined from the blank-sky data included in \textsc{caldb} (Section~\ref{subsec:chandra}). We estimate the count rate of the blank-sky data in the spectral range of $9$--$12$~keV dominated by the particle background \citep{Hickox06}. Using the ratio between the count rate observed in A370 and the background one in the $9$--$12$~keV band, we rescale the background contribution in each energy band to match the observations of A370, accounting for the difference in exposure times. We then construct the azimuthally averaged radial profile of the background map in each energy band. Similarly, we create azimuthally averaged radial profiles of exposure maps in the 10 energy bands.

We simultaneously fit the observed X-ray surface brightness profiles in the 10 energy bands with our model using affine-invariant MCMC sampling \citep{Goodman10} implemented by the \textsc{emcee} python package \citep{Foreman-Mackey13}. The log-likelihood function for the data is defined by (up to a normalization constant)
\begin{equation}
-2 \ln \mathcal{L} = \sum_{i,j} \frac{\left[d_{ij} - \left( w_{ij} T_i \widehat{S}_{\mathrm{X},ij} + {\cal N}_i \times \mathrm{BGD}_{ij} \right)  \right]^{2}}{\sigma_{ij}^{2}},
\end{equation}
where $i$ and $j$ run over all energy bands and all radial bins, respectively, $d_{ij}$ is the binned X-ray brightness measured in units of counts per pixel, $\sigma_{ij}$ is the statistical uncertainty of the measurement in each bin, $T_i$ represents the Galactic transmission in the $i$th energy band calculated by \textsc{xspec} using the photoionization cross sections of \citet{Verner1996}, $\widehat{S}_{\mathrm{X},ij}$ is the model prediction in each bin for the X-ray surface brightness given by Equation~({\ref{eq:Sx}}),  $w_{ij}$ is the conversion factor proportional to the product of the effective area and the net exposure time in each bin, $\mathrm{BGD}_{ij}$ denotes the background contribution in each bin given in units of counts per pixel, and ${\cal N}_i$ is a dimensionless calibration factor of the background in the $i$th energy band.

\begin{deluxetable}{lc}[tbp]
\tabletypesize{\footnotesize}
\tablecaption{X-ray model parameter constraints derived from forward modeling of \Chandra observations}
\label{tab:gas}
\tablehead{
\multicolumn{1}{c}{Parameter} &
\multicolumn{1}{c}{Posterior summary} }
\startdata
$n_\mathrm{e0}$ ($10^{-3}$~cm$^{-3}$) 	      & $5.03 \pm 0.09$\\
$r_\mathrm{c}$ ($h^{-1}$~kpc) 		      & $190 \pm 7$\\
$\beta$  	                              & $0.693 \pm 0.019$\\
$T_{0}$ (keV) 	        	              & $9.63 \pm 2.38$\\
$r_\mathrm{t}$ ($h^{-1}$~kpc) 		      & $176 \pm 100$\\
$a$		                              & $-0.042 \pm 0.248$\\
$c$		                              & $0.20 \pm 0.44$\\
${\cal N}(0.50\mathrm{-}0.75~\mathrm{keV})$   & $0.94 \pm 0.06$\\
${\cal N}(0.75\mathrm{-}1.00~\mathrm{keV})$   & $0.99 \pm 0.11$\\
${\cal N}(1.00\mathrm{-}1.25~\mathrm{keV})$   & $0.74 \pm 0.10$\\
${\cal N}(1.25\mathrm{-}1.50~\mathrm{keV})$   & $0.76 \pm 0.10$\\
${\cal N}(1.50\mathrm{-}2.00~\mathrm{keV})$   & $0.90 \pm 0.06$\\
${\cal N}(2.00\mathrm{-}3.00~\mathrm{keV})$   & $1.01 \pm 0.04$\\
${\cal N}(3.00\mathrm{-}4.00~\mathrm{keV})$   & $0.98 \pm 0.05$\\
${\cal N}(4.00\mathrm{-}5.00~\mathrm{keV})$   & $1.02 \pm 0.04$\\
${\cal N}(5.00\mathrm{-}6.00~\mathrm{keV})$   & $1.01 \pm 0.04$\\
${\cal N}(6.00\mathrm{-}7.00~\mathrm{keV})$   & $1.00 \pm 0.03$				
\enddata
\end{deluxetable}

Our model has a total of 17 parameters, of which seven parameters describe the cluster X-ray emission (see Equation~(\ref{eq:3dmodel})), namely $(n_\mathrm{e0},  r_\mathrm{c}, \beta, T_{0}, r_\mathrm{t}, a, c)$, and the rest are calibration nuisance parameters, $\{{\cal N}_i\}_{i=1}^{N_\mathrm{spec}}$. For the parameters describing the cluster X-ray emission, we use uninformative uniform priors of $n_\mathrm{e0} \in [0, 1]$~cm$^{-3}$, $r_\mathrm{c} \in [0, 350]\kpch$, $\beta \in [0,2]$, $T_{0} \in [1, 20]$~keV, $r_\mathrm{t} \in [0, 350]\kpch$, $a \in [-0.5, 0.5]$, and $c \in [-1, 1]$. For each calibration parameter $\{{\cal N}_i\}_{i=1}^{N_\mathrm{spec}}$, we adopt a Gaussian prior of $1.0\pm 0.2$. We sample the posterior PDFs of all model parameters over the full parameter space allowed by the priors. Posterior summaries of the model parameters are listed in Table~\ref{tab:gas}. In Appendix~\ref{appendix:sx}, we show the \Chandra X-ray brightness profiles along with the best-fit model. 

Finally, we use the posterior samples obtained with the MCMC algorithm to derive constraints on the gas mass $M_\mathrm{gas}(<r)$ enclosed within the spherical radius $r$ and the hydrostatic equilibrium mass $\Mhse(<r)$ of A370. The hydrostatic mass $\Mhse(<r)$ is given by
\begin{equation}
\label{eq:HE}
\Mhse(< r) = - \frac{k_\mathrm{B} T_\mathrm{gas}(r) r}{G \mu_\mathrm{g} m_\mathrm{p}} \left[ \frac{d \ln \rho_\mathrm{gas}(r)}{d \ln r} + \frac{d \ln T_\mathrm{gas}(r)}{d \ln r} \right], 
\end{equation}
where $k_\mathrm{B}$ is the Boltzmann constant and $\mu_\mathrm{g}\approx 0.60$ is the mean molecular weight. We will compare the resulting hydrostatic mass profile, $\Mhse(<r)$, with our weak-lensing results in Section~\ref{subsec:bhse}.

In the top panel of Figure~\ref{fig:xmodel}, we show the marginalized $1\sigma$ confidence region of $T_\mathrm{3D}(r)$ obtained from our forward modeling, along with the deprojected temperatures inferred from our spectral deprojection analysis (Section~\ref{subsec:deproj}). Similarly, we compare in the bottom panel of Figure~\ref{fig:xmodel} our determinations of $M_\mathrm{gas}(<r)$ from both methods. For both comparisons, we find that the two complementary approaches yield highly consistent results.


\section{Discussion}
\label{sec:discussion}

\subsection{Hydrostatic Mass Bias}
\label{subsec:bhse}


\begin{figure}[tbp]
 \begin{center}
   \includegraphics[width=0.45\textwidth, angle=0, clip]{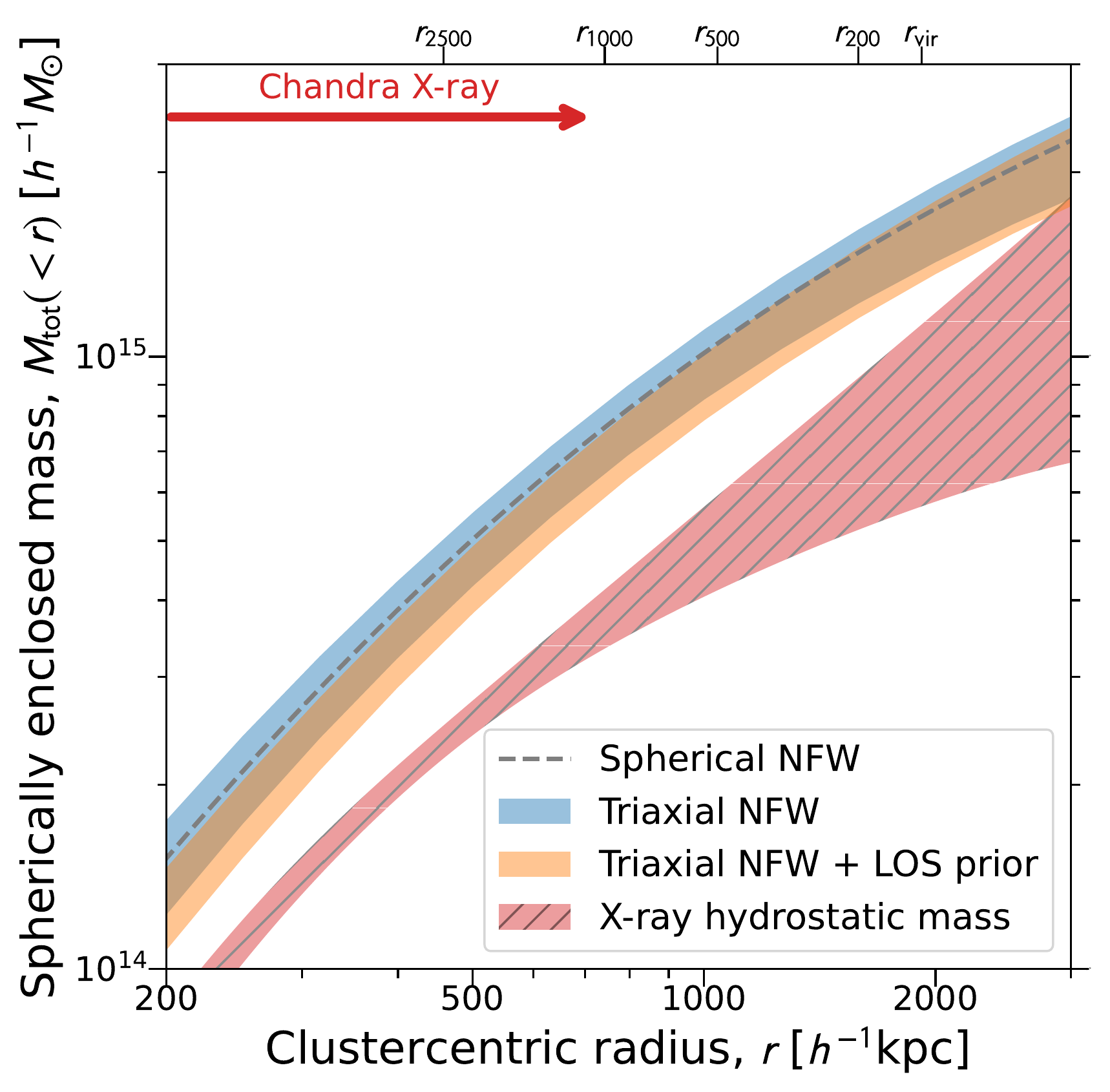}
 \end{center}
\caption{Comparison of spherically enclosed total mass profiles $M_\mathrm{tot}(<r)$ of A370 as a function of spherical radius $r$. The orange and blue shaded areas represent the marginalized $1\sigma$ confidence regions from triaxial NFW modeling of the $\Sigma(\btheta)$ map (Figure~\ref{fig:kmap}) with and without using the LOS prior, respectively. The dashed line shows the posterior mean from spherical NFW modeling of the $\Sigma(\btheta)$ map. The red hatched area represents the marginalized $1\sigma$ confidence region of the total hydrostatic mass $\Mhse(<r)$ obtained from forward modeling of \Chandra X-ray data that cover the radial range $\theta\in [10\arcsec,200\arcsec]$.
\label{fig:m3d} }
\end{figure}


\begin{figure}[tbp]
 \begin{center}
  \includegraphics[width=0.45\textwidth, angle=0, clip]{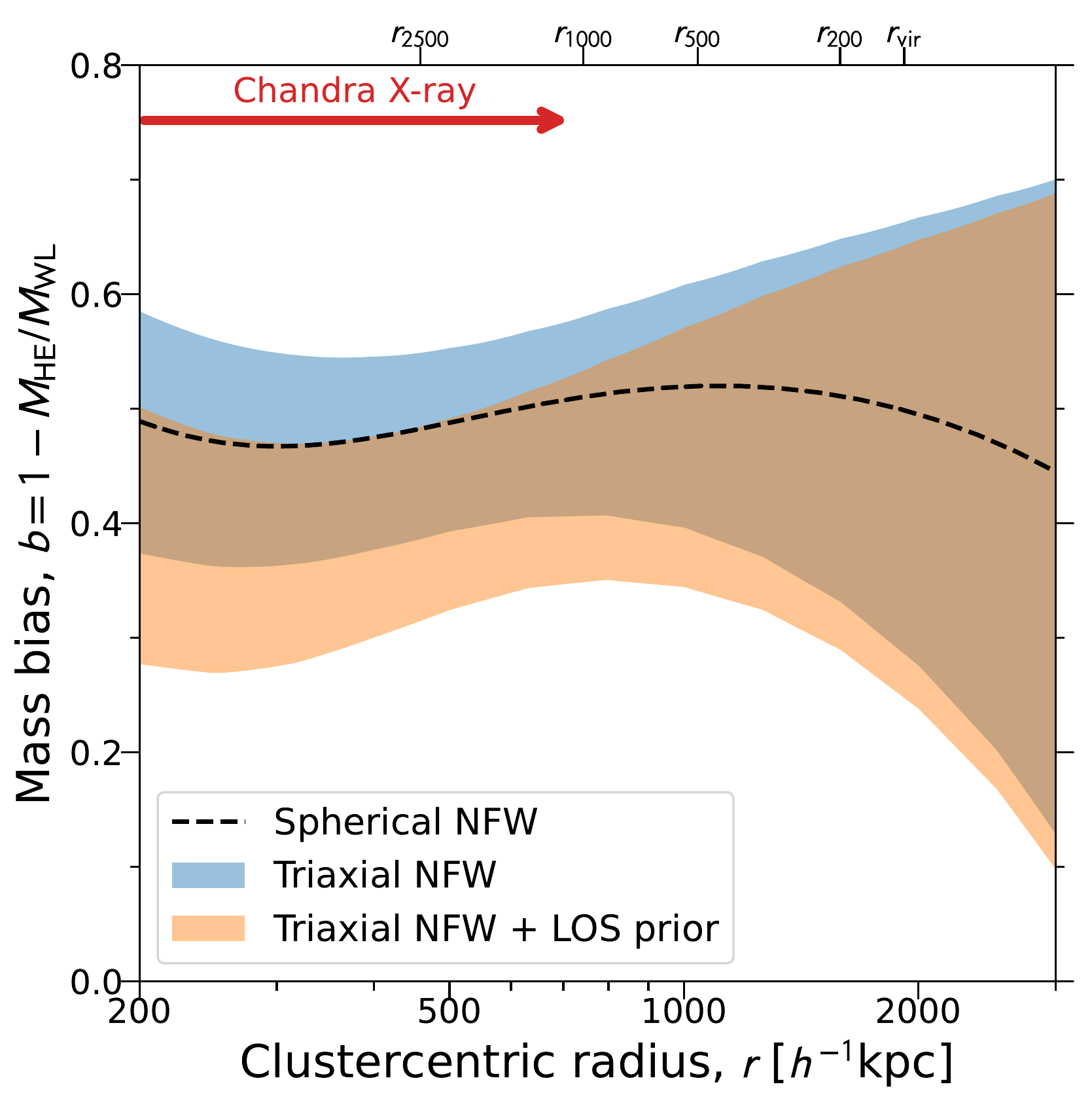}
 \end{center}
\caption{
X-ray hydrostatic mass bias $b(r)=1-\Mhse(<r)/M_\mathrm{WL}(<r)$ in A370 as a function of spherical radius $r$ derived from combined \Chandra X-ray and Subaru weak-lensing observations. The orange and blue shaded areas represent the marginalized $1\sigma$ confidence regions based on triaxial NFW modeling of the $\Sigma(\btheta)$ map (Figure~\ref{fig:kmap}) with and without using the LOS prior, respectively. The dashed line shows the posterior mean based on spherical NFW modeling of the $\Sigma(\btheta)$ map. 
\label{fig:bHE} }
\end{figure}

Hydrostatic mass estimates $\Mhse$ are expected to be biased low, depending on details of nongravitational processes and the level of residual gas motions in the ICM. Determining the level of hydrostatic mass bias for a representative sample of galaxy clusters has important implications for both cluster cosmology and astrophysics \citep{Planck2013XX,Planck2015XXIV,Pratt2019}. Cosmological hydrodynamical simulations suggest a modest level of hydrostatic mass bias for an ensemble of galaxy clusters, $b_\mathrm{HE}\equiv 1-\langle\Mhse/M_\mathrm{true}\rangle \sim 5\percent$--$20\percent$ at $r\le r_{500}$ \citep[][]{Nagai2007cosmology,Lau+2009,Meneghetti2010a,Nelson2012,Angelinelli2020,Ansarifard2020}, defined with respect to the true enclosed mass $M_\mathrm{true}$. Since A370 is a highly disturbed system, the cluster is likely to exhibit a higher than typical value of mass bias, which could serve as an extreme limit expected for galaxy clusters.

With the aim of characterizing the level of hydrostatic mass bias in A370, we compare our lensing-based determinations of the cluster mass profile (Section~\ref{sec:model}) to the hydrostatic mass profile $\Mhse(<r)$ derived from \Chandra X-ray data (Section~\ref{subsec:forward}). For this purpose, we compute the total mass $M_\mathrm{tot}(<r)$ of a triaxial halo enclosed within a sphere of radius $r$:
\begin{equation}
M_\mathrm{tot}(<r)=\iiint\limits_{\cal V}\!\rho(X,Y,Z) dV=\int\limits_{4\pi}\!d\Omega\int_0^{r}\! r^{\prime 2} \rho dr^\prime,
\end{equation}
where $\rho(X,Y,Z)$ is the density function (Equations~(\ref{eq:triNFW}) and (\ref{eq:ellipsoid})), the region of integration ${\cal V}$ is $\sqrt{X^2+Y^2+Y^2}\le r$, and $d\Omega=\sin\vartheta d\vartheta d\phi$ is the solid angle element in spherical coordinates.

In Figure~\ref{fig:m3d}, we compare the spherically enclosed total mass profiles $M_\mathrm{tot}(<r)$ obtained from our WL-2D and X-ray analyses. Here we have extrapolated the X-ray forward model beyond the range of the fitted data ($\simlt 720\kpch$) to compute $\Mhse(<r)$ out to larger cluster radii. We note that in contrast to the triaxial lensing constraints on $M_\mathrm{tot}(<r)$, the hydrostatic mass $\Mhse(<r)$ obtained assuming spherical symmetry is not corrected for the projection effect due to the LOS elongation of the gas distribution. Since the shape of the collisional gas is rounder than the underlying matter \citep[e.g.,][]{Suto2017}, the level of projection bias in the gas distribution is expected to be less than $\sim 10\percent$ found in the total mass distribution (see Section~\ref{subsec:posterior}).


In Figure~\ref{fig:bHE}, we show the hydrostatic mass bias as a function of spherical radius $r$, defined with respect to the total mass $M_\mathrm{WL}(<r)$ determined from weak lensing:
\begin{equation}
 b(r) = 1 - \frac{\Mhse(<r)}{M_\mathrm{WL}(<r)}.
\end{equation}
The results are shown for the three different priors on the halo shape employed in our mass modeling of the WL-2D data (see Table~\ref{tab:nfw}). We find no significant evidence for a strong variation of $b(r)$ both within and beyond the radial range probed by the \Chandra data ($\simlt 720\kpch$). At each radius $r$, we find similar central values of the distributions from the spherical and the fiducial triaxial cases (see Table~\ref{tab:nfw}).

From the triaxial lens modeling, we obtain mass ratios of $1-b(r) = 0.56 \pm 0.09$ and $0.51 \pm 0.09$ at $r=0.7\Mpch\sim 0.7r_{500}$, with and without using the LOS prior, respectively. When the X-ray forward model is extrapolated out to $r_{500}\sim 1\Mpch$, we find $1-b(r_{500}) = 0.54 \pm 0.12$ and $0.50 \pm 0.11$ with and without the LOS prior, respectively. The range of mass bias inferred for A370, $b\in [0.34,0.61]$ at the $1\sigma$ level, is on the high side of the distribution expected from cosmological cluster simulations \citep[see][]{Nagai2007cosmology,Lau+2009,Ansarifard2020} and is in better agreement with the value of $1-b_\mathrm{HE}=\langle\Mhse/M_\mathrm{true}\rangle=0.58\pm 0.04$ required to bring the \Planck CMB and cluster constraints into full agreement in the base \LCDM cosmology of \citet{Planck2015XXIV}. However, it should be noted again that the mass bias found for this cluster should be considered as an extreme value expected for galaxy clusters.


\subsection{Gas Mass Fraction}
\label{subsec:fgas}


\begin{figure}[tbp]
 \begin{center}
  \includegraphics[width=0.45\textwidth, angle=0, clip]{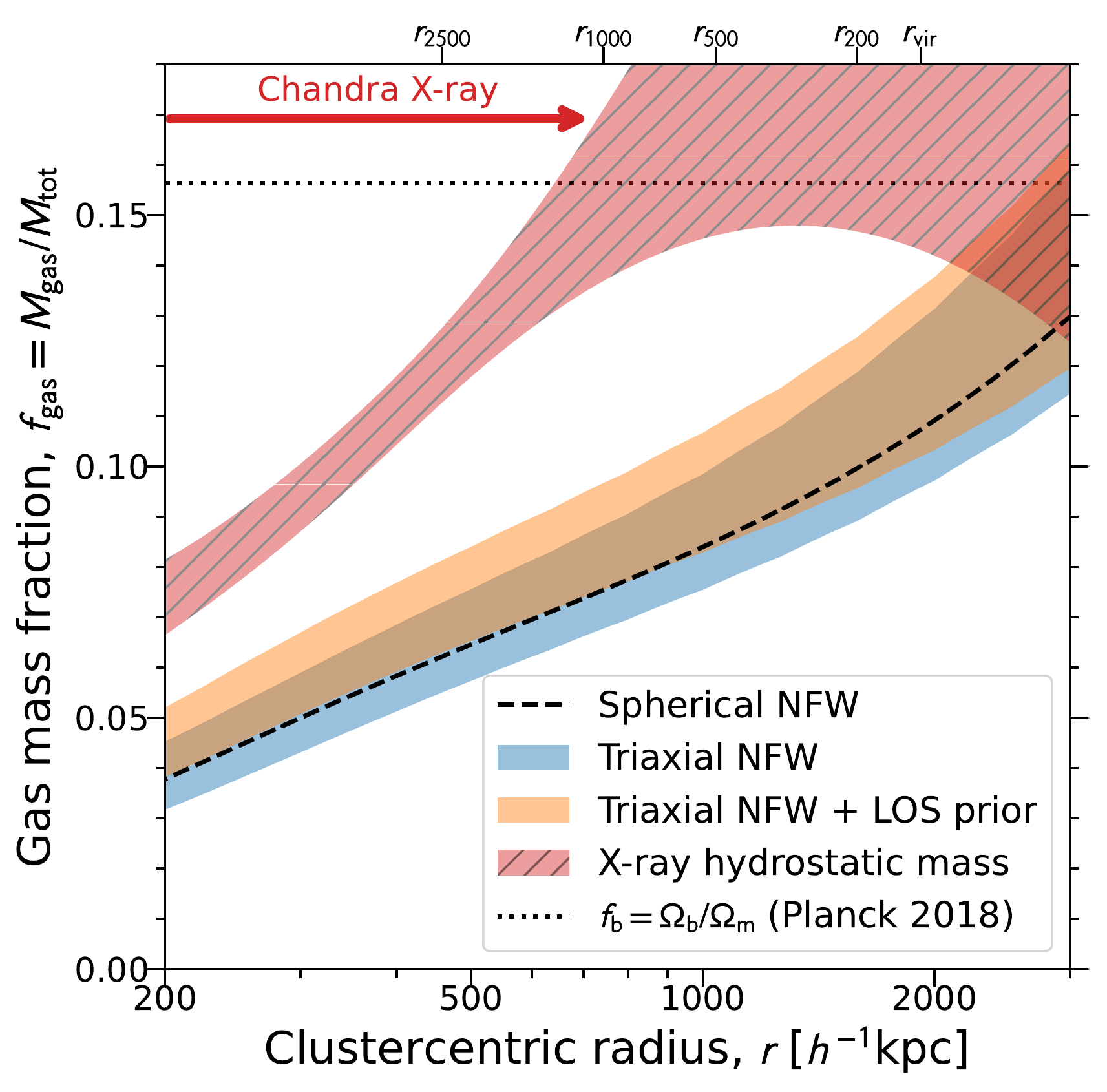}
 \end{center}
\caption{
Ratio of spherically enclosed gas mass $M_\mathrm{gas}$ to total mass $M_\mathrm{tot}$, $\fgas(r)=M_\mathrm{gas}(<r)/M_\mathrm{tot}(<r)$, as a function of spherical radius $r$ derived from combined \Chandra X-ray and Subaru weak-lensing observations. The orange and blue shaded areas represent the marginalized $1\sigma$ confidence regions based on triaxial NFW modeling of the $\Sigma(\btheta)$ map (Figure~\ref{fig:kmap}) with and without using the LOS prior, respectively. The dashed line shows the posterior mean based on spherical NFW modeling of the $\Sigma(\btheta)$ map. The red hatched area represents the marginalized $1\sigma$ confidence region based on the hydrostatic total mass $\Mhse(<r)$ obtained from forward modeling of \Chandra X-ray data. The horizontal dotted line shows the cosmic baryon fraction $f_\mathrm{b}=\Ob/\Om$ determined by \citet{Planck2018VI}.
\label{fig:fgas} }
\end{figure}

In Figure~\ref{fig:fgas}, we show the ratio of spherically enclosed gas mass $M_\mathrm{gas}(<r)$ to total mass $M_\mathrm{tot}(<r)$ as a function of spherical radius $r$:
\begin{equation}
\label{eq:fgas}
 \fgas(<r) = \frac{M_\mathrm{gas}(<r)}{M_\mathrm{tot}(<r)}.
\end{equation}
Here the total mass $M_\mathrm{tot}$ is taken to be either the weak-lensing mass $M_\mathrm{WL}$ or the hydrostatic mass $\Mhse$ (see Section~\ref{subsec:bhse}) and the gas mass $M_\mathrm{gas}$ is derived from the X-ray forward model (Section~\ref{subsec:forward}). We find that the gas mass fraction $\fgas(<r)$ increases progressively outward, indicating that the hot gas is more extended than the underlying matter distribution. 

Our lensing results, when combined with the X-ray gas mass measurements, yield a direct estimate for $\fgas(<r)$, free from the assumption of hydrostatic equilibrium. Using the total mass derived from triaxial lens modeling, the gas mass fraction enclosed within a sphere of radius $r=0.7\Mpch \sim 0.7 r_{500}$ is found to be $\fgas(<r)=(8.4\pm 1.0)\percent$ and $(7.6 \pm 1.0)\percent$, with and without using the LOS prior, respectively. Extrapolating the gas mass measurements out to $r_{500}$, we find $\fgas(<r_{500})=(9.6\pm 1.2)\percent$ and $(9.0 \pm 1.2)\percent$ with and without the LOS prior, respectively. 

When compared to the cosmic baryon fraction $f_\mathrm{b}\equiv \Ob/\Om = 0.156\pm 0.002$ determined by the \Planck mission \citep[][]{Planck2018VI}, our constraint on the gas mass fraction indicates $\fgas(<r_{500})/f_\mathrm{b} = 0.62\pm 0.08$ and $0.58 \pm 0.08$, with and without using the LOS prior, respectively. These are significantly lower than the typical values of $\fgas/f_\mathrm{b}\sim 0.8$--$0.9$ observed for high-mass galaxy clusters \citep{Chiu+2018,Tian2020,Akino2022}. Such a high degree of gas depletion can be caused by the adiabatic expansion of the post-shock gas \citep{Ricker+Sarazin2001,Umetsu+2010CL0024}. It would take of the order of Gyrs for the gas to fall back into the gravitational potential well of the cluster.

By contrast, the gas mass fraction based on the X-ray hydrostatic mass, $\fgas(<r)=M_\mathrm{gas}(<r)/\Mhse(<r)$, reaches the cosmic baryon fraction $f_\mathrm{b}$ at $r\approx 0.8\Mpch$ and increasingly exceeds it at larger cluster radii.


\section{Summary and Conclusions}
\label{sec:summary}

The Frontier Fields cluster A370 is a superlens characterized by a large Einstein radius ($\Rein=33.9\arcsec\pm 1.1\arcsec$ for $\zs=2$; Table~\ref{tab:cluster}) and is one of the most massive known lenses on the sky. Recent dedicated numerical simulations of binary cluster mergers constrained by multi-probe observations suggest that the cluster is a post-major merger of two similar-mass clusters \citep{Molnar2020}. These results also suggest that A370 is in a highly disturbed dynamical state and is elongated along the current direction of the collision axis, which is closely aligned with the line of sight in their best-matching simulation.

In this paper, we have carried out a detailed weak-lensing and X-ray study of A370 using wide-field \BRz Subaru/Sprime-Cam (Section~\ref{sec:data}) and \Chandra X-ray (Section~\ref{sec:xray}) observations. By combining 2D shear and azimuthally averaged magnification constraints derived from the Subaru data, we have performed a lensing mass reconstruction in a free-form manner (Section~\ref{sec:mrec}; Figures~\ref{fig:rgb} and \ref{fig:kmap}), which allows us to determine both radial structure and 2D morphology of the cluster mass distribution. 

In a parametric triaxial framework assuming an NFW density profile, we have constrained the intrinsic structure, shape, and orientation of the cluster halo by forward modeling the reconstructed $\Sigma(\btheta)$ map (Section~\ref{sec:model}; Tables~\ref{tab:nfw} and \ref{tab:mass}). We obtain a halo mass $M_{200}=(1.54\pm 0.29)\times 10^{15}\Msunh$ and a halo concentration $c_{200}=5.27\pm 1.28$
with uninformative uniform priors. Using a prior on the LOS alignment of the halo major axis derived from the binary merger simulations of \citet{Molnar2020}, we find that the data favor a more prolate geometry with lower mass and lower concentration, $M_{200}=(1.38\pm 0.20)\times 10^{15}\Msunh$ and $c_{200}=4.45\pm 0.93$.

When compared to the hydrostatic mass estimate $\Mhse$ from \Chandra observations (Section~\ref{subsec:bhse}), our triaxial weak-lensing analysis yields spherically enclosed mass ratios $\Mhse/M_\mathrm{WL}$ of $1-b(r)=0.56\pm 0.09$ and $0.51\pm 0.09$ at $r=0.7\Mpch\sim 0.7r_{500}$, with and without using the LOS prior, respectively (Figure~\ref{fig:bHE}). Extrapolating our X-ray forward model to $r_{500}$, we find $1-b(r_{500})=0.54\pm 0.12$ and $0.50\pm 0.11$ with and without the LOS prior, respectively. Since the cluster is in a highly disturbed dynamical state (Section~\ref{subsec:dynamical}), this represents the likely maximum level of hydrostatic bias expected in galaxy clusters. 

Our lensing results, when combined with the X-ray gas mass measurements, yield a direct estimate for the gas mass fraction, free from the assumption of hydrostatic equilibrium. From triaxial lens modeling with the LOS prior, the gas mass fraction enclosed within a sphere of radius $r=0.7\Mpch \sim 0.7 r_{500}$ is found to be $\fgas(<r)=(8.4\pm 1.0)\percent$ (Section~\ref{subsec:fgas}). When the gas mass measurements are extrapolated to $r_{500}$, $\fgas(<r_{500})=(9.6\pm 1.2)\percent$, or $\fgas(<r_{500})/f_\mathrm{b}=0.62\pm 0.08$ relative to the cosmic baryon fraction, $f_\mathrm{b}=\Ob/\Om$ (Figure~\ref{fig:fgas}). These are significantly lower than the typical values of $\fgas/f_\mathrm{b}\sim 0.8$--$0.9$ found in high-mass galaxy clusters \citep{Chiu+2018,Tian2020,Akino2022}. The high degree of gas depletion observed for A370 is in line with the post-major merger scenario of \citet{Molnar2020}.

We have also constructed the projected radial mass profile from an optimally weighted projection of the $\Sigma(\btheta)$ map (Table~\ref{tab:m2d}), obtaining a model-independent constraint on the projected total mass of $\Mproj(<r_\perp)=(3.11\pm 0.47)\times 10^{15}\Msunh$ at $r_\perp\approx 2.3\Mpch\sim 1.2r_\mathrm{vir}$ for the projected mass of the whole system, including any currently unbound material around the cluster.

Combining the data products presented in this work with \HST strong- and weak-lensing data sets available from the Frontier Fields and BUFFALO programs will allow us to conduct a multi-scale lensing reconstruction in the cluster of exceptional projected mass. Such a full-lensing analysis can then be used to detect and study mass substructures in the unique merging environment \citep[e.g.,][]{Jauzac2016a2744,Jauzac2018m0717,Tam+2020wlsl}, for a detailed comparison with the distribution of intracluster baryons. It will also allow us to perform a detailed characterization of the mass profile shape and its deviation from the equilibrium form over a wide radial range, for an improved determination of the total mass bound to the cluster.


\begin{acknowledgements}
We thank the anonymous referee for providing insightful comments and suggestions. We acknowledge fruitful discussions with Mauro Sereno, Mathilde Jauzac, Liliya L.~R. Williams, Nobuhiro Okabe, Jose M. Diego, and Renyue Cen. This work is supported by the Ministry of Science and Technology of Taiwan (grants MOST~106-2628-M-001-003-MY3 and MOST~109-2112-M-001-018-MY3) and by the Academia Sinica Investigator award (grant AS-IA-107-M01). M.N. acknowledges support from grants INAF 1.05.01.86.20 and PRIN MIUR 2017 "Zooming into Dark Matter and proto-galaxies with massive lensing clusters". M.O. acknowledges support from JSPS KAKENHI Grant Nos. JP18K03693, JP20H00181, and JP20H05856.

This paper is based on data collected at Subaru Telescope, which is operated by the National Astronomical Observatory of Japan. This work has made use of data from the European Space Agency (ESA) mission Gaia (\url{https://www.cosmos.esa.int/gaia}), processed by the Gaia Data Processing and Analysis Consortium (DPAC, \url{https://www.cosmos.esa.int/web/gaia/dpac/consortium}). Funding for the DPAC has been provided by national institutions, in particular the institutions participating in the Gaia Multilateral Agreement.
\end{acknowledgements}

\software{IMCAT package \citep{1995ApJ...449..460K}, xspec \citep[v12.11.1;][]{Arnaud96}, Sherpa \citep{Freeman01, Doe07, sherpa_2021_5554957}, CIAO \citep[v4.13;][]{Fruscione06}, PYATOMDB \citep{Foster20}, SCAMP software \citep{SCAMP}, SWARP \citep{Bertin+2002Swarp}, EMCEE \citep{Foreman-Mackey13}, LEPHARE \citep{Ilbert2006}}


\begin{appendix}

\section{The Effect of the Brighter Magnitude Cut on Magnification Bias}
\label{appendix:magslope}

In this study, we applied both bright and faint magnitude cuts to define a magnitude-limited sample of background galaxies (Table~\ref{tab:musample}). Applying an additional bright magnitude cut is to reduce the contamination from unlensed foreground galaxies \citep[][]{Medezinski+2010,Medezinski+2011,Medezinski2018src}. However, it will also modify the signal of magnification bias because magnified source galaxies near the bright cut will be removed from the observed sample. As a result, the net effect of magnification bias includes the contribution from the bright cut as well as from the faint cut \citep{Chiu2020hscmag}.

Following \citet{Chiu2020hscmag}, we obtain the expression for the magnification bias signal expected for a background sample defined in the magnitude range $m_\mathrm{bright}\le m < m_\mathrm{faint}$ as
\begin{equation}
\frac{N_\mu(m_\mathrm{bright}\le m<m_\mathrm{faint})}{\overline{N}_\mu(m_\mathrm{bright}\le m<m_\mathrm{faint})} \simeq 1+\left(5s_\mathrm{eff}-2\right)\kappa,
\end{equation}
where we have used the weak-lensing limit ($\mu\simeq 1+2\kappa$); the quantity $s_\mathrm{eff}$ denotes the effective count slope for a background sample defined in the magnitude range $m_\mathrm{bright}\le m<m_\mathrm{faint}$:
\begin{equation}
 s_\mathrm{eff} = \frac{s(m_\mathrm{faint})-f_\mathrm{bright} s(m_\mathrm{bright})}{1-f_\mathrm{bright}},
\end{equation}
with $f_\mathrm{bright} = \overline{N}_\mu(<m_\mathrm{bright})/\overline{N}_\mu(<m_\mathrm{faint})$.

In typical observations of magnification bias based on deep multi-band imaging \citep{Umetsu2014clash,Chiu2016magbias}, we require $m_\mathrm{bright}$ to be $2$--$3$ magnitudes brighter than $m_\mathrm{faint}$ (Table~\ref{tab:musample}). In this work, we have $f_\mathrm{bright}\approx 1.0\percent$ for the lensing-cut sample and $f_\mathrm{bright}\approx 3.3\percent$ for the null-test sample, so that we can safely ignore the correction terms proportional to $f_\mathrm{mask}$. In the limit $f_\mathrm{bright}\to 0$, we have
\begin{equation}
  \frac{N_\mu(m_\mathrm{bright}\le m<m_\mathrm{faint})}{\overline{N}_\mu(m_\mathrm{bright}\le m<m_\mathrm{faint})} \to \frac{N_\mu(<m_\mathrm{faint})}{\overline{N}_\mu(<m_\mathrm{faint})}.
\end{equation}
In this study, we interpret the magnification bias signal using the approximation $s_\mathrm{eff}\simeq s(m_\mathrm{faint})$.


\section{Photometric Zero Point Calibration}
\label{appendix:CCcalib}


\begin{figure}[tbp]
 \begin{center}
  \includegraphics[width=0.35\textwidth,clip]{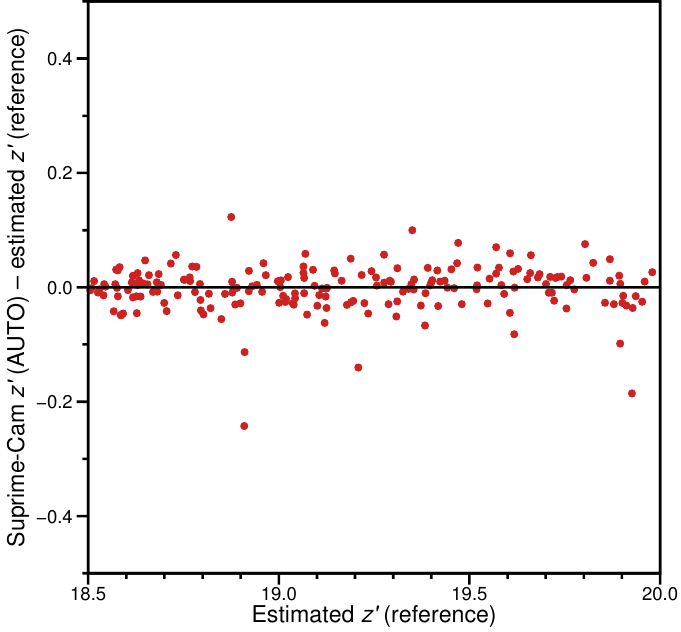}
 \end{center}
\caption{Magnitude difference between the Suprime-Cam $z^\prime$ and the reference $z^\prime$ derived from Pan-STARRS $grizy$ photometry, shown for a calibration sample of point sources with reference $z^\prime$ magnitudes in the range $\in [18.5,20.0]$.}
\label{fig:zcomp}
\end{figure}


\begin{figure*}[tbp]
 \begin{center}
  \includegraphics[width=0.7\textwidth,clip]{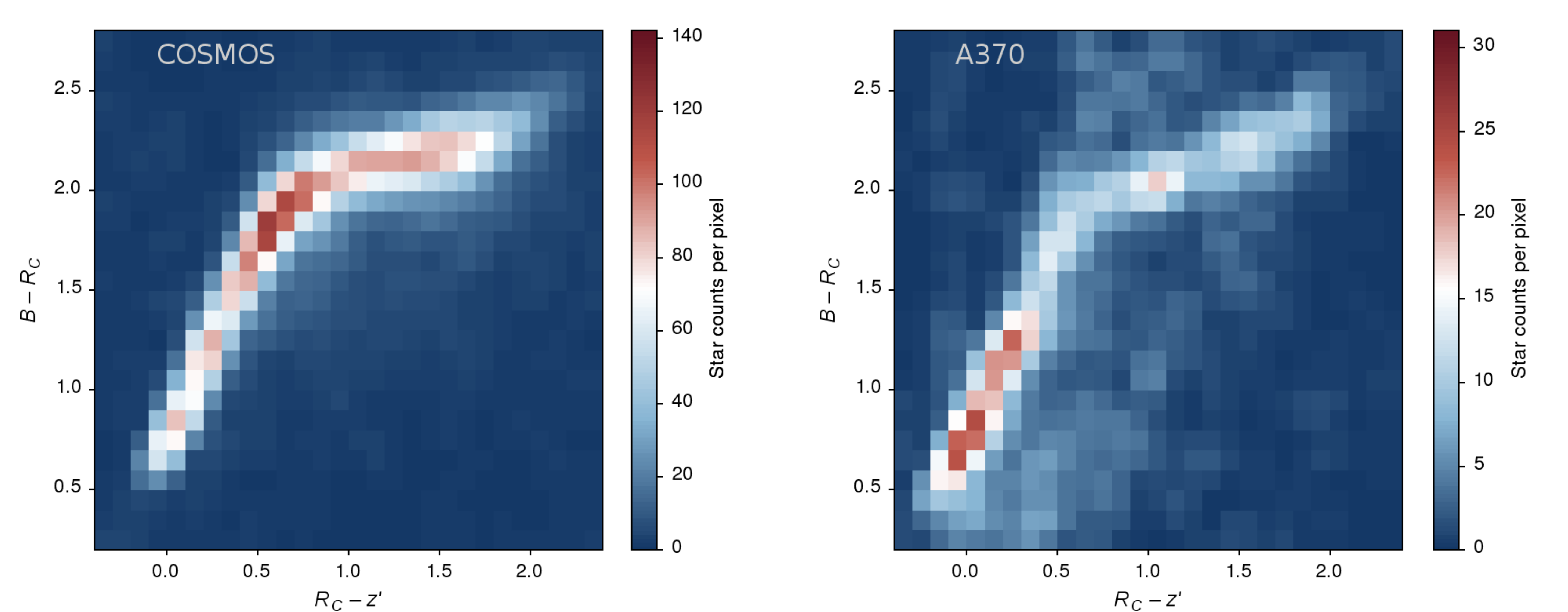}
 \end{center}
\caption{Binned distribution of stars in color--color space for the COSMOS field (left) and A370 (right), after the color matching. The locations of the peak density (redder colors) in the two fields are different because of the field-to-field variations of stellar populations as well as of different magnitude cuts applied to both samples. By contrast, the shapes of the stellar locus in the two fields are similar. Therefore, the cross-correlation technique can be used to match the color distribution of stars in the cluster field to the reference COSMOS photometry.}
\label{fig:slocus}
\end{figure*}

The zero point for the Suprime-Cam $z^\prime$ filter was calibrated by matching \textsc{sextractor}'s AUTO fluxes for point sources to their PSF fluxes from the Pan-STARRS DR1 catalog \citep{PS1DR1}. Since the transmission curves of the Pan-STARRS $z$ filter and the Suprime-Cam $z^\prime$ filter are different, we followed the procedure of \citet{Umetsu+2010CL0024} to infer $z^\prime$-band fluxes from the Pan-STARRS photometry. We use the \textsc{hyperz} code \citep[New-Hyperz ver.~11;][]{hyperz} to perform a spectral energy distribution (SED) fitting to Pan-STARRS $grizy$ photometry, with stellar templates from the Pickles library \citep{Pickles1998}. Pan-STARRS's $z^\prime$ fluxes were obtained using the transmission curve of the Suprime-Cam $z^\prime$ filter. Point sources with Pan-STARRS's $z^\prime$ magnitudes in the range $[18.5,20.0]$ are used for calibration, because stars with $z^\prime$ brighter than $18.5$~mag are saturated. Figure~\ref{fig:zcomp} compares the calibrated Suprime-Cam $z^\prime$ magnitudes and the reference $z^\prime$ magnitudes derived from Pan-STARRS $grizy$ photometry. The residual rms scatter is 0.034~mag. 

The zero points for the Suprime-Cam $B$ and $R_\mathrm{C}$ filters were first derived by matching the stellar locus in the $B-R_\mathrm{C}$ vs. $R_\mathrm{C}-z^\prime$ diagram to the COSMOS2020 photometry \citep{COSMOS2020}. The uniformity of the colors of Galactic stars permits a reliable color calibration between fields with $|b| > 30\ensuremath{^\circ}$ \citep[for detail, see][]{Gilbankd2011}. To this end, we use isophotal fluxes for better color measurements. Since the COSMOS photometry does not cover the Suprime-Cam $R_\mathrm{C}$ band, we needed to estimate $R_\mathrm{C}$ magnitudes for COSMOS field objects. We use again the \textsc{hyperz} code with SEDs from the Pickles library \citep{Pickles1998} to obtain the best-fit model for each star using COSMOS2020 isophotal photometry in 4 Suprime-Cam intermediate bands $(IB574, IA624, IA679, IB709)$. The $R_\mathrm{C}$ photometry was derived using the transmission curve of the Suprime-Cam $R_\mathrm{C}$ filter. Since the wavelength coverage of the Suprime-Cam $R_\mathrm{C}$ band is well sampled by the 4 Suprime-Cam intermediate filters, this $R_\mathrm{C}$ estimation is regarded as an interpolation of data. The $R_\mathrm{C}$ magnitudes obtained with this method are thus model independent and sufficiently accurate for our purpose \citep[see][]{Umetsu+2010CL0024}. The matching was performed using the cross correlation in CC space between our Suprime-Cam data in A370 and the COSMOS2020 data. Figure~\ref{fig:slocus} shows the result of color matching. Once the color offsets are determined, the zero points for the Suprime-Cam $B$ and $R_\mathrm{C}$ filters were derived from the color offsets.

We repeat the same procedure to obtain Suprime-Cam $R_\mathrm{C}$ magnitudes for galaxies in the COSMOS field. The SED fitting for each COSMOS galaxy was performed with the \textsc{hyperz} code by using spectral templates from the GALAXEV library \citep{2003MNRAS.344.1000B} and by fixing the redshift to each photometric redshift (computed with the \LePhare code; \citealt{Ilbert2006}) from the COSMOS2020 \textsc{farmer} catalog. A Galactic extinction correction was applied to galaxies in both data sets according to \citet{1998ApJ...500..525S}. By matching the distributions of galaxies in the $B-R_\mathrm{C}$ vs. $R_\mathrm{C}-z^\prime$ diagram for both data sets, we find that additional offsets of $-0.02$~mag and $+0.10$~mag need to be added to the Suprime-Cam $B$ and $R_\mathrm{C}$ magnitudes, respectively. We have corrected for the residual offsets by adding $-0.02$~mag and $+0.10$~mag to the respective magnitudes in our data set. These additional offsets may be due to the bias in our Suprime-Cam $R_\mathrm{C}$ estimation based on the SED fitting and the uncertainty of the cross correlation matching.

\section{Weak-lensing Magnification Analysis}
\label{appendix:magbiasdata}

Here we detail our magnification analysis. Following the procedure outlined in \citet{Umetsu2014clash,Umetsu2016clash}, we account for the Poisson, intrinsic clustering, and additional systematic contributions to the total uncertainty $\sigma_\mu$ (Section~\ref{subsec:magbias}). First, we estimate $\sigma_{\mu,i}^\mathrm{int}$ dominated by intrinsic clustering from the azimuthal variation of the counts in cell. A positive tail of $>\nu\sigma$ cells is then removed in each bin using iterative $\sigma$ clipping with $\nu=2$. This is to alleviate the bias due to angular clustering of red galaxies. The Poisson noise term $\sigma_{\mu,i}^\mathrm{stat}$ is estimated from the clipped mean counts in each annular bin. The difference between the mean counts estimated with and without $\sigma$ clipping is taken as a systematic error,  $\sigma_{\mu,i}^\mathrm{sys}=|n_{\mu,i}^{(\nu)}-n_{\mu,i}^{(\infty)}|/\nu$, with $n_{\mu,i}^{(\nu)}$ and $n_{\mu,i}^{(\infty)}$ the clipped and unclipped mean counts in the $i$th annulus, respectively. Finally, these errors are combined in quadrature as 
\begin{equation}
 \label{eq:sigma_mu}
 \sigma_{\mu,i}^2 = (\sigma_{\mu,i}^\mathrm{int})^2 +
  (\sigma_{\mu,i}^\mathrm{stat})^2 +(\sigma_{\mu,i}^\mathrm{sys})^2.
\end{equation}
Our magnification analysis is insensitive to the particular choice of $\nu$ because of the inclusion of the $\sigma_\mu^\mathrm{sys}$ term \citep{Umetsu2014clash,Umetsu2016clash}. Note that by including the $\sigma_\mu^\mathrm{stat}$ term in Equation~(\ref{eq:sigma_mu}), we are in effect double counting the contribution of Poisson fluctuations in estimating the errors. We find that including the $\sigma_\mu^\mathrm{stat}$ term increases the estimated total uncertainty by $10\percent$--$20\percent$. This slight overestimate of the uncertainty is not expected to significantly affect our joint mass reconstruction, because our lensing constraints are dominated by the shear measurements.

Masking of observed sky is corrected for using the method of \citet[][Method B of Appendix A]{Umetsu+2011}, which is fully automated once the configuration parameters of \textsc{sextractor} \citep{SExtractor} are  optimally tuned \citep{Umetsu2014clash,Umetsu2016clash,Chiu2016magbias}. We find that the masked area fraction $f_\mathrm{mask}$ is $\sim 4\percent$ of the sky at $\theta>10\arcmin$, increasing toward the cluster center up to $\sim 10\percent$ at $\theta\simlt 2\arcmin$. The masked area fraction averaged over the radial range $\theta\in [\theta_\mathrm{min},\theta_\mathrm{max}]$ is $\approx 4.6\percent$. This is similar to the results of \citet[][]{Umetsu2016clash} for the CLASH sample at the median redshift $\overline{\zl}\approx 0.35$. 

The mask-corrected magnification bias profile $b_{\mu,i}=n_{\mu,i}/\overline{n}_\mu$ is proportional to $(1-f_\mathrm{mask,back})/(1-f_{\mathrm{mask},i})\equiv 1+\Delta f_{\mathrm{mask},i}$ with $f_{\mathrm{mask,back}}$ estimated in the reference background region at $\theta\in [12\arcmin,16\arcmin]$. Thus, the mask correction essentially depends on the difference of the $f_\mathrm{mask}$ values, $\Delta f_{\mathrm{mask},i}\simeq f_{\mathrm{mask},i}-f_\mathrm{mask,back}$, which is insensitive to the particular choice of the configuration parameters for source extraction.
Accordingly, the systematic uncertainty on the mask correction is not expected to significantly bias our magnification measurements.

\section{Two-dimensional to One-dimensional Projection}
\label{appendix:2dto1d}

To enable a direct comparison between the results from 1D and 2D mass reconstructions, we construct a surface mass density profile $\Sigma(\theta)$ from an optimally weighted projection of the $\Sigma(\btheta)$ field as \citep{Umetsu2015} 
\begin{equation}
 \bSigma_{(1)}=
\left[
 A^t C_{(2)}^{-1}A
\right]^{-1} A^t C_{(2)}^{-1} \bSigma_{(2)},
\end{equation}
where $\bSigma_{(2)}=\{\Sigma(\btheta_m)\}_{m=1}^{\Npix}$ is a pixelized mass map, $C_{(2)}$ is the pixel-to-pixel covariance matrix of $\bSigma_{(2)}$, $\bSigma_{(1)}$ is a data vector containing radially binned $\Sigma$ values, and $A$ is a mapping matrix whose elements $A_{mi}$ represent the area fraction of the $m$th pixel lying within the $i$th clustercentric radial bin (Section \ref{subsec:magbias}). The bin-to-bin covariance matrix for $\bSigma_{(1)}$ is given by
\begin{equation}
C_{(1)}=\left[A^tC_{(2)}^{-1}A\right]^{-1}.
\end{equation}

\section{Chandra X-ray Brightness Profiles}
\label{appendix:sx}

\begin{figure*}[tbp]
 \begin{center}
  \includegraphics[width=0.24\textwidth,clip]{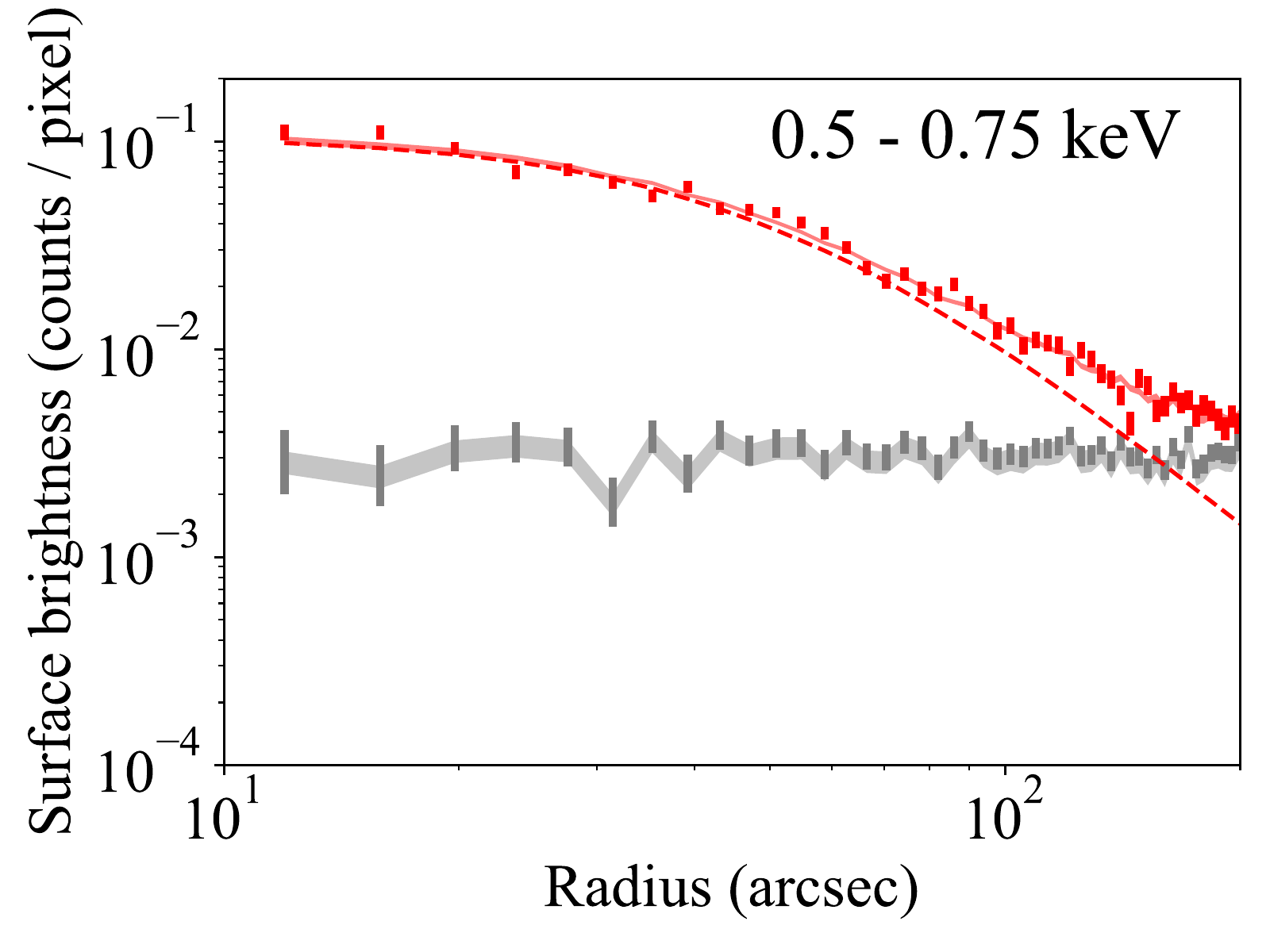}
  \includegraphics[width=0.24\textwidth,clip]{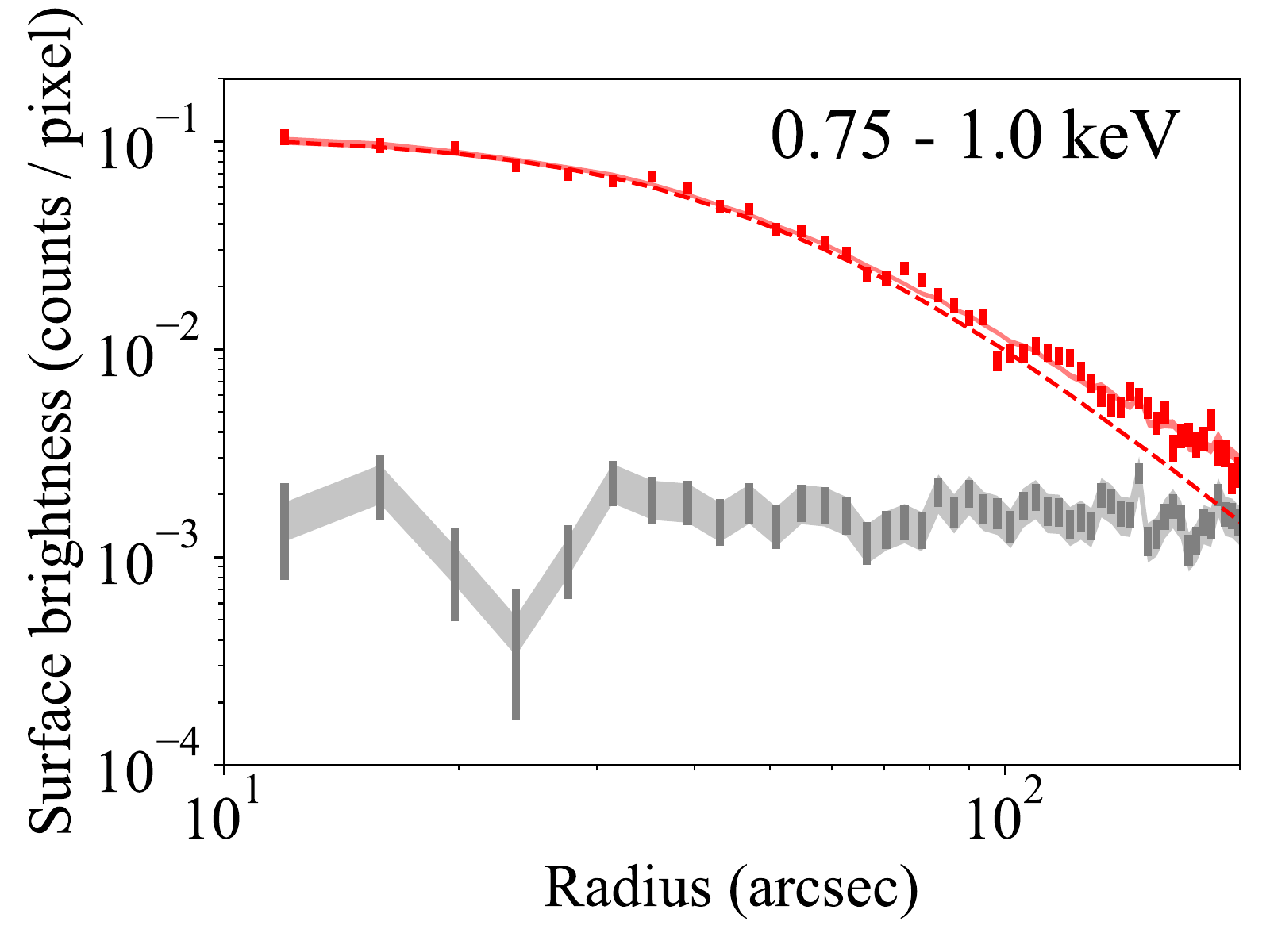}
  \includegraphics[width=0.24\textwidth,clip]{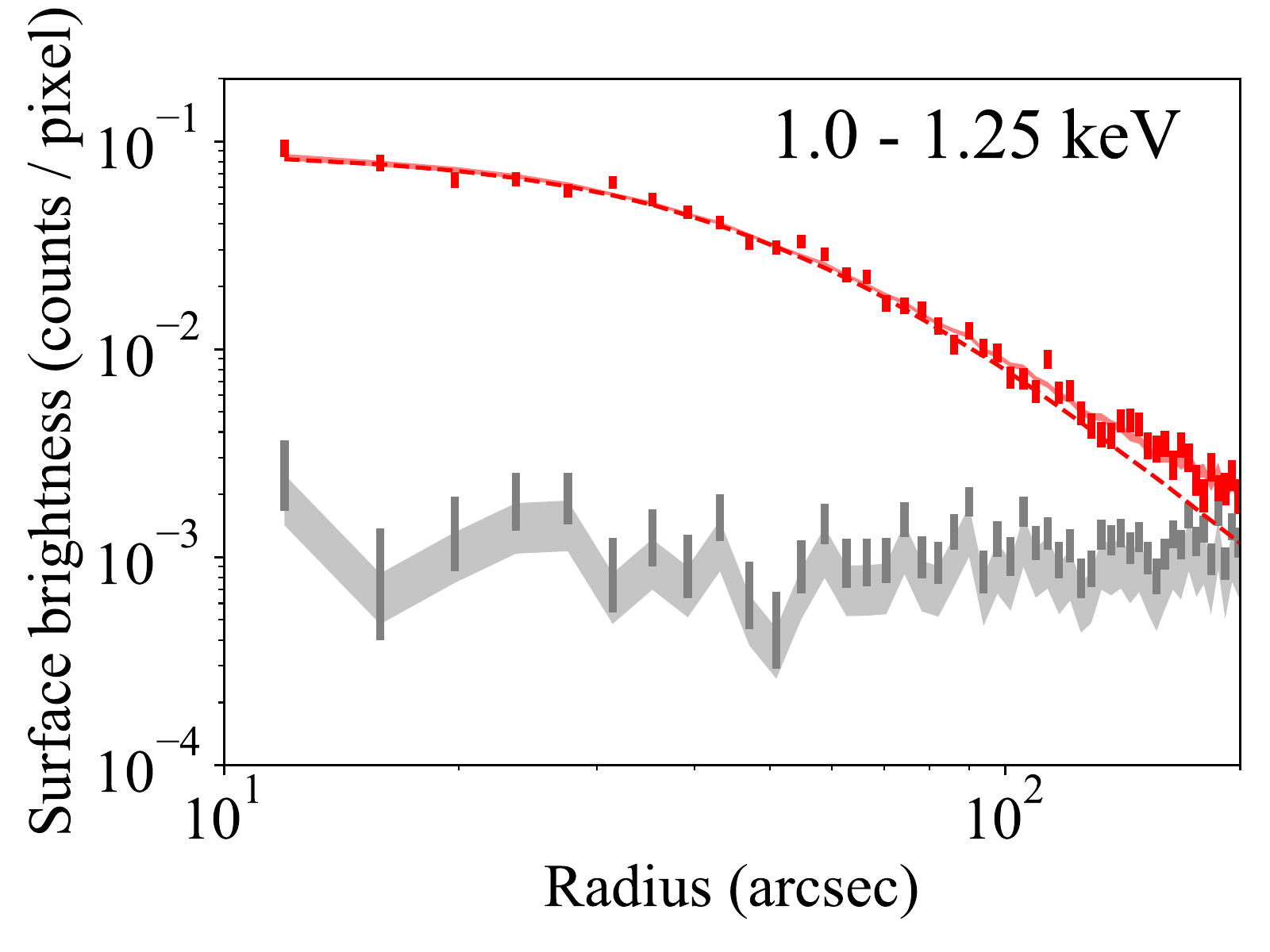}
  \includegraphics[width=0.24\textwidth,clip]{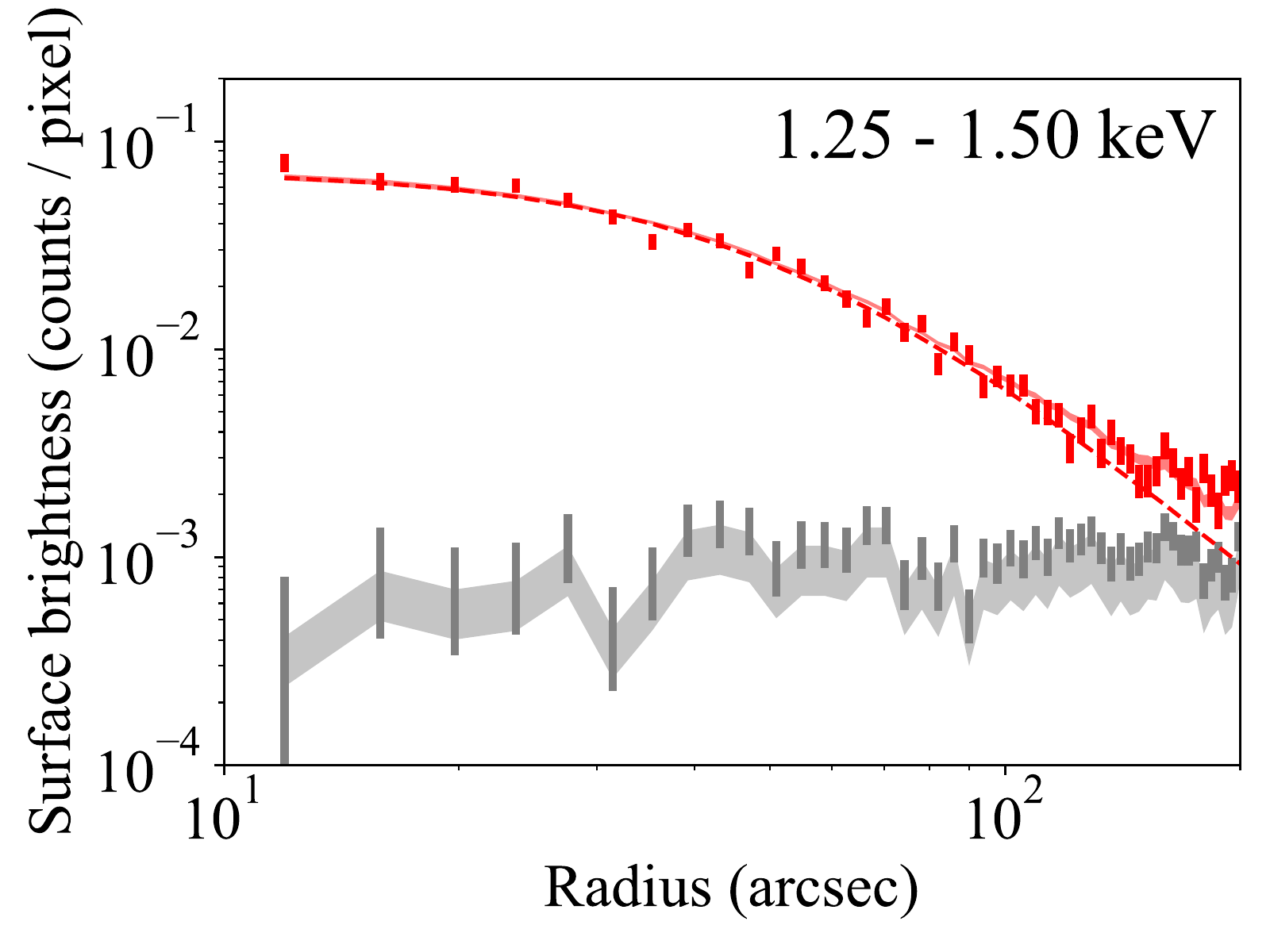}
  \includegraphics[width=0.24\textwidth,clip]{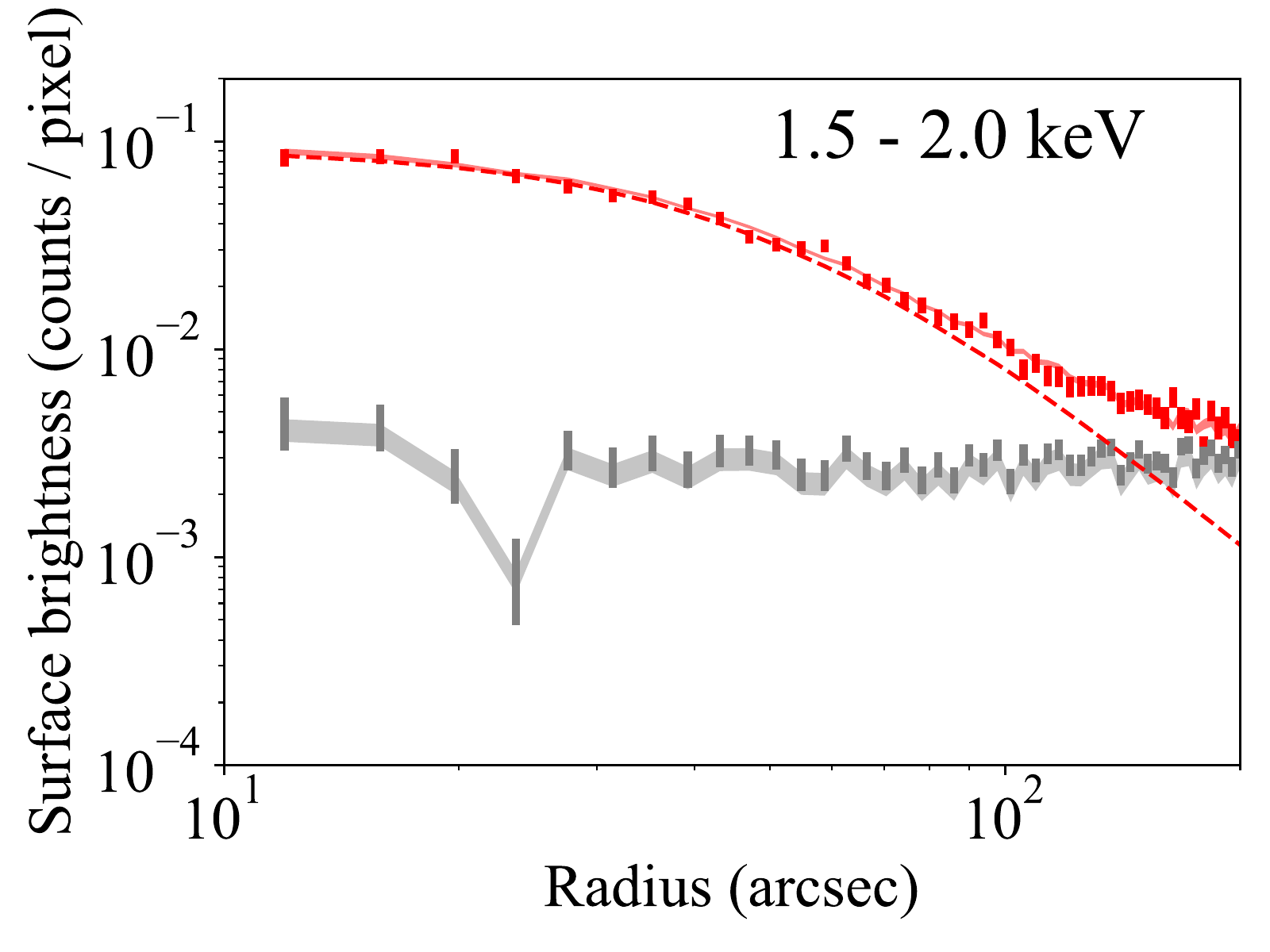}
  \includegraphics[width=0.24\textwidth,clip]{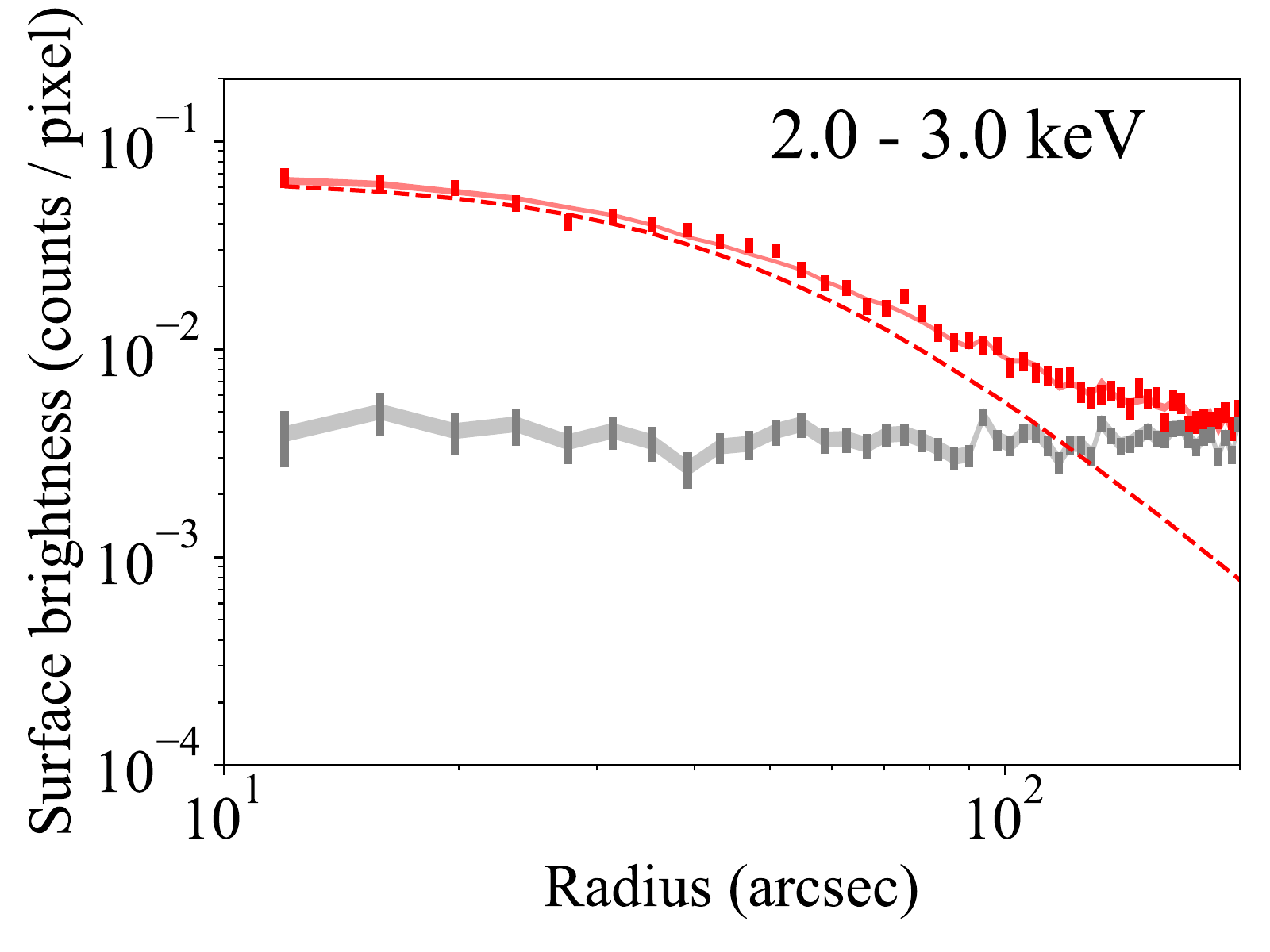}
  \includegraphics[width=0.24\textwidth,clip]{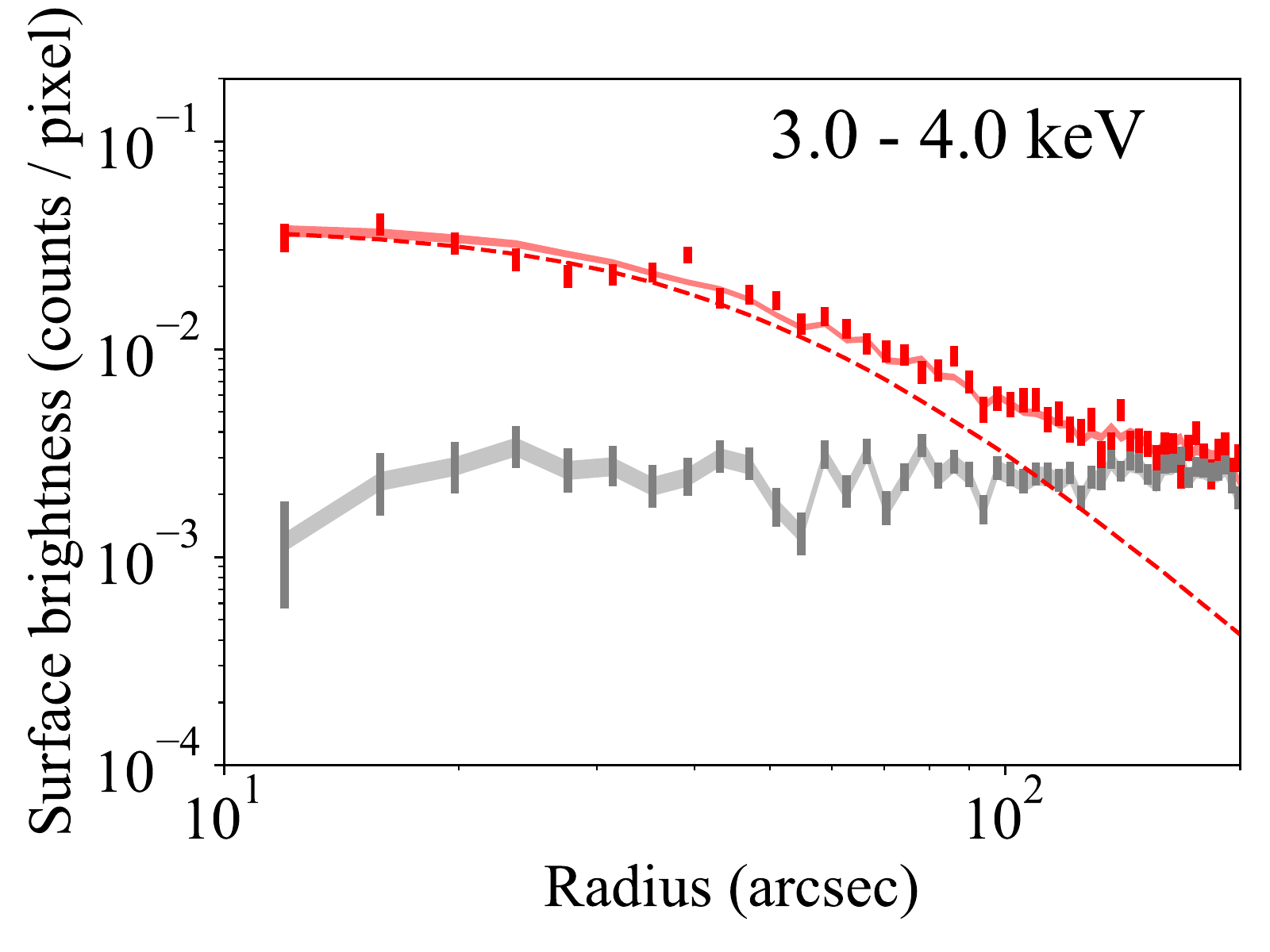}
  \includegraphics[width=0.24\textwidth,clip]{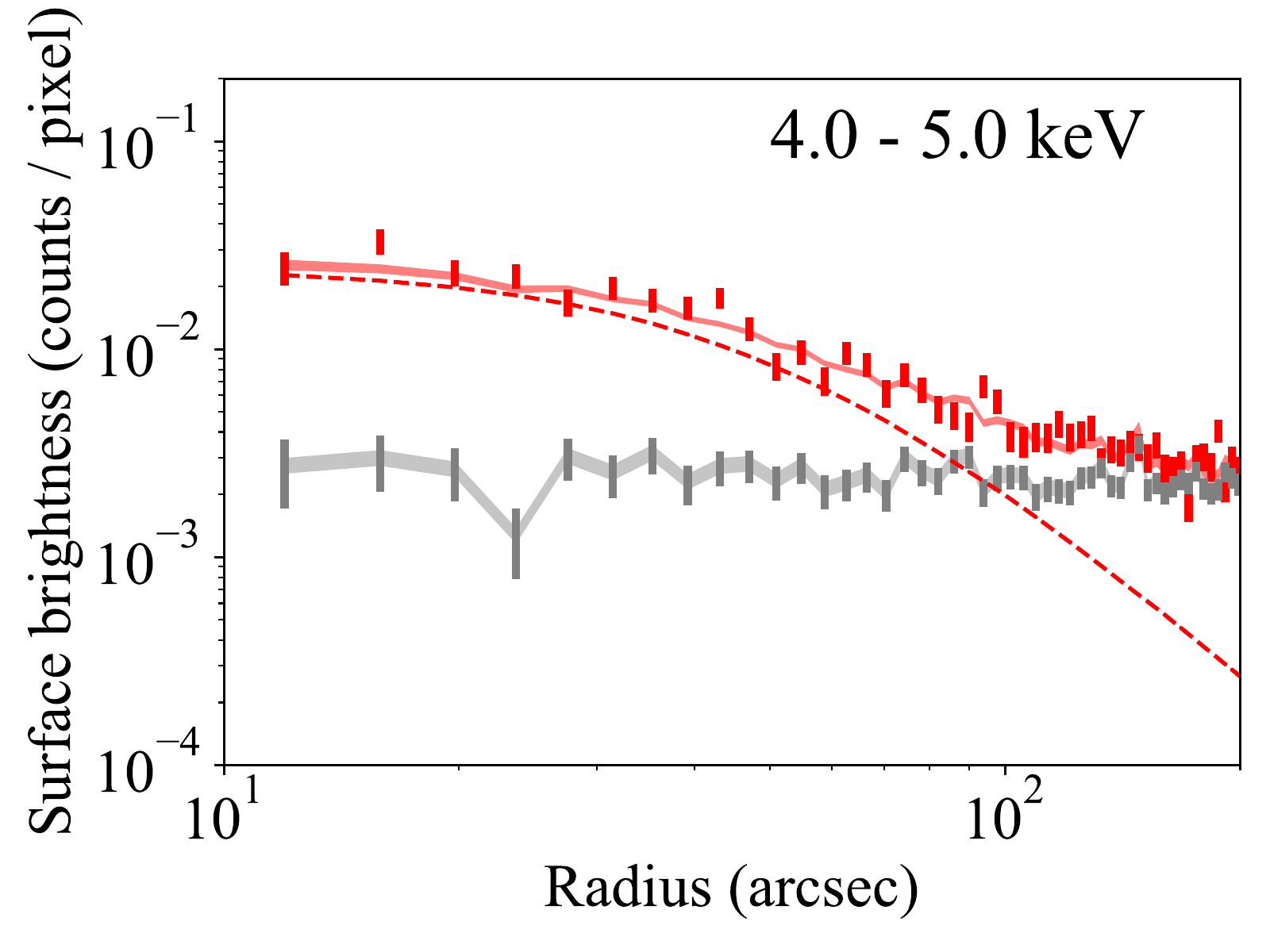}
  \includegraphics[width=0.24\textwidth,clip]{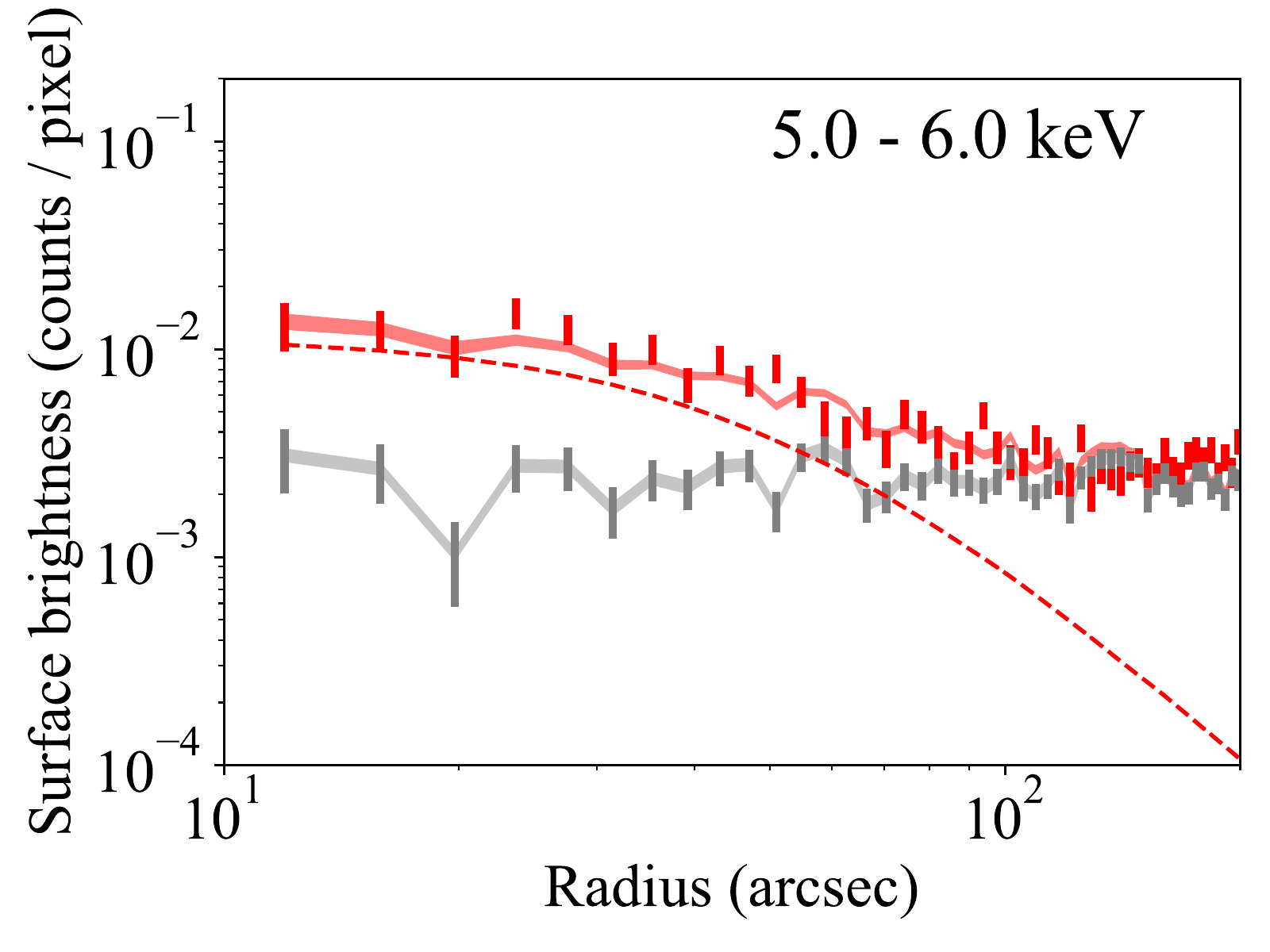}
  \includegraphics[width=0.24\textwidth,clip]{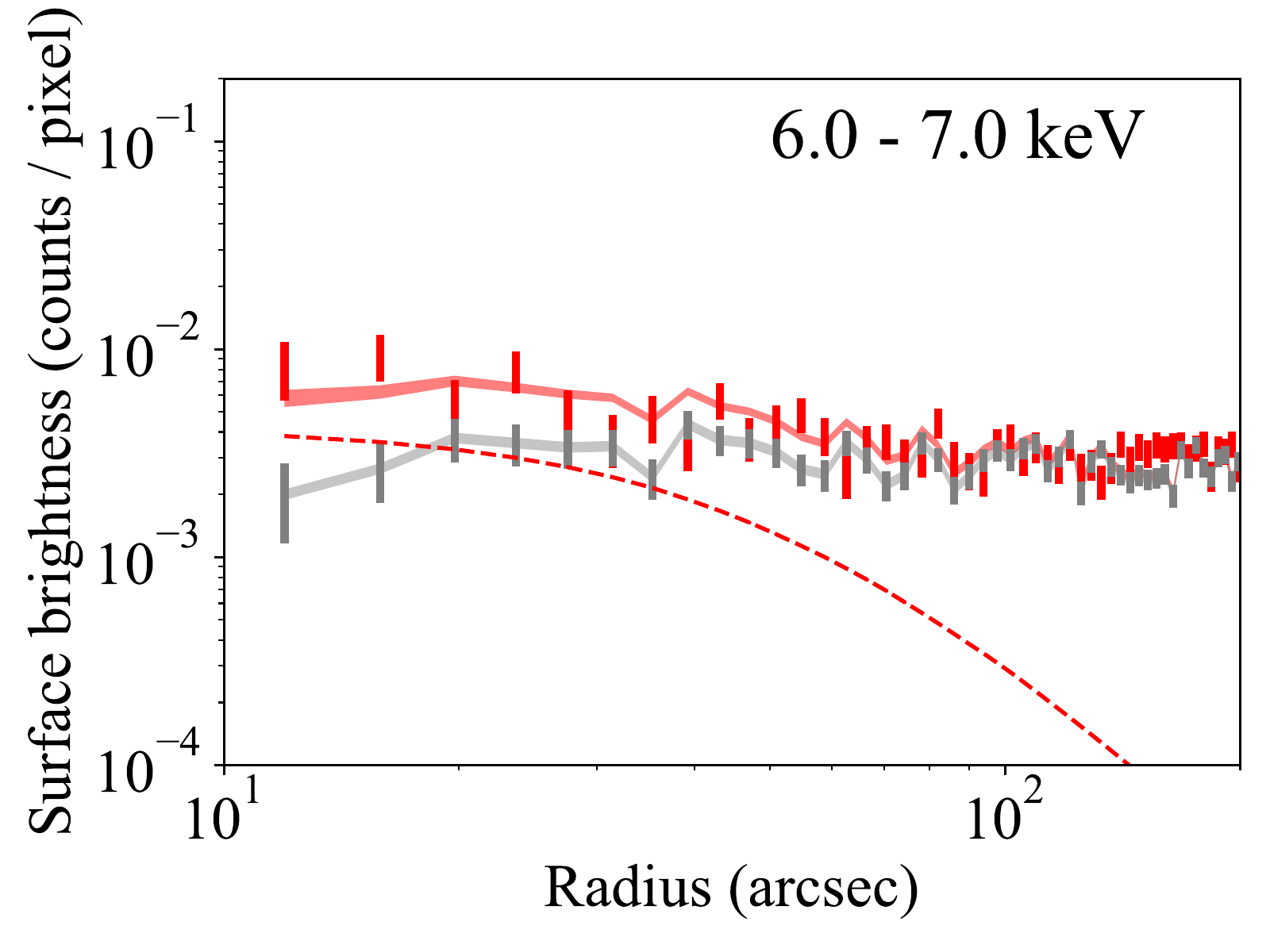}
 \end{center}
\caption{Radial X-ray surface brightness profiles of A370 measured in 10 energy bands from \Chandra observations. Red (gray) vertical bars in each panel show the $1\sigma$ confidence range of the total (background) counts in radial bins. The red (gray) shaded area in each panel shows the marginalized $2\sigma$ confidence region of the model for the total (background) counts obtained from simultaneous forward fitting of the 10-band brightness profiles. The red dashed line in each panel shows the posterior mean profile of the cluster contribution.}
\label{fig:sx}
\end{figure*}

Figure~\ref{fig:sx} shows the radial X-ray surface brightness profiles of A370 measured in 10 energy bands ($0.5$ to $7$~keV) from \Chandra observations. The binned total and background X-ray brightness profiles in each energy band are plotted in each panel, along with the best-fit model derived from simultaneous forward modeling of the 10 energy bands.


\end{appendix}

\clearpage




\end{document}